%% file: ApJ.tex
\documentclass[twocolumn]{aastex701}

\newcommand{\addref}[1]{\textcolor{blue}{[ADD REF]}}
\newcommand{\checknote}[1]{\textcolor{blue}{[CHECK]}}
\newcommand{\complete}[1]{\textcolor{blue}{[COMPLETE]}}

\newcommand{\review}[1]{#1}
\usepackage{booktabs}
\usepackage{amsmath}
\usepackage{subcaption}

\begin{document}

\title{Tabular foundation models for the estimation of probabilistic quasar photometric redshifts in S-PLUS}

\author[orcid=0000-0001-8847-0047]{Raquel R. Valen\c{c}a}
\affiliation{Instituto de Astronomia, Geof\'isica e Ci\^encias Atmosf\'ericas, Universidade de S\~ao Paulo, S\~ao Paulo, SP, Brazil}
\email[show]{raquelrvalenca@gmail.com}

\author[orcid=0000-0001-6480-1155]{Lilianne Nakazono}
\affiliation{Observat\'orio Nacional/MCTI, Rio de Janeiro, RJ, Brazil}
\email{liliannenakazono@on.br}

\author[orcid=0000-0003-0379-9690]{Rafael Izbicki}
\affiliation{Departamento de Estat\'istica, Universidade Federal de S\~ao Carlos, S\~ao Carlos, SP, Brazil}
\email{rizbicki@ufscar.br}

\author[orcid=0000-0002-6865-5404]{Marco Henrique de Almeida In\'acio}
\affiliation{Cloudflare, London, United Kingdom}
\email{m@marcoinacio.com}

\author[orcid=0009-0008-0034-7983]{Bruno Marcondes e Resende}
\affiliation{Departamento de Estat\'istica, Universidade Federal de S\~ao Carlos, S\~ao Carlos, SP, Brazil}
\email{bruno.resende@estudante.ufscar.br}

\author[orcid=0009-0006-7823-225X]{Maycon Jorge Del\'{a}qua da Silva}
\affiliation{Engenharia da Computa\c{c}\~ao, Instituto Federal Fluminense, Campus Bom Jesus do Itabapoana, Bom Jesus do Itabapoana, RJ, Brazil}
\email{maycon.delaqua@gsuite.iff.edu.br}

\author[orcid=0009-0007-2849-9025]{Kiana Coimbra Buin Lins}
\affiliation{Observat\'orio Nacional/MCTI, Rio de Janeiro, RJ, Brazil}
\email{kianalins@on.br}

\author[orcid=0000-0002-2238-9665]{Natanael M. Cardoso}
\affiliation{Escola Politécnica, Universidade de São Paulo, São Paulo, 05508-010, SP, Brasil}
\email{natanael.mc@usp.br}




\author[orcid=0000-0002-7736-4297]{Claudia Mendes de Oliveira}
\affiliation{Instituto de Astronomia, Geof\'isica e Ci\^encias Atmosf\'ericas, Universidade de S\~ao Paulo, S\~ao Paulo, SP, Brazil}
\email{}

\collaboration{all}{The S-PLUS collaboration}

\begin{abstract}
\input{abstract}
\end{abstract}


\section{Introduction}
\input{introduction}
\section{Data}
\input{data}
\section{Methods}
\input{methods}
\section{Results}
\input{results}
\section{Conclusions and future work}
\input{conclusions}


\begin{acknowledgments}

The S-PLUS project, including the T80-South robotic telescope and the S-PLUS scientific survey, was founded as a partnership between the Fundação de Amparo à Pesquisa do Estado de São Paulo (FAPESP), the Observatório Nacional (ON), the Federal University of Sergipe (UFS), and the Federal University of Santa Catarina (UFSC), with important financial and practical contributions from other collaborating institutes in Brazil, Chile (Universidad de La Serena), and Spain (Centro de Estudios de Física del Cosmos de Aragón, CEFCA). We further acknowledge financial support from the São Paulo Research Foundation (FAPESP) grant 2019/263492-3, the Brazilian National Research Council (CNPq), the Coordination for the Improvement of Higher Education Personnel (CAPES), the Carlos Chagas Filho Rio de Janeiro State Research Foundation (FAPERJ), and the Brazilian Innovation Agency (FINEP). The members of the S-PLUS collaboration are grateful for the contributions from CTIO staff in helping in the construction, commissioning and maintenance of the T80-South telescope and camera. This research uses services or data provided by the SPectra Analysis and Retrievable Catalog Lab (SPARCL), which is part of the Community Science and Data Center (CSDC) program at NSF National Optical-Infrared Astronomy Research Laboratory. NOIRLab is operated by the Association of Universities for Research in Astronomy (AURA), Inc. under a cooperative agreement with the National Science Foundation. Funding for the SDSS and SDSS-II has been provided by the Alfred P. Sloan Foundation, the Participating Institutions, the National Science Foundation, the U.S. Department of Energy, the National Aeronautics and Space Administration, the Japanese Monbukagakusho, the Max Planck Society, and the Higher Education Funding Council for England. The SDSS Web Site is http://www.sdss.org/.

The SDSS is managed by the Astrophysical Research Consortium for the Participating Institutions. The Participating Institutions are the American Museum of Natural History, Astrophysical Institute Potsdam, University of Basel, University of Cambridge, Case Western Reserve University, University of Chicago, Drexel University, Fermilab, the Institute for Advanced Study, the Japan Participation Group, Johns Hopkins University, the Joint Institute for Nuclear Astrophysics, the Kavli Institute for Particle Astrophysics and Cosmology, the Korean Scientist Group, the Chinese Academy of Sciences (LAMOST), Los Alamos National Laboratory, the Max-Planck-Institute for Astronomy (MPIA), the Max-Planck-Institute for Astrophysics (MPA), New Mexico State University, Ohio State University, University of Pittsburgh, University of Portsmouth, Princeton University, the United States Naval Observatory, and the University of Washington.

R. R. V. acknowledges the support from CAPES (grant 88887.821818/2023-00) and FAPESP (grants 2023/05003-0 and 2024/16592-9). L. N. acknowledges FAPESP grant number 2024/07281-0. R. I. is grateful for the financial support of CNPq (grants  305065/2023-8 and 403458/2025-0) and FAPESP (grant 2023/07068-1). B. M. e R.  is grateful for the financial support of FAPESP (grant 2025/04853-5). N. M. C. thanks the Coordenação de Aperfeiçoamento de Pessoal de Nível Superior -- Brasil (CAPES) -- Finance Code 88887.133104/2025-00.

\review{The spectroscopic compilation from \citet{erik_v_2025_15127060} used in this work is publicly available and S-PLUS DR6 photometry will be made public at \url{https://www.splus.cloud} following the survey's data release schedule. The code used to preprocess the data, run the benchmark, and produce the tables and figures in this paper is available at \url{https://github.com/rizbicki/qsoSplusShift/}.}

\end{acknowledgments}

\appendix
\input{appendix}

\bibliography{ApJ}{}
\bibliographystyle{aasjournalv7}



\end{document}

%% file: abstract.tex
We assess whether tabular foundation models can be used as off-the-shelf probabilistic photometric-redshift estimators for quasars in the 12-band S-PLUS DR6 survey, where colour-redshift degeneracies produce multi-modal posteriors and spectroscopic training sets are shifted relative to the photometric population. TabPFN 2.5, RealTabPFN 2.5, and TabICL are benchmarked against eight task-specific baselines, including linear conditional Gaussians, FlexZBoost, mixture-density networks, normalising flows, random forests, and gradient-boosted trees, with training sets from 500 to 121,626 quasars, using both density and point-prediction metrics, together with importance-weighted scores that approximate deployment on the photometric target sample. \review{TabPFN 2.5 is best or statistically tied for best on all metrics except the unweighted CDE loss, on which the normalising flow is statistically tied and attains the lowest mean value}; its largest gains occur for small training sets and in difficult regimes (very bright and faint sources, high redshift), while retaining near-nominal calibration under covariate shift. \review{Its main practical cost is inference: with frozen weights, large support and target catalogues require substantial GPU/accelerator memory, and full-catalogue deployment may need support-set subsampling or distillation.} SHAP attributions identify WISE W1/W2 as the strongest individual predictors, with UV and optical bands offering non-negligible refinements. We conclude that TabPFN 2.5 is a strong default for probabilistic quasar photo-z estimation, \review{particularly when training data are limited or when calibration under covariate shift is critical.}

%% file: introduction.tex
\label{sec:intro}

Redshift is the single most important quantity attached to an extragalactic source: without it, observed photometry cannot be turned into rest-frame luminosities, distances, or cosmic ages.  This is especially true for quasars, the luminous active galactic nuclei powered by accretion onto supermassive black holes, which can be detected over essentially the full observable Universe and therefore underpin studies of black-hole growth, the intergalactic medium, large-scale structure, and the high-redshift Universe \citep[e.g.][]{2001ARA&A..39...19L, 2014ARA&A..52..589H, 2016ASSL..423..187M, 2023ARA&A..61..373F}.  Spectroscopic redshifts are accurate but expensive to obtain at survey scale, so most quasars in modern wide-field imaging surveys must be assigned redshifts photometrically.

For quasars, however, returning a single point estimate $\hat{z}(\mathbf{x})$ from the observed photometric feature vector $\mathbf{x}$ (typically broad- and narrow-band magnitudes, colours, and their associated uncertainties) is often inadequate \citep{2016arXiv160808016P}.  Broad-band colours of quasars are shaped by a smooth continuum, broad emission lines moving through the filters, host-galaxy contamination, dust reddening, variability, and photometric noise; as a result, the mapping from $\mathbf{x}$ to $z$ ($z$ referring to redshift) is genuinely multi-valued, with quasars at different redshifts populating overlapping regions of colour space and quasars at the same redshift showing appreciably different observed colours \citep[e.g.][]{2001AJ....121.2308R, 2017ApJ...849...53H, 2021MNRAS.507.5847N}.  
 \review{A concrete S-PLUS illustration is SDSS J232043.35-003049.3, for which FlexCoDE produces two well-separated redshift modes, with the spectroscopic solution associated with the secondary rather than the primary peak \citep[Fig.~6]{2024MNRAS.531..327N}.}

The natural object to estimate is therefore the full conditional density $p(z \mid \mathbf{x})$, which can represent multimodality, asymmetric uncertainty, and catastrophic-redshift risk \citep[e.g.][]{Brammer+2008, Hildebrandt+2010} in a single object.  \review{At catalogue level, retaining these densities allows ambiguous objects to
contribute probabilistically across redshift, rather than being assigned to a
single bin, thereby propagating photo-$z$ uncertainty into quasar clustering,
lensing statistics, and luminosity-function inference
\citep{Weinstein2004,2009MNRAS.399.2279M,mandelbaum2008precision,2009MNRAS.399.2279M, 2016MNRAS.459.1293A,2023MNRAS.524.5109R}.}

The standard route to $p(z \mid \mathbf{x})$ has been task-specific machine-learning models trained from scratch on each survey's spectroscopic sample.  Across the last decade of photo-$z$ (namely, photometric redshift) benchmarks, a relatively stable picture has emerged: tree-based ensembles---random forests and gradient-boosted trees---tend to dominate point-prediction accuracy, while specialised conditional density estimators such as FlexZBoost and mixture density networks tend to dominate density-quality and calibration metrics, and no single method wins on every figure of merit \citep{Bishop1994,2008ApJ...683...12B, IzbickiLee2017}. This pattern has been confirmed in survey-scale comparisons including the LSST-DESC photo-$z$ data challenges \citep{schmidt2020evaluation}, applications to the Sloan Digital Sky Survey (SDSS; \citealt{2000AJ....120.1579Y}) and the Dark Energy Spectroscopic Instrument (DESI;  \citealt{2016arXiv161100036D}, \citealt{2026AJ....171..285D}) quasar samples, and dedicated quasar photo-$z$ pipelines for narrow-band surveys.  In particular, the QuCatS analysis of the Southern Photometric Local Universe Survey (S-PLUS) data release 4 (DR4; \citealt{, 2024A&A...689A.249H}) systematically compared random forests, FlexCoDE, and mixture density networks for quasar photo-$z$ estimation in this 12-band system \citep{2024MNRAS.531..327N}, and provides the most direct baseline for the present work.

\review{Early applications of machine learning to quasar photometric redshift estimation date back to neural network approaches, such as the one developed by \citet{Brescia2013}, who demonstrated competitive performance on multi-band surveys while already highlighting the challenges posed by colour--redshift degeneracies and the value of full probabilistic outputs.}

At the same time, the recent emergence of tabular foundation models
changes the set of plausible tools for this problem.  Models such as TabPFN and
TabICL use transformer architectures pre-trained on large collections of
synthetic tabular tasks and are then applied in an in-context mode: the
labelled training set is supplied as support, the model weights remain
frozen, and predictions are produced without survey-specific training
\citep{Hollmann2025,Qu2025}. Although these models were initially
evaluated mainly as point predictors, both prior-fitted networks and
tabular in-context learners can return object-by-object distributional
outputs. Recent work has therefore begun to use them as conditional
density estimators for generic regression problems
\citep{izbicki2026benchmarking,landsgesell2026distributional}. This makes them
natural candidates for probabilistic photo-$z$ estimation, but their
suitability for quasar redshifts is not yet established.

Quasar photo-$z$ estimation differs from standard tabular regression in
ways that directly test the assumptions behind off-the-shelf foundation
models. The first issue is population shift. The spectroscopic sample is
not a random draw from the photometric catalogue: it is shaped by survey
footprints, magnitude limits, colour cuts, target-selection algorithms,
and follow-up priorities \citep{cunha2009estimating,freeman2017unified}. As a result, the
feature distribution of labelled quasars can differ substantially from
that of the photometric quasar-candidate population on which the model
will be deployed. Standard test-set averages over spectroscopic objects
therefore estimate performance under the spectroscopic selection
function, not necessarily under the target photometric population.
Importance weighting under a covariate-shift assumption provides a
principled way to approximate target-population performance
\citep{Shimodaira2000,Bickel2009,izbicki2017photo}, but the behaviour of
tabular in-context learners under this type of shift remains largely
unexplored for conditional density estimation.

The second issue is the loss structure of the scientific problem. In
generic regression benchmarks, methods are often compared using point
losses such as mean squared error. For quasar photo-$z$, however, the
quality of the full predictive distribution is central: colour--redshift
degeneracies produce multimodal posteriors, sparse regions of colour
space can dominate high-redshift science cases, and miscalibrated tails
can lead to catastrophic redshift failures. A method that performs well
on average point error can still be inadequate if it assigns too little
probability to plausible secondary redshift solutions or produces
overconfident tails.

In this paper, we therefore ask whether frozen, off-the-shelf tabular
foundation models can deliver calibrated, multimodal redshift densities
for quasars in the 12-band S-PLUS system, and whether their apparent
performance remains stable when evaluation is shifted from the
spectroscopic test distribution toward the photometric target
population. We benchmark TabPFN 2.5, RealTabPFN 2.5, and TabICL against
linear, tree-based, and neural conditional-density baselines, including
the methods used in the previous S-PLUS quasar photo-$z$ pipeline. We
evaluate both full predictive densities and the corresponding point
predictions, and we report metrics in two regimes: the usual
spectroscopic-test-set evaluation and an importance-weighted evaluation
designed to approximate deployment on the photometric quasar-candidate
population.

Our main contributions are threefold. First, we provide the first
systematic benchmark of tabular foundation models for probabilistic
quasar photo-$z$ estimation in S-PLUS DR6, comparing frozen
in-context predictors with established task-specific conditional-density
estimators. Second, we evaluate the methods as probabilistic estimators,
not only as point predictors, using density quality, calibration,
coverage, and catastrophic-outlier diagnostics. Third, we quantify how
method rankings change when the evaluation is reweighted toward the
photometric target population, separating performance on the labelled
spectroscopic sample from the performance expected for the catalogue on
which the models will ultimately be used.

The paper is organised as follows.  Section~\ref{sec:data} describes the S-PLUS DR6 photometry, the spectroscopic compilation, the engineered features, and the photometric target sample used for weighting.  Section~\ref{sec:methods} presents the benchmarked methods, the covariate-shift framework, and the evaluation metrics.  Section~\ref{sec:results} reports the density-estimation, point-prediction, scaling, and weighted-evaluation results.  Section~\ref{sec:conclusions} summarises the implications for probabilistic quasar photo-$z$ estimation in S-PLUS and for future applications of tabular foundation models in astronomical surveys.

%% file: data.tex
\label{sec:data}
\subsection{Photometric data: S-PLUS}
\label{sec:splus}
In this work, we test foundation models on data from the Southern Photometric Local Universe Survey (S-PLUS; \citealt{splus}). S-PLUS is a multi-band optical survey designed to map $\sim$9,300 square degrees of the southern sky using a dedicated 0.8m robotic telescope at Cerro Tololo Inter-American Observatory (CTIO) in Chile. It employs the Javalambre photometric system \citep{cenarro+19} consisting of 12 filters: five broad-band filters similar to those of the SDSS ($u$, $g$, $r$, $i$, $z$), and seven narrow-band filters strategically centered on key stellar features, such as H$\alpha$. Full details of the survey can be found in \cite{splus} and on the survey website\footnote{\url{https://www.splus.iag.usp.br/?q=instrumentation}}.

We use the sixth data release\footnote{Available internally for collaboration members at the time of paper submission. Data will be available at \url{https://www.splus.cloud}} (DR6; Oliveira-Schwarz et al. in preparation), reduced with the  Multiband Astronomical Data Reduction package \citep{mar}. The depths of each band are shown in Table \ref{tab:depths}.

\begin{table}[ht]
\begin{tabular}{@{}ccc@{}}
\toprule
Filter & Effective Wavelength (\AA) & Depth \\ \midrule 
u      & 	3536& 21.6  \\ 
g      & 	4751& 22.3  \\ 
r      & 	6258& 22.1  \\ 
i      & 	7690& 21.7  \\ 
z      & 	8831& 21.2  \\ 
J0378  & 	3770& 21.3  \\ 
J0395  & 	3940& 20.8  \\ 
J0410  &4094& 20.9  \\ 
J0430  & 	4292& 21.0  \\ 
J0515  & 5133& 21.2  \\ 
J0660  &6614& 21.8  \\ 
J0861  &	8611& 20.9  \\ \bottomrule 
\end{tabular}
\caption{Median depth for each S-PLUS band considering the PSF
photometry at a signal-to-noise threshold of $S/N \geq 3$. }
\label{tab:depths}
\end{table}


\subsection{Complementary photometric data}
\label{sec:complementary}

The S-PLUS optical photometry is complemented with ultraviolet (UV) data from the \textit{Galaxy Evolution Explorer} \citep[GALEX;][]{Martin2005} and mid-infrared data from the \textit{Wide-field Infrared Survey Explorer} \citep[WISE;][]{Wright2010}.

GALEX was a NASA Small Explorer mission that surveyed the sky simultaneously in two UV broad bands using a dichroic beam splitter: the far-UV (FUV; \hbox{$\lambda_\mathrm{eff}\approx1528$\,\AA}) and the near-UV (NUV; $\lambda_\mathrm{eff}\approx2310$\,\AA), with an angular resolution of $4\farcs2$ (FUV) and $5\farcs3$ (NUV) over a $1\fdg25$-diameter field of view \citep{Morrissey2007}. Photometry is on the AB system. The All-Sky Imaging Survey (AIS) reaches typical $5\sigma$ depths of $\mathrm{FUV}\approx19.9$ and $\mathrm{NUV}\approx20.8$\,mag ($\sim$100\,s exposures), while the Medium Imaging Survey (MIS) extends to $\approx22.6$ and $22.7$\,mag in FUV and NUV, respectively ($\sim$1500\,s exposures) \citep{Morrissey2007}. GALEX covered approximately 75\% of the sky, avoiding the Galactic plane. We incorporate the FUV and NUV magnitudes (\texttt{FUVmag}, \texttt{NUVmag}) from cross-matches with the GALEX source catalog \citep{Bianchi2017}.

WISE is a NASA space observatory that mapped the entire sky in four mid-infrared bands. We use only the two shortest-wavelength bands: W1 ($\lambda_\mathrm{eff}=3.4\,\mu\mathrm{m}$, FWHM $\approx6\farcs1$) and W2 ($\lambda_\mathrm{eff}=4.6\,\mu\mathrm{m}$, FWHM $\approx6\farcs4$), which remain active through the ongoing NEOWISE-Reactivation mission and are the most sensitive WISE channels for extragalactic point sources, reaching $5\sigma$ depths of $\approx6.8$ and $9.8\,\mu\mathrm{Jy}$ in W1 and W2, respectively \citep{Wright2010}. We use the unWISE forced-photometry fluxes \citep{Schlafly2019}, converting to Vega magnitudes via $W = 22.5 - 2.5\log_{10}(F_\nu/\mathrm{nMgy})$, where the zero-point offset of 22.5\,mag is tied to the Vega system through the AllWISE calibration. To convert to AB magnitudes, offsets of $+2.699$ and $+3.339$\,mag must be applied to W1 and W2, respectively. The W1$-$W2 colour is a well-established discriminator for AGN at all redshifts \citep{stern2012}, and SHAP attributions in Section \ref{sec:res_shap} confirm that these two features carry the largest predictive weight in our models.

\subsection{Spectroscopic data}
For the ground-truth table, we use the compilation of spectroscopic confirmed sources from \citet{erik_v_2025_15127060}, including sources from SDSS, DESI, among others. We select all sources that are classified as quasars and crossmatch within 1\arcsec{} with S-PLUS data. \review{Because the crossmatch relies on a fixed 1\arcsec{} radius, faint-end results may be affected by an increase in mismatches or missed counterparts driven by larger positional uncertainties as sources approach the S-PLUS depth limits (Table \ref{tab:depths}).} A total of 135\,164 quasars are in our ground-truth sample spanning $0.002<z<6.981$ and no flagging during photometry process. 

\subsection{Feature engineering}
\label{sec:features}
We construct a 39-dimensional feature vector for each quasar, combining the S-PLUS PSF photometry from Section \ref{sec:splus} with the WISE and GALEX magnitudes from Section \ref{sec:complementary}. \review{Non-detections are recorded with a sentinel value in the underlying tables, and this convention is propagated to every benchmarked method, as detailed in Appendix \ref{app:methods}.} The feature vector consists of:
\begin{enumerate}
    \item 12 PSF magnitudes in the S-PLUS bands ($u$, J0378, J0395, J0410, J0430, $g$, J0515, $r$, J0660, $i$, J0861, $z$);
    \item 11 colours defined as the magnitude difference relative to the $r$-band (e.g. $u - r$, J0378 $- r$, \ldots);
    \item 12 magnitude errors corresponding to each S-PLUS band;
    \item 2 WISE magnitudes (W1 and W2)
    \item 2 GALEX magnitudes (FUVmag and NUVmag).
\end{enumerate}

\subsection{Train-test split and sample sizes}
\label{sec:split}

The spectroscopic sample is divided into a training set of 121\,626 objects and a test set of 13\,538 objects, following the split defined in the original S-PLUS quasar pipeline \citep{2024MNRAS.531..327N}.  The training set merges the original training and validation partitions.  No validation holdout is used; hyperparameter tuning is performed via cross-validation on the training set.

To study how performance scales with training-set size, we draw random subsets of $n_{\mathrm{train}} \in \{500, 1\,000, 5\,000, 50\,000\}$ from the full training pool, in addition to using the complete training set ($n_{\mathrm{train}} = 121\,626$).  For each configuration we perform five independent repetitions with different random seeds for the subsampling.

\subsection{Photometric sample for covariate shift correction}
\label{sec:photosample}

For the importance-weight estimation described in Sect.~\ref{sec:covshift}, we require a representative sample drawn from the photometric (target) population.  We use the catalogue of photometric quasar candidates from S-PLUS DR6 that lack spectroscopic confirmation. Details of the quasar classification are provided in the DR6 documentation\footnote{\url{https://splus.cloud/documentation/dr6}}. The documentation follows the strategy of \cite{2021MNRAS.507.5847N}.  After removing objects that overlap with the spectroscopic sample, we draw a random subsample of 121\,626 objects (matching the spectroscopic training-set size) to serve as the unlabelled class in the weight-estimation classifier.

\review{The spectroscopic and photometric samples differ substantially in both their
measured feature distributions and their non-detection patterns.
Appendix~\ref{app:feature_shift} compares three representative features
($r$, $u-r$, and $W1-W2$) in the two samples; these one-dimensional
comparisons indicate that it is indeed important to take selection-bias into account, which is what we do in Sect.~\ref{sec:covshift}.}

%% file: methods.tex
\label{sec:methods}

\subsection{Photo-\texorpdfstring{$z$}{z} estimation methods}
\label{sec:models}

We benchmark eleven machine-learning methods for quasar photometric-redshift
estimation using the S-PLUS DR6 feature set described in Sect.~\ref{sec:data}. The benchmark
is designed to compare tabular foundation models against both simple statistical
baselines and established photo-\(z\) methods, while keeping the train/test split,
input features, redshift grid, and evaluation protocol fixed across methods.

We focus on data-driven methods rather than template-fitting approaches.
Template-based photo-\(z\) estimators remain useful in many astronomical settings,
but their performance depends on the adopted spectral libraries, priors, and
calibration choices \citep{2019NatAs...3..212S}. They are therefore outside the scope of this controlled
comparison of learned conditional-density estimators.

The eleven methods span four families: linear conditional-Gaussian models,
tree-based ensembles, neural density estimators, and tabular foundation models.
The linear models provide transparent parametric baselines; the tree-based and
neural methods represent strong task-specific photo-\(z\) alternatives; and the
foundation models test whether frozen, pre-trained tabular transformers can be
used off-the-shelf for probabilistic quasar redshift inference.

Nine methods return a full conditional density \(\hat p(z \mid x)\) evaluated on
a grid of 200 redshift values. The two remaining methods, RF-Point and
GBM-Point, are point estimators only and are included only in the
point-prediction comparisons. For density-producing methods, the corresponding
point prediction is the posterior mean,
\[
\hat z = \int z \hat p(z \mid x)\, dz .
\]
All methods use the same train/test split and the 39-dimensional feature vector
described in Sects.~\ref{sec:features} and~\ref{sec:split}; cross-validation grids and tuning protocols are
given in Appendix~\ref{app:methods}.

The full list is:
\begin{itemize}
    \item \textbf{LinGauss-Homo-Ridge.}  A ridge-regularised linear regression $\mu(\mathbf{x}) = \mathbf{x}^\top\boldsymbol{\beta}$ paired with a single residual variance $\sigma^2$ estimated from out-of-fold residuals, giving the homoscedastic Gaussian density $p(z \mid \mathbf{x}) = \mathcal{N}\!\left(z;\, \mathbf{x}^\top\boldsymbol{\beta},\, \sigma^2\right)$.  Used as a parametric baseline.

    \item \textbf{LinGauss-Hetero-Ridge.}  A heteroscedastic Gaussian extension that jointly estimates the mean and an input-dependent log-variance via penalised maximum likelihood, $p(z \mid \mathbf{x}) = \mathcal{N}\!\left(z;\, \mathbf{x}^\top\boldsymbol{\beta},\, \exp(\mathbf{x}^\top\boldsymbol{\gamma})\right)$.

    \item \textbf{FlexZBoost} \citep{IzbickiLee2017}.  A non-parametric conditional density estimator that expands $p(z \mid \mathbf{x}) = \sum_{j=0}^{J-1} \beta_j(\mathbf{x})\,\phi_j(z)$ on a cosine basis, with coefficient functions $\beta_j(\mathbf{x}) = \mathbb{E}[\phi_j(Z) \mid \mathbf{x}]$ estimated by XGBoost regressors trained on the transformed responses $\phi_j(z_i)$.  The number of basis terms and a post-hoc sharpening exponent are selected on the CDE loss.

    \item \textbf{MDN} \citep{Bishop1994}.  A shallow (one hidden layer) mixture density network that outputs the parameters of a Gaussian-mixture posterior $p(z \mid \mathbf{x}) = \sum_{k=1}^{K} \pi_k(\mathbf{x})\,\mathcal{N}(z;\, \mu_k(\mathbf{x}), \sigma_k^2(\mathbf{x}))$.  The number of mixture components, hidden width, learning rate, and number of epochs are tuned by cross-validated CDE loss.

    \item \textbf{MDN-deep.}  A deeper variant of the MDN that retains the same Gaussian-mixture likelihood but uses a fixed three-hidden-layer architecture trained with AdamW and a reduce-on-plateau learning-rate schedule.

    \item \textbf{Flow-Spline} \citep{DurkanEtAl2019}.  A neural rational-quadratic spline normalising flow that models the conditional density as an invertible coupling-based transformation of a base Gaussian.  Flow layers, spline bins, hidden width, learning rate, and weight decay are tuned by cross-validated CDE loss.

    \item \textbf{RF-Point} \citep{Breiman2001RandomForests} \emph{(point estimator only)}.    A scikit-learn random forest regressor used without hyperparameter tuning; included as a robust tree baseline.

    \item \textbf{GBM-Point} \citep{ChenGuestrin2016XGBoost} \emph{(point estimator only)}.  An XGBoost   regressor  with hyperparameters (number of trees, depth, learning rate, subsample fraction) tuned by cross-validated MSE.

    \item \textbf{TabPFN~2.5} \citep{Hollmann2025,Grinsztajn2025TabPFN25}.  A tabular foundation model: a transformer pre-trained on a very large corpus of synthetically generated tabular tasks to approximate conditional densities for tabular regression.  There is no training in the traditional sense -- the network's weights are frozen and reused on every dataset.  At inference time the labelled training set is supplied as ``in-context support'' and a test row is predicted in a single forward pass; the model yields a piecewise-constant ``bar'' distribution over the response, which we interpolate onto the 200-point redshift grid and renormalise to obtain $\hat{f}(z \mid \mathbf{x})$.  We use the public version-2.5 regressor checkpoint and an ensemble of 8 estimators.

    \item \textbf{RealTabPFN~2.5} \citep{Garg2025RealTabPFN}.  The same TabPFN~2.5 backbone with weights fine-tuned by the authors on a curated collection of real-world tabular datasets, distributed as the checkpoint \texttt{tabpfn-v2.5-regressor-v2.5\_real.ckpt}.  The inference protocol and density extraction are identical to TabPFN~2.5; the comparison isolates the effect of fine-tuning the prior on real data.

    \item \textbf{TabICL} \citep{Qu2025}.  A separately developed transformer-based in-context learner for tabular prediction.  Like TabPFN, no per-dataset training is performed: the labelled training set is supplied in context and predictions follow in a single forward pass.  Conditional densities are extracted from quantile predictions: 199 quantile levels in $[0.005, 0.995]$ define an empirical CDF that is linearly interpolated on the redshift grid, numerically differentiated, clipped to be non-negative, and renormalised.  We use an ensemble of 8 estimators.
\end{itemize}

\subsection{Covariate shift correction}
\label{sec:covshift}

Photo-$z$ models are trained on spectroscopic samples, but their ultimate purpose is to predict redshifts for photometric sources.  If the distribution of photometric features $\mathbf{x}$ differs between these two populations, then standard evaluation on the spectroscopic test set can be misleading.  Specifically, a metric computed as a simple average over spectroscopic test objects reflects the performance under the spectroscopic distribution, not the photometric distribution to which the model will be applied.

Formally, let $p_s(\mathbf{x})$ denote the feature distribution of spectroscopic (labelled) objects and $p_t(\mathbf{x})$ the feature distribution of the photometric (target) population.  In photometric redshift prediction problems it is usual to make the  covariate shift assumption,
which states that the conditional distribution of redshift given features is the same in both populations,
\begin{equation}
\label{eq:covshift_assumption}
    p(z \mid \mathbf{x}) = p_s(z \mid \mathbf{x}) = p_t(z \mid \mathbf{x}),
\end{equation}
but the marginal feature distributions differ: $p_s(\mathbf{x}) \neq p_t(\mathbf{x})$ \citep{cunha2009estimating,freeman2017unified}.  In the quasar photo-$z$ context, this assumption is reasonable:
spectroscopic
targeting may depend on the observed photometric covariates $\mathbf{x}$, such as
colors, magnitudes, and other quantities used in selection, but once these
covariates are fixed, the event of receiving a spectroscopic label is assumed not
to carry additional information about the true redshift. This implies the covariate shift property \citep{izbicki2017photo}.

\review{Because the target sample lacks spectroscopic redshifts, the equality
$p_s(z\mid\mathbf{x})=p_t(z\mid\mathbf{x})$ cannot be tested from these
data without additional assumptions \citep{BenDavid2010,GulrajaniHashimoto2022}.
It may be violated if targeting uses information not represented in
$\mathbf{x}$ (e.g. variability, morphology, or ancillary detections), if
obtaining a secure spectrum remains redshift-dependent at fixed $\mathbf{x}$,
or if the photometric candidate catalogue contains non-QSO contaminants.}

Under this assumption, the expected value of any loss function $L(z, \hat{p})$ under the target distribution can be rewritten as an importance-weighted expectation under the source distribution:
\begin{equation}
\label{eq:importance_weighting}
    \mathbb{E}_{p_t}\!\left[L(z, \hat{p})\right] = \mathbb{E}_{p_s}\!\left[\beta(\mathbf{x})\,L(z, \hat{p})\right],
\end{equation}
where the importance weight is
\begin{equation}
\label{eq:weight_ratio}
    \beta(\mathbf{x}) = \frac{p_t(\mathbf{x})}{p_s(\mathbf{x})}.
\end{equation}
This identity shows that, by reweighting each spectroscopic test-set observation by $\beta(\mathbf{x}_i)$, we can estimate the performance that would be achieved on the photometric population using only the spectroscopic test set.  Objects that are under-represented in the spectroscopic sample relative to the photometric population receive higher weight, and vice versa.

\subsubsection{Estimating importance weights}
\label{sec:weight_estimation}
 Importance weighting requires estimating $\beta(\mathbf x)$, but direct estimation of   $p_t(\mathbf{x})$ and $p_s(\mathbf{x})$ is difficult in higher dimensions \citep{izbicki2014high}.  A widely used alternative  is to reformulate the problem as a classification task \citep{Bickel2009}.  Define a binary label $S = 1$ for spectroscopic objects and $S = 0$ for photometric objects.  By Bayes' theorem,
\begin{equation}
\label{eq:weight_classifier}
    \beta(\mathbf{x}) = \frac{P(S=0 \mid \mathbf{x})}{P(S=1 \mid \mathbf{x})},
\end{equation}
assuming equal prior probabilities (i.e. balanced classes in the training set). Because the weights are renormalised, exact balance is required only approximately. Thus, one can estimate the weights via
$$\widehat \beta(\mathbf{x}) = \frac{\widehat  P(S=0 \mid \mathbf{x})}{\widehat  P(S=1 \mid \mathbf{x})},$$
where $\widehat  P(S=1 \mid \mathbf{x})$ is the output of a probabilistic classifier.
In this paper, we train a TabICL classifier \citep{Qu2025} with 8 estimators on a balanced dataset formed by the spectroscopic training sample ($S=1$) and a same-size random subsample of the photometric catalogue ($S=0$; see Sect.~\ref{sec:photosample}).  The estimated class probabilities $\hat{P}(S=1 \mid \mathbf{x})$ are clipped to $[0.01, 0.99]$ before computing the ratio, to prevent extreme weights.  The resulting weights are normalised to sum to one for use in weighted metric computation.
\review{In Appendix \ref{app:weight_estimator}, we also show the results we get if we replace the TabICL classifier with a boosting procedure; the results are almost identical.}

\subsubsection{Tempered weights and the exponent \texorpdfstring{$\alpha$}{alpha}}
\label{sec:alpha_weights}

Raw density-ratio weights $\beta(\mathbf{x})$ can be heavy-tailed: a small subset of test objects can carry the bulk of the total weight, which inflates the variance of any weighted average and may collapse the effective sample size to a small fraction of $n$ when estimating the performance of each method $\hat p$.  We therefore evaluate weighted metrics with a one-parameter family of tempered weights,
\begin{equation}
\label{eq:alpha_weights}
    w_i \;=\; \frac{\beta(\mathbf{x}_i)^{\alpha}}{\sum_{j=1}^{n} \beta(\mathbf{x}_j)^{\alpha}},
    \qquad \alpha \in [0,1],
\end{equation}
normalised to sum to one.  This is the standard exponential tempering used in importance-weighted learning under covariate shift \citep{Shimodaira2000}: $\alpha = 0$ recovers the unweighted (spectroscopic) estimator, $\alpha = 1$ uses the raw density-ratio weights and points exactly at the photometric population, and intermediate $\alpha$ trades shift correction against variance.  The bias--variance trade-off is governed by the effective sample size $\mathrm{ESS}(\alpha) = \left(\sum_i w_i\right)^2 / \sum_i w_i^2$, which is monotonically decreasing in $\alpha$.

We report weighted results for two settings: $\alpha = 1$, which is unbiased for the photometric target  (if the weights are well estimated) but has a small ESS, and a tuned $\alpha^{\star}$ chosen so that $\mathrm{ESS}(\alpha^{\star})$ equals $30\%$ of the test-set size; for our test set this gives $\alpha^{\star} \simeq 0.36$.  At $\alpha = 1$ the ESS collapses to $\sim 1.6\%$ of the test set, which is reflected in noticeably larger standard errors in the corresponding tables.



\subsection{Evaluation metrics}
\label{sec:metrics}

We evaluate every method using a combination of density-based and point-prediction metrics.  Density metrics measure the quality of the full predictive distribution -- both how well it concentrates around the true redshift and how honestly it represents its own uncertainty -- while point metrics evaluate only the central prediction $\hat{z}$.  Each metric is reported in a standard (unweighted) and, where applicable, importance-weighted form; the weighted version targets the photometric population that the survey will actually deliver and is therefore the more practically relevant of the two, while the unweighted version is included both for comparability with the existing literature and as a control on the magnitude of the shift correction.

For a test set of $n$ objects, let  $\hat{p}_i(z) \equiv \hat{p}(z \mid \mathbf{x}_i)$ be the estimated conditional density. In what follows, we describe the unweighted metrics that are used in the experiments; but we also compute the weighted versions.   Standard errors for all metrics, along with the bold-set rule used in the headline tables, are described in Appendix~\ref{app:se}.

\subsubsection{Density metrics}
\label{sec:density_metrics}

\paragraph{CDE loss}  The CDE loss \citep{schmidt2020evaluation,Izbicki2025} is a proper scoring rule for conditional density estimation:
\begin{equation}
    \text{CDE loss} = \frac{1}{n}\sum_{i=1}^{n} \left[\int \hat{p}_i(z)^2\,\mathrm{d}z - 2\,\hat{p}_i(z_i)\right].
\end{equation}
It approximates the integrated squared error $\int (\hat{p} - p)^2\,\mathrm{d}z dp(\mathbf x)$ between the estimated and true conditional densities up to an additive constant that does not depend on $\hat{p}$, and decomposes into an integrated squared density (penalising overly peaked estimates) minus twice the density at the true redshift (rewarding mass placed near the truth).  Lower values indicate better density estimation.   

\paragraph{Log-likelihood} The mean log-density at the true redshifts:
\begin{equation}
    \text{LL} = \frac{1}{n}\sum_{i=1}^{n} \log \hat{p}_i(z_i).
\end{equation}
LL rewards methods that concentrate probability mass on the actual redshift and is highly sensitive to severe miscalibration, since a single $\hat{p}_i(z_i) \to 0$ drives LL to $-\infty$.   Higher values indicate better probabilistic predictive performance, reflecting both how well calibrated and how sharp the predictive densities are.
 The negative LL estimates the cross-entropy between the true conditional redshift distribution and the predicted density, so maximizing LL corresponds to minimizing this cross-entropy.

\paragraph{CRPS} The continuous ranked probability score \citep{GneitingRaftery2007} measures the integrated squared difference between the estimated CDF $\hat{F}_i$ (the integrated $\hat p(\cdot \mid \mathbf{x}_i)$)  and the Heaviside function at $z_i$:
\begin{equation}
    \text{CRPS} = \frac{1}{n}\sum_{i=1}^{n} \int \left[\hat{F}_i(z) - \mathbf{1}(z \geq z_i)\right]^2\,\mathrm{d}z.
\end{equation}
CRPS is a proper score that simultaneously rewards predictions that are sharp and that are located near the truth; compared to LL it is less aggressive toward outlying observations, since the squared CDF discrepancy is bounded.  Lower values indicate better predictions.

\paragraph{PIT KS statistic} The probability integral transform (PIT) value for object $i$ is $\hat{F}_i(z_i)$.  If the density estimates are well-calibrated, the PIT values follow a uniform distribution on $[0,1]$ \citep{bordoloi2010photo,zhao2021diagnostics,dey2025towards}.  We summarise departures from uniformity by the Kolmogorov--Smirnov statistic, i.e. the supremum distance between the empirical PIT CDF and the identity.  Lower values indicate better calibration. We define the weighted Kolmogorov–Smirnov statistic as the supremum distance between the weighted empirical PIT CDF and the identity.

\paragraph{$90\%$ credible interval coverage.}  We compute the $90\%$ equal-tailed credible interval $[q_{0.05}, q_{0.95}]$ from each estimated density and report its empirical \emph{coverage}: the fraction of true redshifts that fall inside the interval \citep{izbicki2016nonparametric}.  Values close to the nominal level of $0.90$ indicate well-calibrated uncertainty; values much below $0.90$ reflect over-confident predictive distributions, while values much above $0.90$ signal under-confidence.

\paragraph{Multimodal posteriors.}
\review{PIT and coverage remain valid for multimodal distributions because they use
the complete CDF. However, they do not test whether probability is correctly
distributed among modes, and equal-tailed intervals may span low-density
regions. We therefore interpret them as global calibration diagnostics.}

\subsubsection{Point-prediction metrics}
\label{sec:point_metrics}
We adopt the metrics used in the QuCatS benchmark \citep{2024MNRAS.531..327N}:

\paragraph{RMSE} Root mean squared error, $\text{RMSE} = \sqrt{\tfrac{1}{n}\sum_i (z_i - \hat{z}_i)^2}$.  Penalises large per-object errors quadratically and is therefore dominated by catastrophic failures.

\paragraph{Bias} Mean signed residual, $\text{bias} = \tfrac{1}{n}\sum_i (z_i - \hat{z}_i)$.  Diagnoses systematic over- or under-prediction; an unbiased method has $\text{bias} \approx 0$, but can still be inaccurate when individual errors are large.

\paragraph{NMAD} The normalised median absolute deviation of the normalised residual $\Delta z_{\mathrm{norm}} = (z_i - \hat{z}_i)/(1+z_i)$:
\begin{equation}
    \text{NMAD} = 1.48 \times \mathrm{median}\!\left(|\Delta z_{\mathrm{norm}} - \mathrm{median}(\Delta z_{\mathrm{norm}})|\right).
\end{equation}
A robust scale of the typical residual that is insensitive to a small fraction of catastrophic outliers.  Combined with the RMSE it gives a sense of how much of the RMSE is driven by the tails of the error distribution.  Lower values are better.  The weighted NMAD replaces both medians with weighted medians (i.e. linear interpolation of the weighted ECDF at the 0.5 quantile).

\paragraph{Outlier fraction} The fraction of objects with $|\Delta z_{\mathrm{norm}}| > \eta$, reported at the standard photo-$z$ thresholds $\eta = 0.15$ and $\eta = 0.30$.  Directly counts catastrophic redshift assignments; lower is better.

%% file: results.tex
\label{sec:results}

We present results from five independent repetitions of the full experiment on S-PLUS DR6 data. Each repetition uses a different random seed, controlling both the training-set subsampling and stochasticity internal to the training algorithms; the test set and the estimated importance weights are held fixed across repetitions. All numbers are reported as mean $\pm$ standard error (SE), computed as described in Appendix~\ref{app:se}.   The largest training-set size is $n_{\mathrm{train}} = 121\,626$, on which we focus the headline tables; full scaling curves from $n_{\mathrm{train}} = 500$ are shown in the figures. Method definitions and references are given in Sect.~\ref{sec:models}, and RF-Point and GBM-Point, being point estimators only, are omitted from the density tables.

\subsection{Density-estimation metrics}
\label{sec:res_density}

\input{table_density.tex}

Table~\ref{tab:density_combined} summarises the density-estimation results at the largest training-set size.  Several patterns emerge.

\begin{figure}[th]
\centering
\includegraphics[width=\linewidth, trim={0 0 0 70pt}, clip]{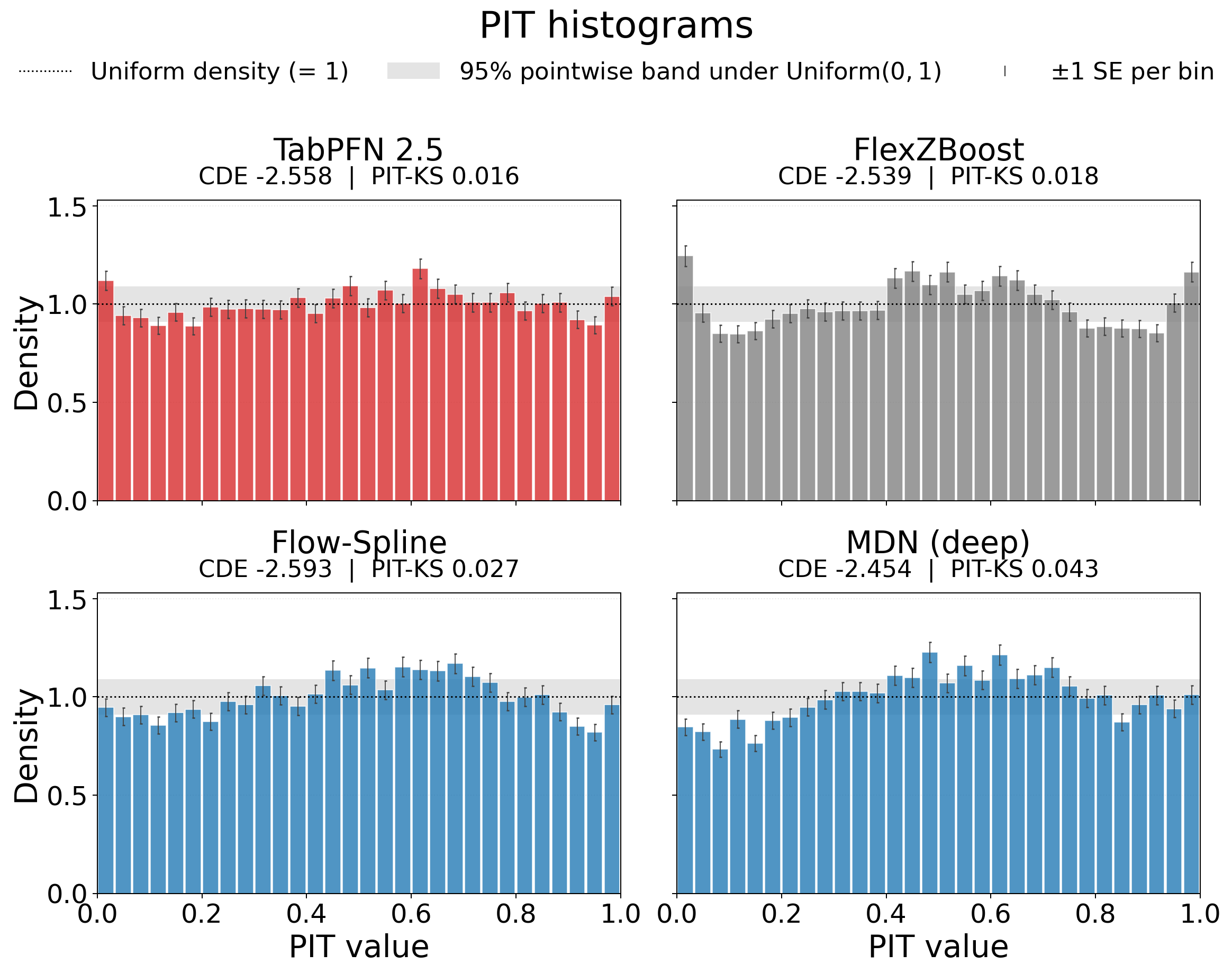}
\caption{PIT histograms for TabPFN~2.5, FlexZBoost, Flow-Spline, and MDN-deep. Panels are ordered by PIT-KS. The dotted line at density $1$ marks the target Uniform$(0,1)$ density; the grey band is the $95\%$ pointwise binomial envelope expected under exact uniformity at this test-set size, and per-bin error bars are $\pm 1$ SE. TabPFN~2.5 shows the smallest aggregate PIT-KS among these methods and no large structured departure from uniformity. FlexZBoost has a comparable PIT-KS but shows excess mass near the boundaries, indicating overconfident tails. MDN-deep and Flow-Spline show milder central excesses, consistent with slight under-confidence.}
\label{fig:pit_hist_full}
\end{figure}

\paragraph{Unweighted CDE loss (panel a).}
Flow-Spline achieves the lowest CDE loss, with TabPFN~2.5 statistically tied with it within one combined SE. This is the one density metric on which a non-foundation method matches the best foundation model. This result is consistent with Flow-Spline being a flexible density estimator whose hyperparameters are selected using the same CDE-loss criterion used in the table. FlexZBoost is also competitive on CDE loss, but remains a step behind the leading pair.
\review{A grid-resolution sensitivity analysis leaves the CDF-based and point-prediction conclusions unchanged, although the CDE-loss ordering of TabPFN 2.5 and Flow-Spline can reverse (Appendix \ref{app:weight_estimator}).}
On log-likelihood, TabPFN~2.5 is the best method, with Flow-Spline also within one combined SE. On CRPS, TabPFN~2.5 and RealTabPFN~2.5 form the bold set, while on $90\%$ coverage TabPFN~2.5 is closest to the nominal level and is the only method in the bold set. On PIT-KS, TabPFN~2.5 and RealTabPFN~2.5 are tied for best, with FlexZBoost also within one combined SE, while TabICL is the only foundation model outside the bold set. Overall, the unweighted density results show that TabPFN~2.5 is either best or statistically indistinguishable from the best method on the main density-quality and calibration summaries.

\begin{figure*}[ht]
\centering
\includegraphics[width=\textwidth, trim={0 0 0 70pt}, clip]{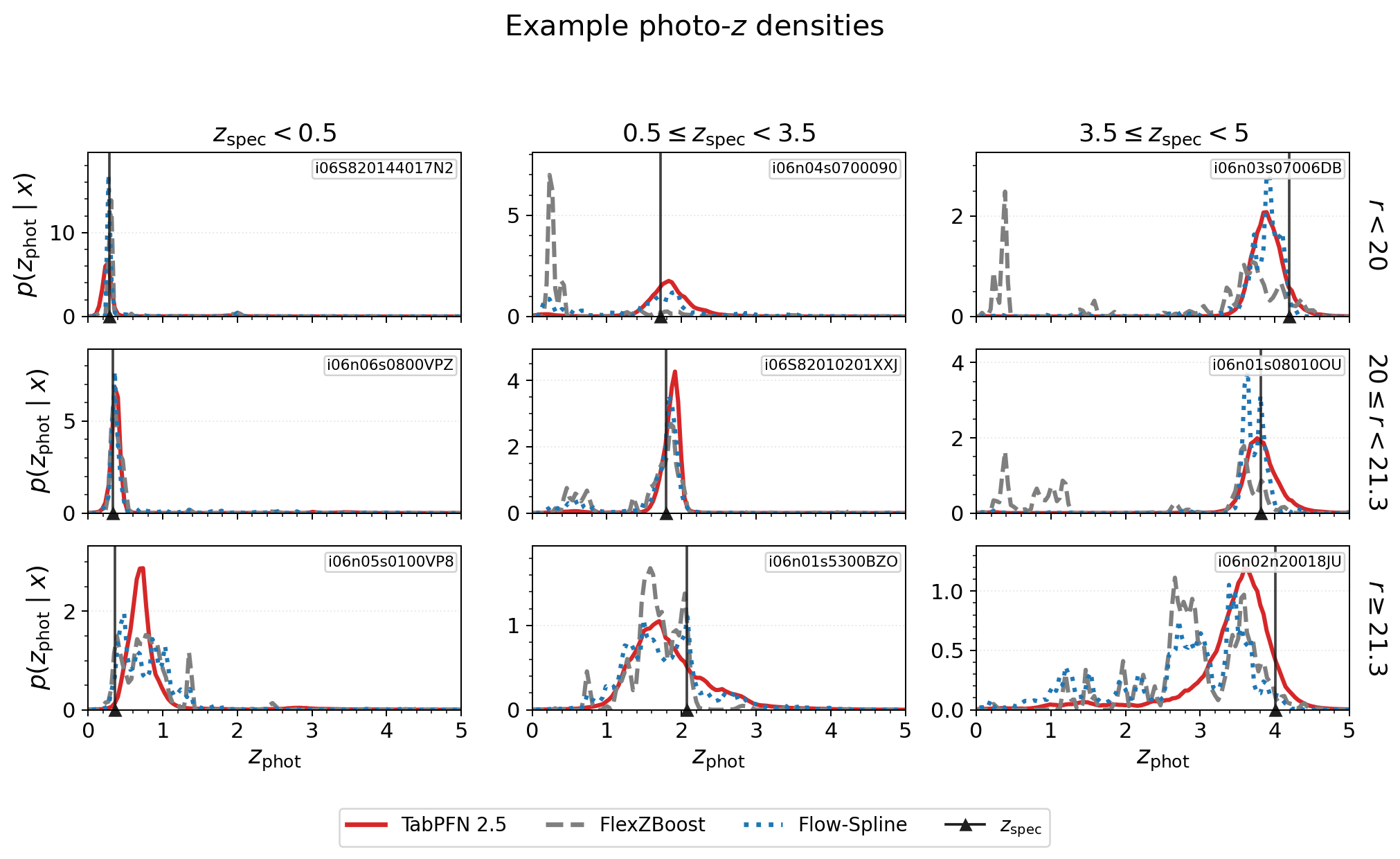}
\caption{Example photo-$z$ conditional densities on the DR6 test set at $n_{\mathrm{train}} = 121\,626$ for the three leading density estimators: TabPFN~2.5 (red, solid; foundation), FlexZBoost (grey, dashed; tree-based), and Flow-Spline (blue, dotted; neural).     The $3 \times 3$ grid cross-tabulates three $r$-magnitude bins (rows) against three $z_{\mathrm{spec}}$ bins (columns); within each cell the displayed object is chosen uniformly at random (fixed seed).  The thin black line and triangle mark the spectroscopic redshift; the object identifier is shown in the top-right of each panel. The bottom row and right column show faint and high-redshift cases, respectively.}
\label{fig:density_examples_spline}
\end{figure*}

\paragraph{Importance-weighted density metrics (panels b and c).}
Under the variance-controlled weighted evaluation in panel~(b), TabPFN~2.5 is the strongest overall density estimator: it has the best CDE loss, the best log-likelihood, the best weighted PIT-KS, and remains in the bold set on CRPS and $90\%$ coverage. Flow-Spline remains competitive on CDE loss, but no non-foundation method is consistently bolded across the density metrics. At $\alpha = 1$ in panel~(c), the same qualitative pattern is visible, but the larger variance of the raw importance weights inflates the SEs and produces broader bold sets; this panel should therefore be interpreted as a higher-variance estimate of target-population performance rather than as a sharply resolved ranking. FlexZBoost provides the clearest example of sensitivity to the shift: its $90\%$ coverage drops from $0.883$ unweighted to $0.842$ in panel~(b) and $0.794$ in panel~(c), while its PIT-KS increases from $0.018 \pm 0.003$ unweighted to $0.038 \pm 0.005$ in panel~(b) and $0.098 \pm 0.022$ in panel~(c). By contrast, TabPFN~2.5 keeps its weighted PIT-KS close to its unweighted value under the tempered weighting scheme.
\review{Figure \ref{fig:pit_weight_comparison} shows the corresponding PIT P–P curves and their sensitivity to importance weighting.}
The two LinGauss baselines remain clearly behind under all three evaluations.

Figure~\ref{fig:pit_hist_full} shows the unweighted PIT histograms at $n_{\mathrm{train}} = 121\,626$ for TabPFN~2.5 and the three strongest competing methods according to CDE loss. TabPFN~2.5 has the smallest PIT-KS among the displayed methods, and its histogram does not show a large systematic departure from uniformity. FlexZBoost has a similar aggregate PIT-KS, but the histogram reveals boundary excesses that are not fully captured by the scalar KS statistic and are consistent with overconfident tails. MDN-deep places comparatively more PIT mass near the centre, indicating mild under-confidence. Flow-Spline is the closest competitor visually, but also shows a small central excess. These diagnostics support the table-level conclusion that TabPFN~2.5 has strong aggregate calibration, while also illustrating why PIT-KS should be interpreted together with the shape of the PIT histogram.

Figure~\ref{fig:density_examples_spline} shows the full conditional density $p(z \mid \mathbf x)$ for nine random test objects that span the $(r, z_{\mathrm{spec}})$ plane.  The $3 \times 3$ grid has rows corresponding to three $r$-magnitude bins ($r < 20$, $20 \leq r < 21.3$, $r \geq 21.3$) and columns to three $z_{\mathrm{spec}}$ bins ($z_{\mathrm{spec}} < 0.5$, $0.5 \leq z_{\mathrm{spec}} < 3.5$, $3.5 \leq z_{\mathrm{spec}} < 5$).
At bright magnitudes and moderate redshift (top-centre, middle-centre panels) all three methods in general produce comparable  densities.  Differences emerge mostly at the faint and high-$z$ extremes. 
These cell-by-cell contrasts are consistent with the calibration diagnostics of Figure~\ref{fig:pit_hist_full} and with the aggregate CDE-loss ranking of Table~\ref{tab:density_combined}: the foundation model is well calibrated everywhere, the neural density estimator is systematically too broad, and FlexZBoost  trades off a tendency to over-confidence against occasional catastrophic mislocalisation. \review{The randomly selected examples in Figure~\ref{fig:density_examples_spline}
are complemented by three targeted cases in Appendix
Figure~\ref{fig:difficult_densities}: a high-$z$ case in which only
TabPFN~2.5 avoids a catastrophic point error, a faint case in which all three
methods fail, and a case in which TabPFN~2.5 fails while the two competitors
do not. These outcome-defined examples illustrate specific failure modes and
should not be interpreted as estimates of their prevalence.}
\vspace{2mm}

We conclude that the density results support TabPFN~2.5 as the strongest overall probabilistic photo-$z$ estimator in this benchmark. Flow-Spline matches it on unweighted CDE loss, and FlexZBoost remains competitive on some unweighted calibration summaries, but TabPFN~2.5 is best or statistically tied for best across the broader set of density metrics. The evidence is strongest under the variance-controlled weighted evaluation, where TabPFN~2.5 retains strong CDE loss, log-likelihood, PIT-KS, CRPS, and coverage simultaneously. The raw $\alpha=1$ weighted panel points in the same direction, but because its effective sample size is much smaller, it should be interpreted more cautiously. The PIT histograms and example densities support the aggregate results by showing that the differences are not only numerical, although the example densities remain illustrative rather than exhaustive.

\subsection{Point-prediction metrics}
\label{sec:res_point}

\input{table_point.tex}

Table~\ref{tab:point_combined} reports the analogous point-prediction summary at the largest training-set size; the bold convention is identical to the density table.

\paragraph{Unweighted point metrics (panel a).}  TabPFN2.5 leads on RMSE (tied within one SE with RealTabPFN2.5 and TabICL), is tied with TabICL on NMAD, is the single best method on $\eta_{0.15}$, and is tied with RealTabPFN~2.5 on $\eta_{0.30}$.  None of the LinGauss, tree, or neural baselines enters the bold set on RMSE or the outlier columns.  The bias column has many bold entries because the absolute biases are within a few thousandths of zero and the SEs are comparable in size; this column should be read as ``most methods are unbiased'', not as a discriminating ranking.

\paragraph{Weighted point metrics (panels b and c).}
Under the variance-controlled weighted evaluation, the foundation models retain their clearest advantage on the outlier columns. TabPFN~2.5 is best or statistically tied for best on RMSE, NMAD, $\eta_{0.15}$, and $\eta_{0.30}$ in the weighted panels. In panel~(b), the RMSE bold set widens to include MDN-deep, RF-Point, and GBM-Point, but the $\eta_{0.15}$ bold set is restricted to the three foundation models, and only MDN-deep joins them on NMAD and $\eta_{0.30}$. At $\alpha = 1$ in panel~(c), the larger SEs induced by the high-variance importance weights make the ranking less sharply resolved, so this panel should be read as broadly consistent with the tempered-weight pattern rather than as decisive evidence for fine-grained ordering among the leading methods. The linear baselines remain visibly worse on the outlier fractions.

\begin{figure}[t]
\centering
\includegraphics[width=\linewidth, trim={0 0 0 70pt}, clip]{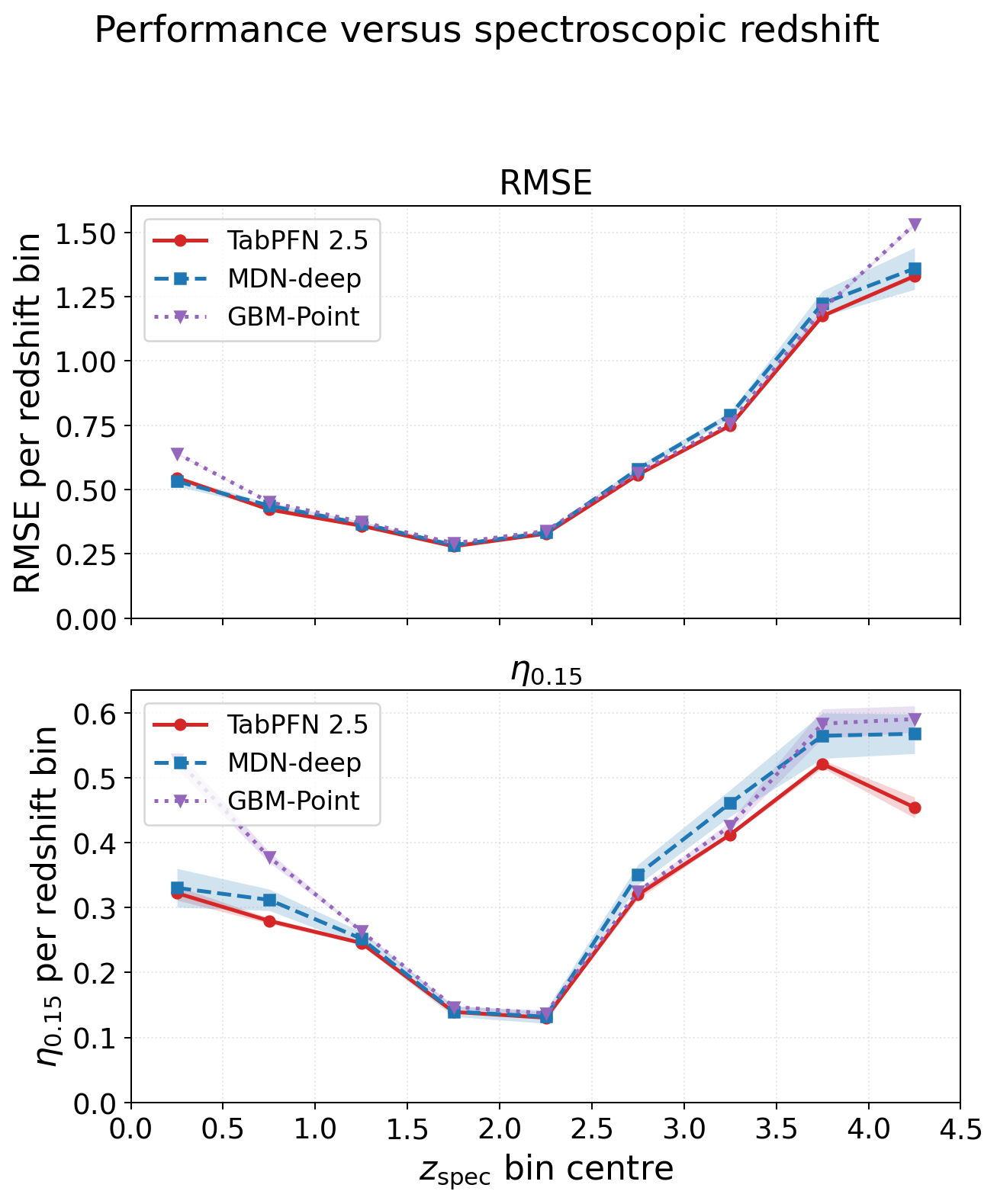}
\caption{Per-redshift-bin RMSE and $\eta_{0.15}$ on the DR6 test set at $n_{\mathrm{train}} = 121\,626$ for the three methods with the lowest overall unweighted RMSE in Table~\ref{tab:point_combined} (panel~a): TabPFN~2.5, MDN-deep, and GBM-Point.  Bins have fixed width $\Delta z_{\mathrm{spec}} = 0.5$; bins with fewer than $30$ test objects are dropped.  Shaded bands are $\pm 1$ standard error.  All three methods agree to within one SE in the mid-redshift range $z_{\mathrm{spec}} \in [0.5, 3.0]$; differences are concentrated at the low- and high-$z$ ends, where the foundation method performs the best.}
\label{fig:rmse_vs_zbin}
\end{figure}

Figure~\ref{fig:rmse_vs_zbin} breaks the unweighted point performance down into $z_{\mathrm{spec}}$ bins of width $\Delta z = 0.5$ for the three methods that achieve the lowest overall RMSE in panel~(a) of the table: TabPFN~2.5, MDN-deep, and GBM-Point.  The RMSE curves are essentially indistinguishable in the well-populated mid-redshift range $0.5 \lesssim z_{\mathrm{spec}} \lesssim 3$.  The differences concentrate at the two ends.  At $z_{\mathrm{spec}} \lesssim 0.5$, GBM-Point sits noticeably above the other two.  At $z_{\mathrm{spec}} \gtrsim 3.5$, the ordering reverses: TabPFN~2.5 holds the lowest per-bin RMSE in every high-$z$ bin, with MDN-deep and GBM-Point progressively above it.  The $\eta_{0.15}$ panel shows the same qualitative pattern in the catastrophic-error rate, with TabPFN~2.5 generally reducing the fraction of outliers at the low- and high-redshift edges.  This behaviour is consistent with the picture that the top-line point-metric differences in Table~\ref{tab:point_combined} are driven primarily by the tails of the redshift distribution, which is consistent with the foundation model's pre-training prior helping in the sparsely sampled tails. \review{The information that the foundation model exploits more effectively in this regime plausibly comes from two channels: the growing incidence of optical non-detections with redshift, driven by Lyman-alpha dropout in the bluer bands, and the WISE mid-infrared colours, which are known to aid quasar photo-z estimation at $z>2$ \citep{DiPompeo_2015}. Disentangling which channel drives the relative advantage would require a dedicated follow-up analysis, which we leave for future work.}

\begin{figure}[ht]
\centering
\includegraphics[width=\linewidth, trim={0 0 0 70pt}, clip]{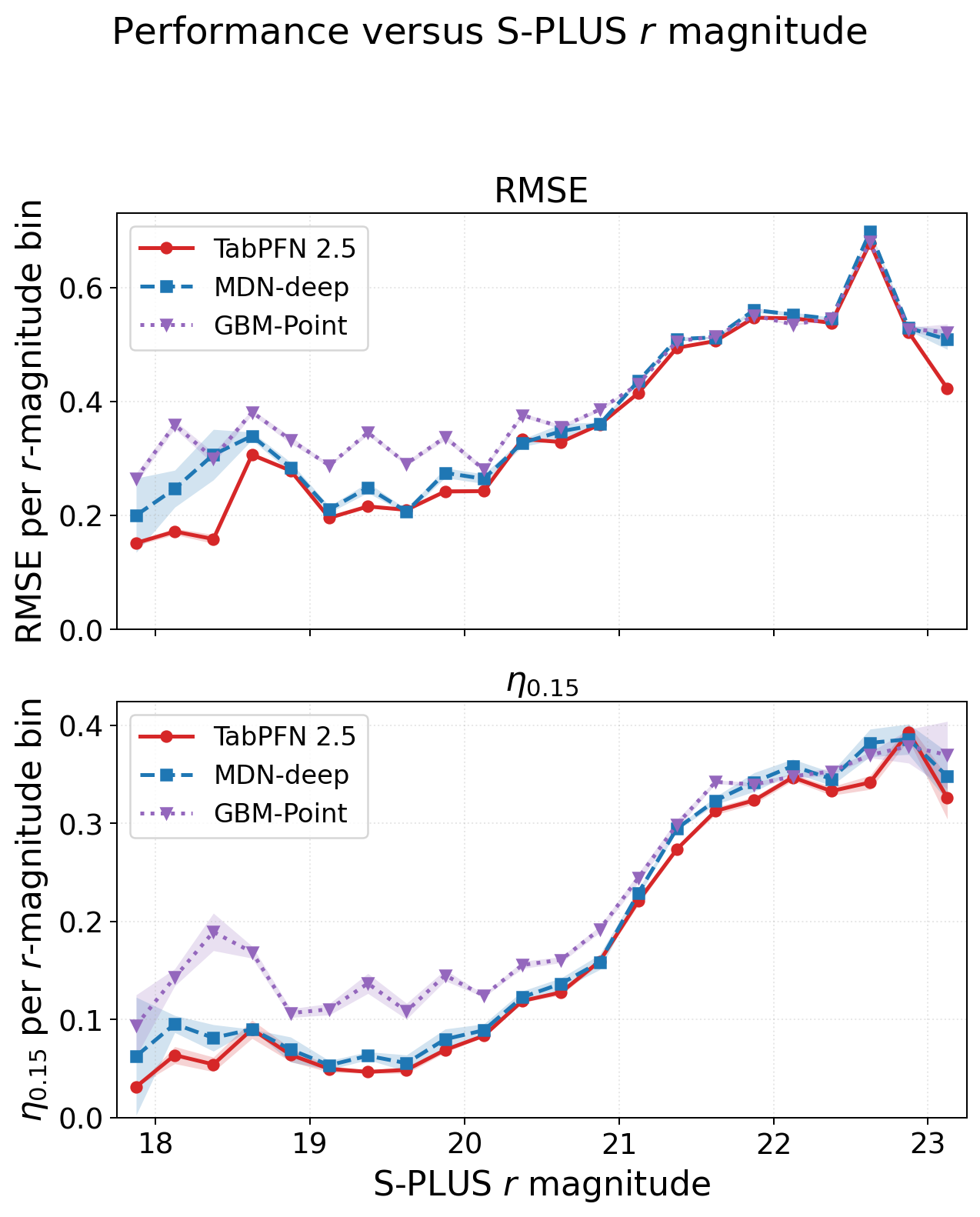}
\caption{Per-$r$-magnitude-bin point performance on the DR6 test set at $n_{\mathrm{train}} = 121\,626$ for the  TabPFN~2.5 and the two other best-performing methods.  Panel (a): per-bin RMSE. Panel (b): per-bin outlier fraction $\eta_{0.15}$. Shaded bands are $\pm 1$ inter-repetition standard deviation across the five seeds. TabPFN~2.5 is strongest at bright magnitudes and retains a modest advantage in the faint tail; GBM-Point is generally worse, particularly for the outlier fraction.}
\label{fig:mag_r_top3}
\end{figure}

Figure~\ref{fig:mag_r_top3} complements the redshift breakdown by re-binning the same three methods against the S-PLUS broad-band $r$ magnitude, with bin width $\Delta r = 0.25$ and a minimum of $30$ test objects per bin.  Two metrics are shown: panel~(a) the per-bin RMSE and panel~(b) the outlier fraction.  As expected, both metrics worsen overall toward fainter magnitudes, although the finite bin counts introduce  bin-to-bin fluctuations.  On RMSE, TabPFN~2.5 is best at the bright end and then tracks MDN-deep closely through the bulk of the sample; GBM-Point is noticeably worse for bright objects and only becomes comparable once the curves enter the noisier intermediate and faint regimes.  The separation is clearer for $\eta_{0.15}$.  TabPFN~2.5 has a substantially lower outlier fraction than MDN-deep for bright objects, remains slightly lower through much of the transition region, and again reduces the number of outliers in the faint tail, apart from a few noisy bins where the curves overlap.  GBM-Point is generally the weakest of the three, especially on $\eta_{0.15}$, where it lies above TabPFN~2.5 across most of the magnitude range.

We conclude that the point-prediction results broadly mirror the density results. TabPFN~2.5 and the other foundation models are best or statistically tied for best on the main point metrics, with their clearest advantage appearing in the catastrophic-outlier fractions and under the variance-controlled weighted evaluation. In the well-populated mid-redshift range, the leading methods are difficult to distinguish. At $\alpha = 1$, the larger SEs broaden the bold sets, so the practical conclusion is not that the exact ranking is resolved, but that the foundation models remain at least competitive with all alternatives under the target-population weighting. Their most visible gains occur near the low- and high-redshift edges, where catastrophic photo-$z$ failures are most frequent.
\review{Figure \ref{fig:metric_rank_heatmap} in Appendix \ref{app:se}  supplements
Tables~\ref{tab:density_combined} and~\ref{tab:point_combined} by displaying
the within-metric ranks of all methods as a heatmap. }

\subsection{Scaling with training-set size}
\label{sec:res_scaling}

\begin{figure*}[t]
\centering
\begin{subfigure}[t]{0.42\linewidth}
    \includegraphics[width=\linewidth, trim={0 0 160pt 0}, clip]{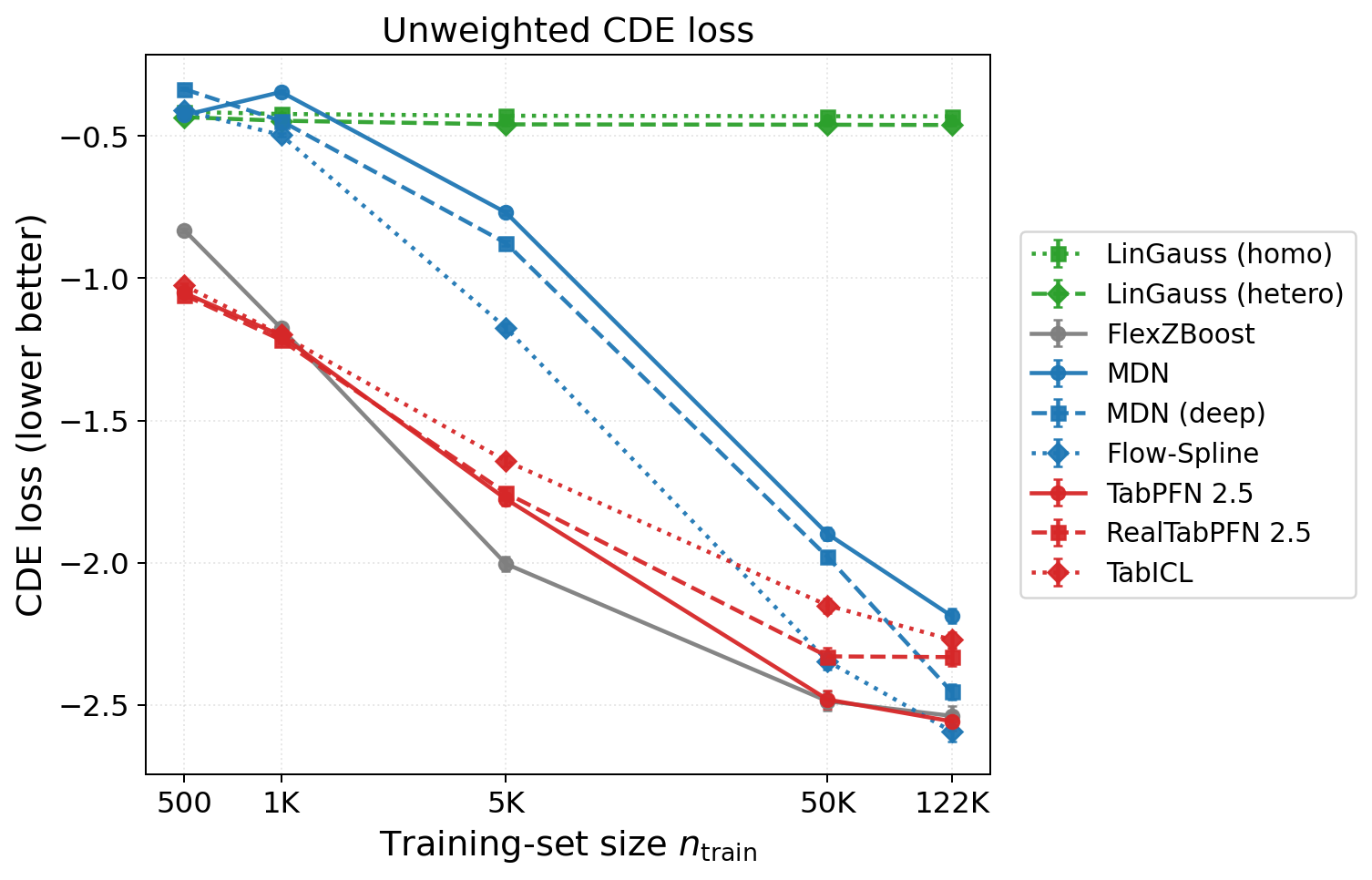}
    \caption{Unweighted CDE loss.}
\end{subfigure}\hfill
\begin{subfigure}[t]{0.58\linewidth}
    \includegraphics[width=\linewidth]{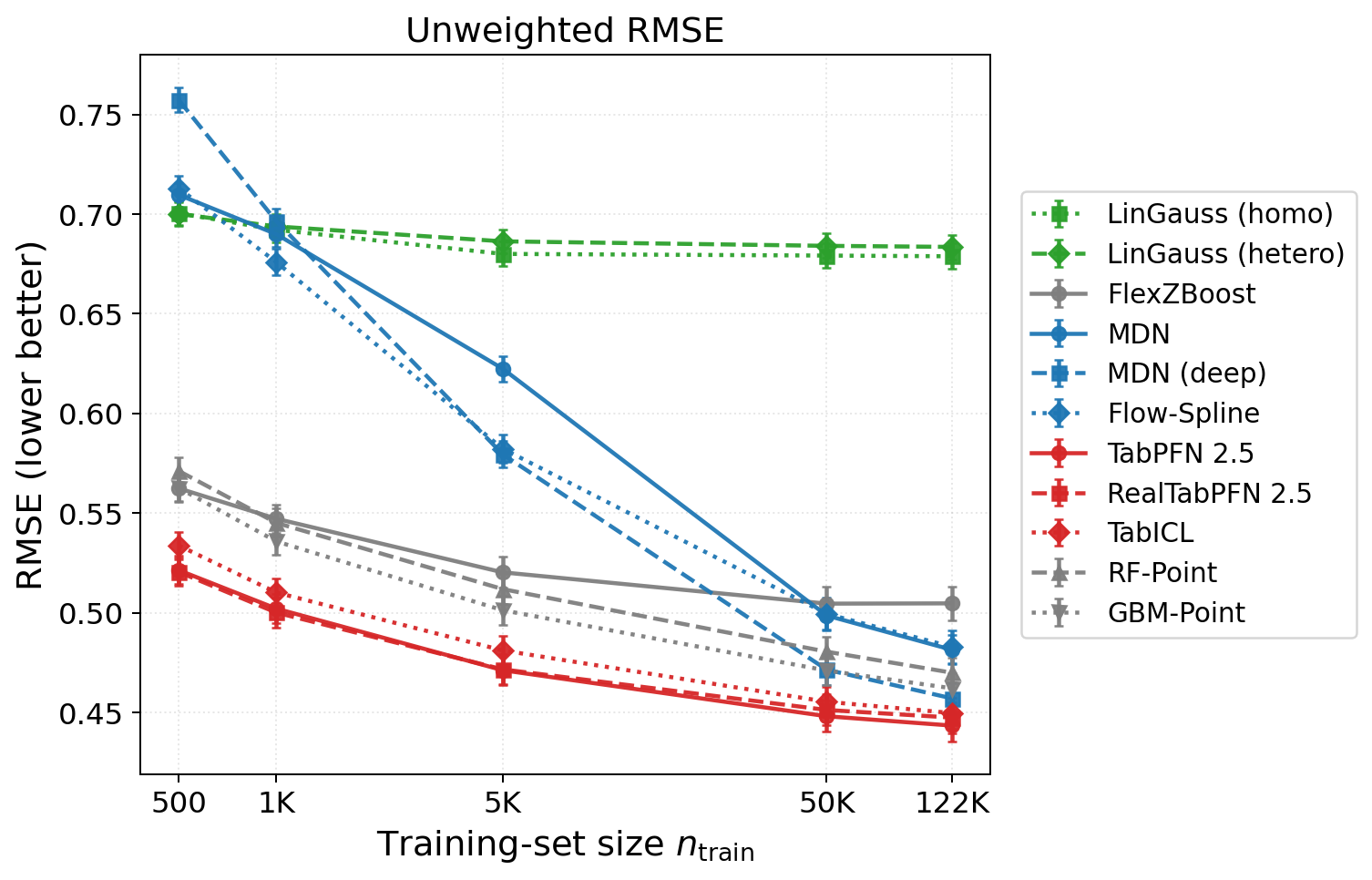}
    \caption{Unweighted RMSE.}
\end{subfigure}
\caption{Performance versus training-set size under the unweighted spectroscopic test distribution.  Each curve shows the mean over five repetitions and error bars indicate $\pm 1$ SE.  Methods are colour-coded by family (linear, tree-based, neural, foundation).  Foundation models (red) lead the RMSE curve at every $n_{\mathrm{train}}$, while on the unweighted CDE loss they are matched at the largest sample size by the density-optimised baselines (Flow-Spline, FlexZBoost).}
\label{fig:scaling_unweighted}
\end{figure*}

\begin{figure*}[t]
\centering
\begin{subfigure}[t]{0.42\linewidth}
\includegraphics[width=\linewidth, trim={0 0 160pt 0}, clip]{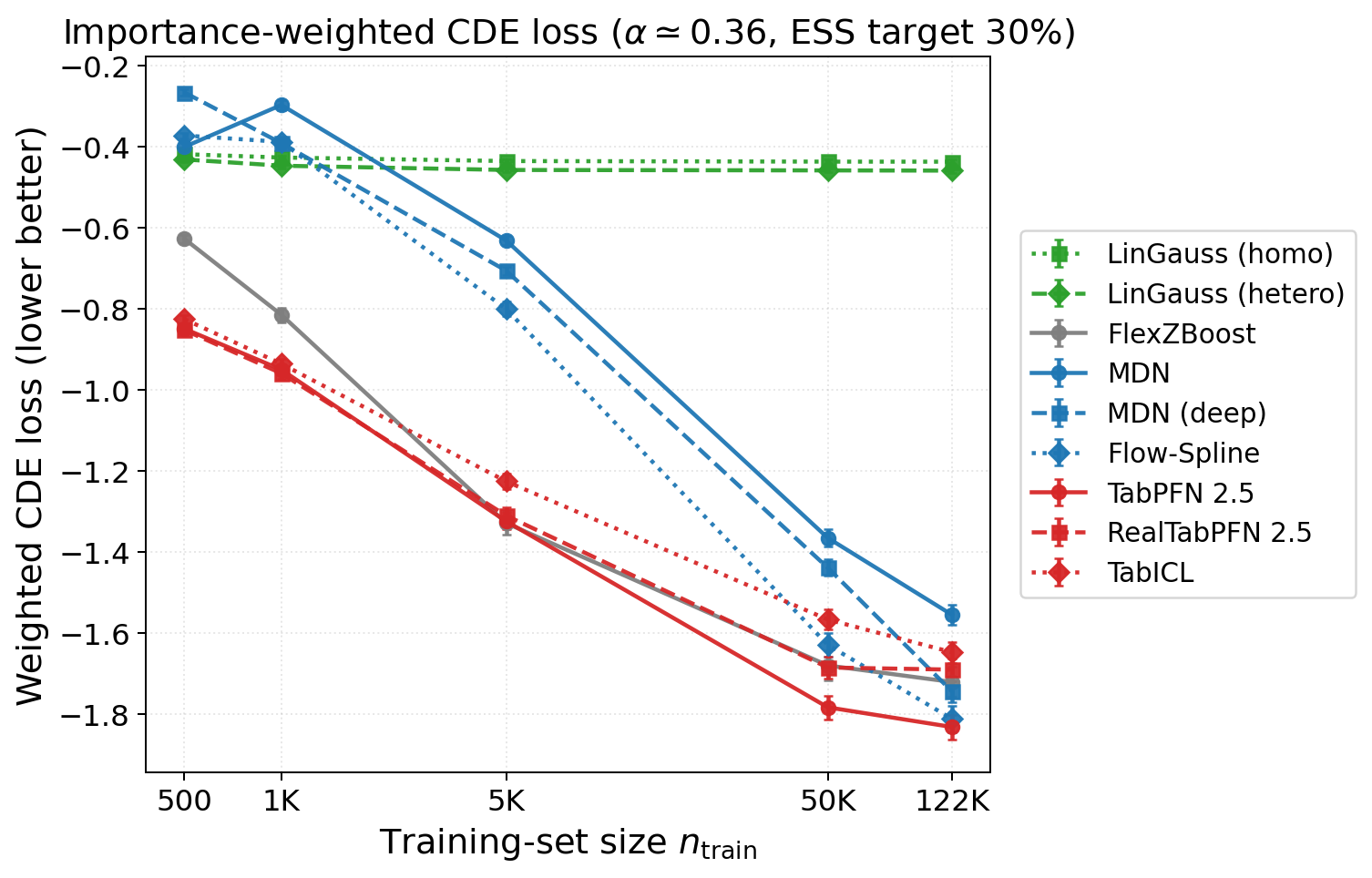}
\caption{Weighted CDE loss ($\alpha \simeq 0.36$).}
\end{subfigure}\hfill
\begin{subfigure}[t]{0.58\linewidth}
\includegraphics[width=\linewidth]{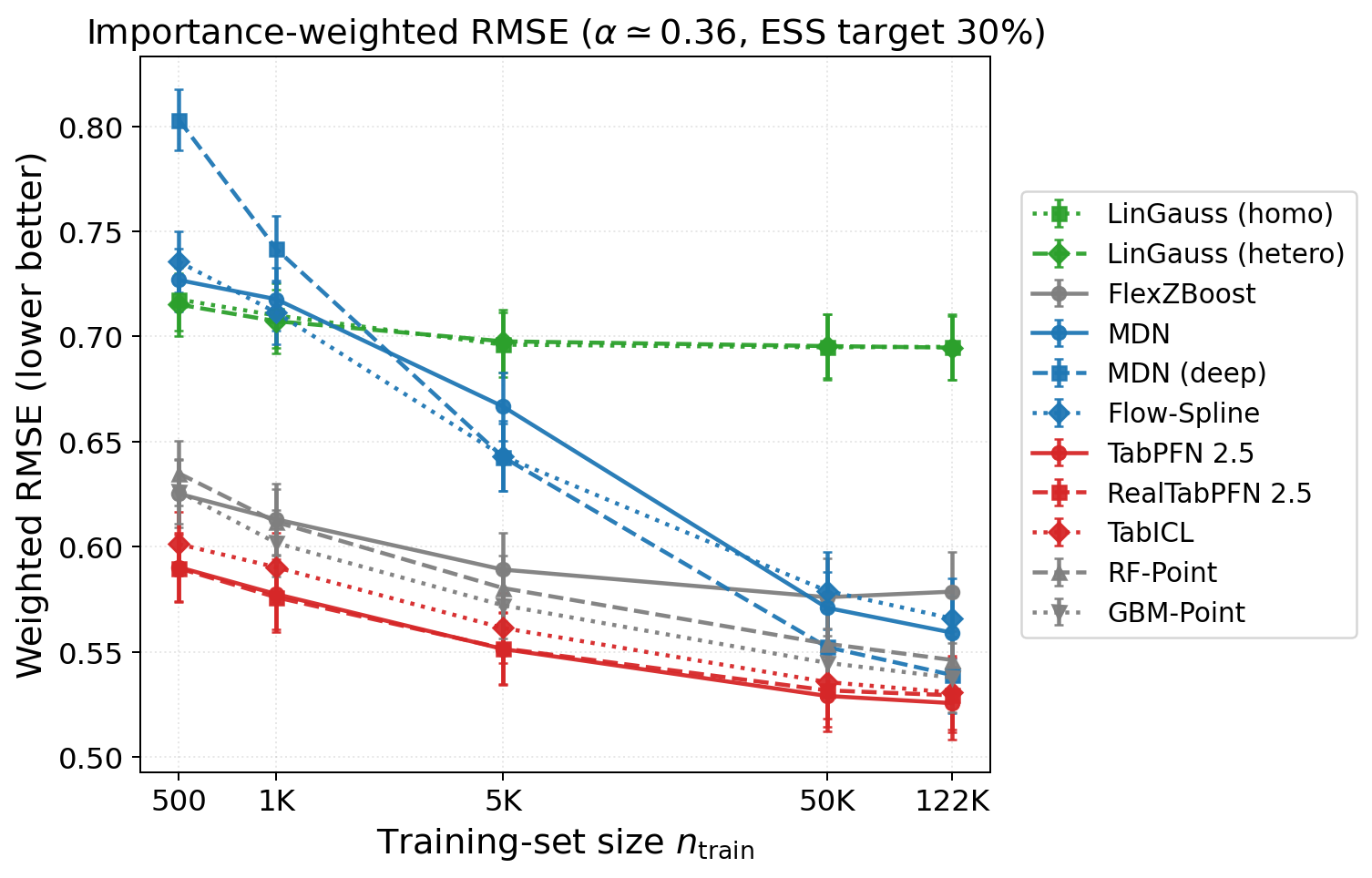}
\caption{Weighted RMSE ($\alpha \simeq 0.36$).}
\end{subfigure}
\caption{Performance versus training-set size under importance-weighted evaluation with the tuned recalibration exponent ($\alpha \simeq 0.36$, ESS target $30\%$).  Conventions match Figure~\ref{fig:scaling_unweighted}.  Under shift-corrected evaluation foundation models lead both metrics across the full range of $n_{\mathrm{train}}$.}
\label{fig:scaling_weighted}
\end{figure*}

Figure~\ref{fig:scaling_unweighted} shows how the CDE loss and RMSE evolve with $n_{\mathrm{train}}$ in the unweighted setting, and Figure~\ref{fig:scaling_weighted} repeats the comparison under the tuned $\alpha$ ($\alpha \simeq 0.36$).

The foundation models (TabPFN~2.5, RealTabPFN~2.5, and TabICL) exhibit a distinctive scaling behaviour. At the smallest training sizes ($n_{\mathrm{train}} \leq 1\,000$), they perform substantially better than the baselines on the displayed CDE-loss and RMSE curves, consistent with a strong inductive bias from pre-training. As the training set grows, the tree-based and neural estimators close part of the gap. At $n_{\mathrm{train}} = 121\,626$, Flow-Spline becomes competitive with TabPFN~2.5 on the unweighted CDE loss specifically. However, TabPFN~2.5 retains a visible advantage on RMSE, both unweighted and weighted, and on the weighted CDE loss under the tempered shift correction. Thus, the main scaling result is not that foundation models dominate every metric at the largest sample size, but that they perform especially well in the small-data regime and remain among the leading methods at the full DR6 training size.

\subsection{SHAP interpretation of TabPFN~2.5}
\label{sec:res_shap}

Finally, we estimated local feature attributions for TabPFN~2.5 using the TabPFN interpretability extension with the \texttt{shapiq} backend. We used an ensemble of eight TabPFN estimators and explained 100 randomly selected objects from the test set, using a subsample of 5,000 training examples for the attribution analysis. All features were standardized using the same scaling procedure fitted on the DR6 training set, and the interpretability analysis was performed in this standardized feature space.

\begin{figure*}[t]
\centering
\includegraphics[width=\textwidth]{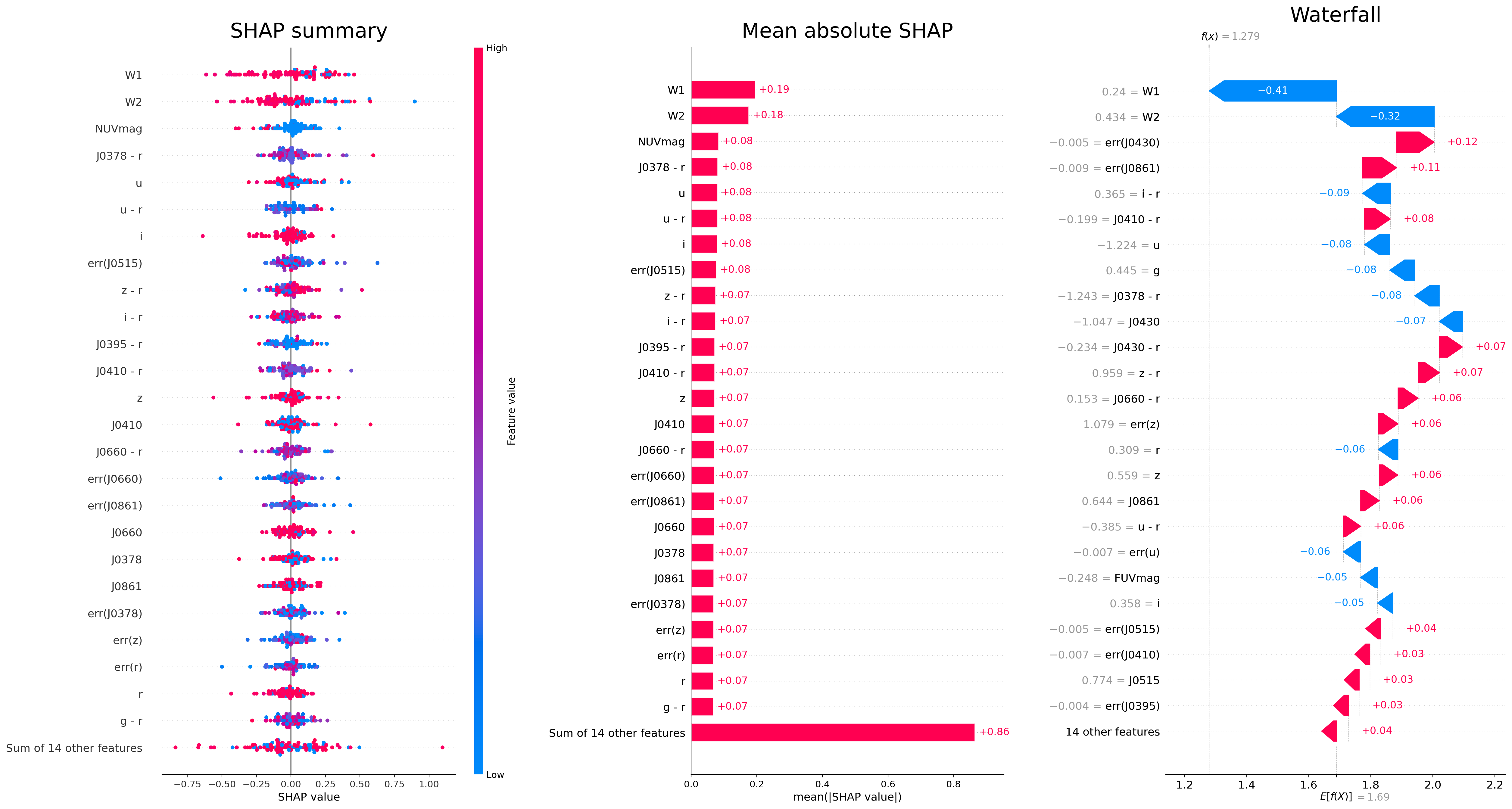}
\caption{SHAP interpretation for TabPFN~2.5.  Left: beeswarm summary over $100$ explained test objects.  Features are ordered by mean absolute SHAP value; each point is one object, the horizontal position gives the contribution to the predicted redshift, and colour encodes the feature value from low (blue) to high (red).  Middle: global feature-importance bar plot, where each bar is the mean absolute SHAP value over the explained objects.  The two global panels show the $25$ highest-ranked named features, with the remaining $14$ features grouped in the final aggregate row.  Right: waterfall explanation for one prediction ($\mathrm{test\_index}=102$), showing how the model moves from $E[f(X)] = 1.69$ to $f(x)=1.279$, compared with $z_{\mathrm{spec}}=1.423$.}
\label{fig:shap_tabpfn_summary_waterfall}
\end{figure*}

We computed first-order Shapley-value attributions, which quantify the contribution of each individual feature to a prediction without including higher-order feature interactions \citep{lundberg2017unified,muschalik2024shapiq}. Each test object was explained separately using a sampling budget of 83 feature coalitions. This budget controls the number of coalitions used to approximate the Shapley values; increasing it generally improves the approximation but also increases the computational cost \citep{muschalik2024shapiq,priorlabs_tabpfn_interpretability}. In this regression setting, positive attributions correspond to features that increase the predicted redshift relative to the reference prediction, while negative attributions correspond to features that lower it. Figure~\ref{fig:shap_tabpfn_summary_waterfall} combines a global beeswarm summary, the standard mean-absolute-SHAP feature-importance plot, and a local waterfall explanation for one representative source.

The global SHAP panels show that the WISE mid-infrared features are the largest individual drivers of the TabPFN~2.5 predictions.  \texttt{W1} and \texttt{W2} have the largest individual mean absolute SHAP values and the broadest beeswarm spreads, and high values of both features mostly push the predicted redshift downward.  The attribution pattern is not limited to these two variables, however: \texttt{NUVmag}, S-PLUS magnitudes, colours relative to the $r$ band, narrow-band colours, and selected magnitude-error terms form a long tail of smaller individual effects.  The aggregate remainder row in the global plots shows that these lower-ranked features collectively carry substantial attribution mass.

The waterfall panel shows the same behaviour for a single object.  Starting from $E[f(X)] = 1.69$, the prediction is lowered to $f(x)=1.279$, mainly by \texttt{W1} ($-0.41$) and \texttt{W2} ($-0.32$).  Other optical, narrow-band, UV, and uncertainty features partly offset this decrease, but not enough to remove the underprediction relative to $z_{\mathrm{spec}}=1.423$.

TabPFN~2.5 therefore appears to use \texttt{W1} and \texttt{W2} as its strongest individual redshift indicators, while combining many weaker optical, UV, colour, and uncertainty features to refine the prediction.

\review{As a cross-model consistency check, we compared the TabPFN~2.5 attribution
ranking with TreeSHAP importances from the GBM-Point baseline.  The feature-level rankings show moderate agreement
($\rho=0.46$), with six
features shared between the two top-10 sets. In particular, both models rank
\texttt{W1}, \texttt{W2}, and \texttt{NUVmag} among their six most important
features. Differences among the optical magnitudes, colours, and uncertainty
variables should be interpreted cautiously because predictive importance can
be redistributed among correlated features.
}

%% file: table_density.tex
\begin{table*}[t]
\centering
\caption{Density-estimation metrics on the DR6 test set at the largest training-set size ($n_{\mathrm{train}} = 121\,626$), reported as mean $\pm$ standard error. The panels correspond to three evaluation settings, and \emph{rankings should be compared only within each panel}: (a) unweighted metrics; (b) importance-weighted metrics with recalibrated weights $\hat{\beta}(\mathbf{x})^{\alpha}$ chosen to target an effective sample size of $30\%$; (c) importance-weighted metrics at $\alpha = 1$ (untransformed weights). For each metric the best method is shown in bold, together with all methods whose value lies within one combined standard error of the best.}
\label{tab:density_combined}
\begin{tabular*}{\textwidth}{@{\extracolsep{\fill}}lccccc}
\toprule
Method & CDE loss & log-likelihood & CRPS & PIT KS & Cov$_{90}$ \\
\midrule
\multicolumn{6}{l}{\textbf{(a)~No weights (unweighted metrics)}} \\
\cmidrule(lr){1-6}
\multicolumn{6}{l}{\quad\textit{\small Foundation models}} \\
\quad TabPFN~2.5 & \textbf{-2.558 $\pm$ 0.032} & \textbf{0.191 $\pm$ 0.013} & \textbf{0.188 $\pm$ 0.002} & \textbf{0.015 $\pm$ 0.004} & \textbf{0.901 $\pm$ 0.002} \\
\quad RealTabPFN~2.5 & -2.332 $\pm$ 0.030 & 0.125 $\pm$ 0.012 & \textbf{0.191 $\pm$ 0.002} & \textbf{0.015 $\pm$ 0.003} & 0.897 $\pm$ 0.002 \\
\quad TabICL & -2.271 $\pm$ 0.027 & -0.090 $\pm$ 0.024 & 0.192 $\pm$ 0.002 & 0.031 $\pm$ 0.004 & 0.907 $\pm$ 0.002 \\
\addlinespace[1pt]
\multicolumn{6}{l}{\quad\textit{\small Other methods}} \\
\quad LinGauss (homo) & -0.431 $\pm$ 0.003 & -1.029 $\pm$ 0.009 & 0.375 $\pm$ 0.003 & 0.026 $\pm$ 0.004 & 0.921 $\pm$ 0.002 \\
\quad LinGauss (hetero) & -0.461 $\pm$ 0.003 & -0.951 $\pm$ 0.008 & 0.368 $\pm$ 0.003 & 0.033 $\pm$ 0.004 & 0.916 $\pm$ 0.002 \\
\quad FlexZBoost & -2.539 $\pm$ 0.034 & -0.034 $\pm$ 0.022 & 0.217 $\pm$ 0.003 & \textbf{0.018 $\pm$ 0.003} & 0.883 $\pm$ 0.003 \\
\quad MDN & -2.188 $\pm$ 0.026 & 0.058 $\pm$ 0.012 & 0.210 $\pm$ 0.002 & 0.024 $\pm$ 0.004 & 0.912 $\pm$ 0.002 \\
\quad MDN (deep) & -2.454 $\pm$ 0.027 & 0.144 $\pm$ 0.012 & 0.194 $\pm$ 0.002 & 0.042 $\pm$ 0.004 & 0.912 $\pm$ 0.002 \\
\quad Flow-Spline & \textbf{-2.593 $\pm$ 0.034} & \textbf{0.175 $\pm$ 0.013} & 0.206 $\pm$ 0.003 & 0.027 $\pm$ 0.005 & 0.905 $\pm$ 0.002 \\
\addlinespace[6pt]
\multicolumn{6}{l}{\textbf{(b)~Importance-weighted, ESS target $30\%$ ($\alpha \simeq 0.36$)}} \\
\cmidrule(lr){1-6}
\multicolumn{6}{l}{\quad\textit{\small Foundation models}} \\
\quad TabPFN~2.5 & \textbf{-1.833 $\pm$ 0.031} & \textbf{-0.181 $\pm$ 0.022} & \textbf{0.239 $\pm$ 0.006} & \textbf{0.010 $\pm$ 0.004} & \textbf{0.896 $\pm$ 0.005} \\
\quad RealTabPFN~2.5 & -1.691 $\pm$ 0.028 & -0.223 $\pm$ 0.022 & \textbf{0.241 $\pm$ 0.006} & 0.017 $\pm$ 0.004 & 0.886 $\pm$ 0.005 \\
\quad TabICL & -1.648 $\pm$ 0.026 & -0.444 $\pm$ 0.040 & \textbf{0.242 $\pm$ 0.006} & 0.025 $\pm$ 0.006 & 0.908 $\pm$ 0.004 \\
\addlinespace[1pt]
\multicolumn{6}{l}{\quad\textit{\small Other methods}} \\
\quad LinGauss (homo) & -0.438 $\pm$ 0.005 & -1.054 $\pm$ 0.024 & 0.376 $\pm$ 0.006 & 0.027 $\pm$ 0.002 & 0.924 $\pm$ 0.003 \\
\quad LinGauss (hetero) & -0.460 $\pm$ 0.005 & -0.990 $\pm$ 0.022 & 0.370 $\pm$ 0.006 & 0.030 $\pm$ 0.003 & 0.923 $\pm$ 0.004 \\
\quad FlexZBoost & -1.722 $\pm$ 0.037 & -0.464 $\pm$ 0.042 & 0.268 $\pm$ 0.007 & 0.038 $\pm$ 0.005 & 0.842 $\pm$ 0.007 \\
\quad MDN & -1.556 $\pm$ 0.025 & -0.300 $\pm$ 0.021 & 0.260 $\pm$ 0.006 & 0.022 $\pm$ 0.007 & \textbf{0.903 $\pm$ 0.004} \\
\quad MDN (deep) & -1.745 $\pm$ 0.026 & -0.220 $\pm$ 0.018 & \textbf{0.246 $\pm$ 0.006} & 0.037 $\pm$ 0.010 & 0.910 $\pm$ 0.004 \\
\quad Flow-Spline & \textbf{-1.813 $\pm$ 0.032} & -0.239 $\pm$ 0.027 & 0.257 $\pm$ 0.006 & 0.021 $\pm$ 0.011 & \textbf{0.896 $\pm$ 0.004} \\
\addlinespace[6pt]
\multicolumn{6}{l}{\textbf{(c)~Importance-weighted, $\alpha = 1$ (untransformed weights)}} \\
\cmidrule(lr){1-6}
\multicolumn{6}{l}{\quad\textit{\small Foundation models}} \\
\quad TabPFN~2.5 & \textbf{-0.652 $\pm$ 0.039} & \textbf{-0.764 $\pm$ 0.063} & \textbf{0.303 $\pm$ 0.018} & \textbf{0.064 $\pm$ 0.023} & \textbf{0.910 $\pm$ 0.015} \\
\quad RealTabPFN~2.5 & \textbf{-0.662 $\pm$ 0.043} & \textbf{-0.750 $\pm$ 0.066} & \textbf{0.301 $\pm$ 0.018} & \textbf{0.057 $\pm$ 0.022} & \textbf{0.882 $\pm$ 0.020} \\
\quad TabICL & \textbf{-0.636 $\pm$ 0.032} & -0.875 $\pm$ 0.101 & \textbf{0.301 $\pm$ 0.017} & \textbf{0.079 $\pm$ 0.023} & 0.932 $\pm$ 0.013 \\
\addlinespace[1pt]
\multicolumn{6}{l}{\quad\textit{\small Other methods}} \\
\quad LinGauss (homo) & -0.487 $\pm$ 0.019 & -0.962 $\pm$ 0.063 & 0.338 $\pm$ 0.018 & 0.091 $\pm$ 0.022 & 0.950 $\pm$ 0.010 \\
\quad LinGauss (hetero) & -0.498 $\pm$ 0.020 & -0.963 $\pm$ 0.073 & 0.336 $\pm$ 0.017 & 0.119 $\pm$ 0.025 & 0.951 $\pm$ 0.011 \\
\quad FlexZBoost & -0.448 $\pm$ 0.096 & -0.997 $\pm$ 0.094 & \textbf{0.316 $\pm$ 0.020} & 0.098 $\pm$ 0.022 & 0.794 $\pm$ 0.029 \\
\quad MDN & -0.576 $\pm$ 0.036 & \textbf{-0.825 $\pm$ 0.059} & \textbf{0.313 $\pm$ 0.017} & \textbf{0.061 $\pm$ 0.021} & \textbf{0.905 $\pm$ 0.015} \\
\quad MDN (deep) & \textbf{-0.619 $\pm$ 0.030} & \textbf{-0.766 $\pm$ 0.050} & \textbf{0.304 $\pm$ 0.017} & \textbf{0.074 $\pm$ 0.024} & 0.923 $\pm$ 0.014 \\
\quad Flow-Spline & -0.575 $\pm$ 0.052 & \textbf{-0.830 $\pm$ 0.073} & \textbf{0.307 $\pm$ 0.018} & \textbf{0.069 $\pm$ 0.023} & \textbf{0.901 $\pm$ 0.016} \\
\bottomrule
\end{tabular*}
\end{table*}

%% file: table_point.tex
\begin{table*}[t]
\centering
\caption{Point-prediction metrics  at the largest training-set size, reported as mean $\pm$ s.e. The panels correspond to three evaluation settings, and \emph{rankings should be compared only within each panel}: (a) unweighted metrics; (b) importance-weighted metrics with recalibrated weights  chosen to target an effective sample size of $30\%$; (c) importance-weighted metrics at $\alpha = 1$. For each metric the best method is shown in bold, together with all methods within one combined s.e. of the best.}
\label{tab:point_combined}
\begin{tabular}{lccccc}
\toprule
Method & RMSE & Bias & NMAD & $\eta_{0.15}$ & $\eta_{0.30}$ \\
\midrule
\multicolumn{6}{l}{\textbf{(a)~No weights (unweighted metrics)}} \\
\cmidrule(lr){1-6}
\multicolumn{6}{l}{\quad\textit{\small Foundation models}} \\
\quad TabPFN~2.5 & \textbf{0.4434 $\pm$ 0.0078} & \textbf{+0.0066 $\pm$ 0.0040} & \textbf{0.0853 $\pm$ 0.0015} & \textbf{0.2222 $\pm$ 0.0032} & \textbf{0.0663 $\pm$ 0.0021} \\
\quad RealTabPFN~2.5 & \textbf{0.4474 $\pm$ 0.0078} & \textbf{+0.0030 $\pm$ 0.0040} & 0.0888 $\pm$ 0.0015 & 0.2278 $\pm$ 0.0032 & \textbf{0.0688 $\pm$ 0.0022} \\
\quad TabICL & \textbf{0.4497 $\pm$ 0.0078} & \textbf{-0.0040 $\pm$ 0.0040} & \textbf{0.0868 $\pm$ 0.0013} & 0.2296 $\pm$ 0.0033 & 0.0707 $\pm$ 0.0022 \\
\addlinespace[1pt]
\multicolumn{6}{l}{\quad\textit{\small Other density methods}} \\
\quad LinGauss (homo) & 0.6788 $\pm$ 0.0064 & \textbf{+0.0005 $\pm$ 0.0065} & 0.2542 $\pm$ 0.0027 & 0.5513 $\pm$ 0.0042 & 0.2294 $\pm$ 0.0035 \\
\quad LinGauss (hetero) & 0.6835 $\pm$ 0.0062 & \textbf{+0.0030 $\pm$ 0.0065} & 0.2463 $\pm$ 0.0030 & 0.5338 $\pm$ 0.0043 & 0.2317 $\pm$ 0.0035 \\
\quad FlexZBoost & 0.5047 $\pm$ 0.0085 & +0.0099 $\pm$ 0.0044 & 0.1179 $\pm$ 0.0014 & 0.2846 $\pm$ 0.0036 & 0.0858 $\pm$ 0.0024 \\
\quad MDN & 0.4814 $\pm$ 0.0074 & \textbf{-0.0012 $\pm$ 0.0043} & 0.1039 $\pm$ 0.0025 & 0.2657 $\pm$ 0.0031 & 0.0844 $\pm$ 0.0022 \\
\quad MDN (deep) & 0.4571 $\pm$ 0.0078 & +0.0118 $\pm$ 0.0041 & 0.0879 $\pm$ 0.0016 & 0.2334 $\pm$ 0.0031 & 0.0706 $\pm$ 0.0021 \\
\quad Flow-Spline & 0.4829 $\pm$ 0.0084 & +0.0096 $\pm$ 0.0041 & 0.0994 $\pm$ 0.0027 & 0.2564 $\pm$ 0.0031 & 0.0805 $\pm$ 0.0022 \\
\addlinespace[1pt]
\multicolumn{6}{l}{\quad\textit{\small Point estimators only}} \\
\quad RF-Point & 0.4700 $\pm$ 0.0075 & \textbf{-0.0071 $\pm$ 0.0044} & 0.1079 $\pm$ 0.0014 & 0.2568 $\pm$ 0.0031 & 0.0795 $\pm$ 0.0024 \\
\quad GBM-Point & 0.4621 $\pm$ 0.0075 & \textbf{-0.0037 $\pm$ 0.0043} & 0.1101 $\pm$ 0.0016 & 0.2539 $\pm$ 0.0033 & 0.0754 $\pm$ 0.0023 \\
\addlinespace[6pt]
\multicolumn{6}{l}{\textbf{(b)~Importance-weighted, ESS target $30\%$ ($\alpha \simeq 0.36$)}} \\
\cmidrule(lr){1-6}
\multicolumn{6}{l}{\quad\textit{\small Foundation models}} \\
\quad TabPFN~2.5 & \textbf{0.5256 $\pm$ 0.0172} & \textbf{+0.0086 $\pm$ 0.0090} & \textbf{0.1156 $\pm$ 0.0033} & \textbf{0.2787 $\pm$ 0.0073} & \textbf{0.0821 $\pm$ 0.0049} \\
\quad RealTabPFN~2.5 & \textbf{0.5293 $\pm$ 0.0177} & \textbf{+0.0052 $\pm$ 0.0092} & \textbf{0.1170 $\pm$ 0.0032} & \textbf{0.2849 $\pm$ 0.0076} & \textbf{0.0849 $\pm$ 0.0050} \\
\quad TabICL & \textbf{0.5306 $\pm$ 0.0177} & \textbf{+0.0020 $\pm$ 0.0093} & \textbf{0.1165 $\pm$ 0.0029} & \textbf{0.2860 $\pm$ 0.0072} & \textbf{0.0842 $\pm$ 0.0048} \\
\addlinespace[1pt]
\multicolumn{6}{l}{\quad\textit{\small Other density methods}} \\
\quad LinGauss (homo) & 0.6950 $\pm$ 0.0156 & +0.0306 $\pm$ 0.0112 & 0.2325 $\pm$ 0.0044 & 0.5154 $\pm$ 0.0076 & 0.2054 $\pm$ 0.0056 \\
\quad LinGauss (hetero) & 0.6947 $\pm$ 0.0151 & +0.0405 $\pm$ 0.0112 & 0.2237 $\pm$ 0.0040 & 0.5017 $\pm$ 0.0076 & 0.2007 $\pm$ 0.0053 \\
\quad FlexZBoost & 0.5786 $\pm$ 0.0187 & +0.0327 $\pm$ 0.0097 & 0.1400 $\pm$ 0.0031 & 0.3257 $\pm$ 0.0074 & 0.0959 $\pm$ 0.0050 \\
\quad MDN & 0.5591 $\pm$ 0.0171 & \textbf{-0.0023 $\pm$ 0.0093} & 0.1335 $\pm$ 0.0045 & 0.3165 $\pm$ 0.0066 & 0.1003 $\pm$ 0.0048 \\
\quad MDN (deep) & \textbf{0.5389 $\pm$ 0.0181} & +0.0206 $\pm$ 0.0091 & \textbf{0.1199 $\pm$ 0.0035} & 0.2936 $\pm$ 0.0071 & \textbf{0.0844 $\pm$ 0.0046} \\
\quad Flow-Spline & 0.5658 $\pm$ 0.0189 & +0.0210 $\pm$ 0.0094 & 0.1276 $\pm$ 0.0038 & 0.3071 $\pm$ 0.0067 & 0.0938 $\pm$ 0.0047 \\
\addlinespace[1pt]
\multicolumn{6}{l}{\quad\textit{\small Point estimators only}} \\
\quad RF-Point & \textbf{0.5461 $\pm$ 0.0159} & \textbf{-0.0105 $\pm$ 0.0097} & 0.1300 $\pm$ 0.0032 & 0.3134 $\pm$ 0.0072 & 0.0960 $\pm$ 0.0051 \\
\quad GBM-Point & \textbf{0.5377 $\pm$ 0.0166} & \textbf{-0.0038 $\pm$ 0.0094} & 0.1347 $\pm$ 0.0034 & 0.3051 $\pm$ 0.0071 & 0.0893 $\pm$ 0.0051 \\
\addlinespace[6pt]
\multicolumn{6}{l}{\textbf{(c)~Importance-weighted, $\alpha = 1$ (untransformed weights)}} \\
\cmidrule(lr){1-6}
\multicolumn{6}{l}{\quad\textit{\small Foundation models}} \\
\quad TabPFN~2.5 & \textbf{0.5830 $\pm$ 0.0424} & \textbf{+0.0116 $\pm$ 0.0326} & \textbf{0.1563 $\pm$ 0.0143} & \textbf{0.3320 $\pm$ 0.0316} & \textbf{0.0807 $\pm$ 0.0155} \\
\quad RealTabPFN~2.5 & \textbf{0.5826 $\pm$ 0.0428} & \textbf{+0.0071 $\pm$ 0.0326} & \textbf{0.1579 $\pm$ 0.0145} & \textbf{0.3422 $\pm$ 0.0326} & \textbf{0.0831 $\pm$ 0.0157} \\
\quad TabICL & \textbf{0.5801 $\pm$ 0.0431} & \textbf{+0.0105 $\pm$ 0.0331} & \textbf{0.1564 $\pm$ 0.0144} & \textbf{0.3375 $\pm$ 0.0318} & \textbf{0.0779 $\pm$ 0.0155} \\
\addlinespace[1pt]
\multicolumn{6}{l}{\quad\textit{\small Other density methods}} \\
\quad LinGauss (homo) & \textbf{0.6308 $\pm$ 0.0438} & +0.0890 $\pm$ 0.0361 & \textbf{0.1761 $\pm$ 0.0149} & 0.4005 $\pm$ 0.0319 & 0.1265 $\pm$ 0.0211 \\
\quad LinGauss (hetero) & \textbf{0.6252 $\pm$ 0.0429} & +0.1101 $\pm$ 0.0355 & \textbf{0.1718 $\pm$ 0.0147} & 0.3902 $\pm$ 0.0302 & \textbf{0.0982 $\pm$ 0.0150} \\
\quad FlexZBoost & \textbf{0.6029 $\pm$ 0.0429} & +0.0609 $\pm$ 0.0322 & \textbf{0.1650 $\pm$ 0.0152} & \textbf{0.3507 $\pm$ 0.0320} & \textbf{0.0839 $\pm$ 0.0156} \\
\quad MDN & \textbf{0.5945 $\pm$ 0.0411} & \textbf{+0.0007 $\pm$ 0.0320} & \textbf{0.1702 $\pm$ 0.0159} & \textbf{0.3513 $\pm$ 0.0285} & \textbf{0.0971 $\pm$ 0.0156} \\
\quad MDN (deep) & \textbf{0.5864 $\pm$ 0.0432} & \textbf{+0.0425 $\pm$ 0.0324} & \textbf{0.1595 $\pm$ 0.0162} & \textbf{0.3625 $\pm$ 0.0288} & \textbf{0.0784 $\pm$ 0.0140} \\
\quad Flow-Spline & \textbf{0.5938 $\pm$ 0.0424} & \textbf{+0.0409 $\pm$ 0.0321} & \textbf{0.1594 $\pm$ 0.0156} & \textbf{0.3443 $\pm$ 0.0272} & \textbf{0.0874 $\pm$ 0.0147} \\
\addlinespace[1pt]
\multicolumn{6}{l}{\quad\textit{\small Point estimators only}} \\
\quad RF-Point & \textbf{0.5997 $\pm$ 0.0407} & \textbf{-0.0230 $\pm$ 0.0359} & \textbf{0.1718 $\pm$ 0.0165} & \textbf{0.3770 $\pm$ 0.0323} & \textbf{0.0873 $\pm$ 0.0160} \\
\quad GBM-Point & \textbf{0.5876 $\pm$ 0.0423} & \textbf{-0.0003 $\pm$ 0.0337} & \textbf{0.1692 $\pm$ 0.0148} & \textbf{0.3498 $\pm$ 0.0316} & \textbf{0.0860 $\pm$ 0.0169} \\
\bottomrule
\end{tabular}
\end{table*}

%% file: conclusions.tex
\label{sec:conclusions}

We have benchmarked three tabular foundation models---TabPFN~2.5, RealTabPFN~2.5, and TabICL---against eight established baselines for quasar photo-$z$ estimation on S-PLUS DR6. Six of the baselines produce full conditional densities---two linear conditional Gaussians, FlexZBoost, two mixture density networks, and a quadratic-spline normalising flow---while two are point estimators only, RF-Point and GBM-Point. The comparison spans density quality, calibration, point-prediction accuracy, catastrophic-outlier rates, training-set sizes from $n_{\mathrm{train}}=500$ to $121\,626$, and both unweighted and importance-weighted evaluations. The importance-weighted evaluation is intended to approximate performance on the photometric quasar-candidate population, rather than only on the spectroscopic test set.

The main result is that tabular foundation models are statistically competitive with, and often outperform, the strongest task-specific baselines. TabPFN~2.5 is the most consistently strong method: it is best or statistically tied for best on the density metrics considered here, including CDE loss, log-likelihood, CRPS, PIT-KS, and $90\%$ coverage, and also on the main point-prediction metrics, including RMSE, NMAD, and the catastrophic-outlier fractions. The main exception is unweighted CDE loss, where Flow-Spline is statistically tied with TabPFN~2.5 and achieves the lowest mean value, as expected for a model trained directly to optimise that type of density score. Thus, the evidence is that TabPFN~2.5 provides the best overall balance of density quality, calibration, point accuracy, and outlier control.

The advantage of TabPFN~2.5 is most apparent in the regimes where quasar photo-$z$ estimation is hardest. The foundation-model lead is largest for catastrophic-outlier fractions and in sparsely populated regions of the data distribution, including the bright end, the faint tail, and high redshift. At small training-set sizes, especially $n_{\mathrm{train}}\leq 1\,000$, the gap between the foundation models and the task-specific baselines is wide on nearly all metrics. This behaviour is consistent with the strong inductive bias supplied by the pre-trained tabular prior, which is most useful when the spectroscopic support set is small or unevenly distributed.

The importance-weighted evaluation changes both the absolute performance levels and the interpretation of method rankings. In this benchmark, weighting the spectroscopic test set toward the photometric quasar-candidate population degrades performance for all methods, consistent with the target sample being more difficult than the labelled spectroscopic sample. The degradation is not uniform across methods. FlexZBoost, for example, has $90\%$ coverage of $0.883$ in the unweighted evaluation but drops to $0.794$ at $\alpha=1$, while its PIT-KS statistic increases from $0.018$ to $0.098$. By contrast, TabPFN~2.5 remains comparatively stable, with PIT-KS values of $0.010$--$0.015$ across the unweighted and tuned-$\alpha$ panels. In this dataset, reporting only unweighted spectroscopic metrics would therefore give an optimistic view of deployed performance and could hide method-level brittleness under covariate shift.

The feature-attribution analysis indicates that the WISE \texttt{W1} and \texttt{W2} features are the strongest individual drivers of the TabPFN~2.5 predictions. This result is consistent with the findings of \citealt{stern2012} and \citealt{assef2013}, who demonstrated that the WISE W1-W2 color is among the most effective individual features for quasar photometric redshift estimation. GALEX \texttt{NUVmag}, the S-PLUS optical and narrow-band magnitudes and colours, and selected magnitude-error terms provide a broad tail of smaller corrections that are collectively non-negligible. This pattern is consistent with the known diagnostic value of mid-infrared information for quasar redshift estimation, while also showing that the model uses multi-band optical, UV, colour, and uncertainty information rather than reducing to a two-feature rule. Because the analysis uses first-order Shapley values on a finite test subset, these attributions should be interpreted as evidence about model behaviour rather than as a physical decomposition of the colour--redshift relation.

The main practical caveat is computational. TabPFN~2.5 shifts the cost from training and model selection to inference, where the labelled spectroscopic sample is supplied as an in-context support set. In our benchmark, using the full support set of $n_{\rm train}=121{,}626$ with $n_{\rm test}=13{,}538$ took approximately 76 minutes on a Strix Halo machine, compared with approximately 14 minutes for $n_{\rm train}=50{,}000$ on the same test set. This is feasible at the scale studied here, but should not be interpreted as a scalability guarantee for full-catalogue deployment with millions of photometric quasars. The total runtime also grows with the number of target objects, since large catalogues must be processed in batches, partly because of memory constraints. Large-scale deployment may therefore require support-set subsampling, distillation into a cheaper survey-specific model, or other acceleration strategies.

\review{Several limitations remain. Although the comparison with gradient boosting in Appendix~\ref{app:weight_estimator} tests sensitivity to the density-ratio estimator, uncertainty in the weights and impurity of the photometric candidate catalogue remain.  A purity-aware extension could use calibrated candidate probabilities $P(\mathrm{QSO}\mid\mathbf{x})$ as target-sample weights, repeat the analysis under different probability thresholds or assumed contamination fractions, and validate the results using representative spectroscopic follow-up.  The covariate-shift assumption may also fail if the observed features do not fully encode spectroscopic selection or if non-detections differ between samples, and cannot be used if the photometric catalogue includes regions with little spectroscopic support. The benchmark is restricted to quasars in S-PLUS DR6; applying the same protocol to galaxies, stars, J-PLUS \citep{cenarro+19}, J-PAS \citep{Benitez2014}, or other photometric surveys is necessary before extending the recommendation beyond S-PLUS-like quasar samples.}

Overall, TabPFN~2.5 is a strong default for S-PLUS-like probabilistic quasar photo-$z$ inference, particularly when calibrated conditional densities, reduced catastrophic-outlier rates, and strong performance at small training-set sizes are priorities.

%% file: appendix.tex
\section{Feature distributions under covariate shift}
\label{app:feature_shift}

\review{Table~\ref{tab:feature_shift} compares three representative features in the
spectroscopic training and photometric target samples used for importance-weight
estimation. The photometric candidates are substantially fainter in $r$ and
have shifted and broader $u-r$ and $W1-W2$ distributions. The differences in
non-detection rates are also pronounced. Thus, both the measured values and
the missingness patterns contribute to the source-to-target covariate shift.}

\begin{deluxetable}{lrrrrrrrrr}
\tablecaption{\review{Comparison of representative feature distributions in the
spectroscopic training and photometric target samples.}
\label{tab:feature_shift}}
\tablehead{
\colhead{} &
\multicolumn{3}{c}{Spectroscopic training} &
\multicolumn{3}{c}{Photometric target} &
\colhead{Missing S} &
\colhead{Missing P} &
\colhead{KS}\\
\colhead{Feature} &
\colhead{P5} &
\colhead{Median} &
\colhead{P95} &
\colhead{P5} &
\colhead{Median} &
\colhead{P95} &
\colhead{(\%)} &
\colhead{(\%)} &
\colhead{$D$}
}
\startdata
$r$       & 19.13 & 21.17 & 22.46 & 21.20 & 22.47 & 23.27 &  6.7 & 55.2 & 0.655 \\
$u-r$     & -0.50 &  0.26 &  1.28 & -1.53 & -0.25 &  2.03 & 41.4 & 94.1 & 0.388 \\
$W1-W2$   &  0.48 &  1.01 &  1.38 & -0.73 &  0.31 &  1.18 & 22.0 & 92.8 & 0.627
\enddata
\tablecomments{\review{Percentiles are calculated
using objects with valid measurements. For $u-r$ and $W1-W2$, an object is
considered missing if either constituent band is missing. ``S'' and ``P''
denote the spectroscopic and photometric samples, respectively. The final
column gives the two-sample Kolmogorov--Smirnov distance between the valid-value
empirical distributions.}}
\end{deluxetable}

\onecolumngrid

\section{Standard errors and the bold convention}
\label{app:se}

Most entries in Tables~\ref{tab:density_combined} and~\ref{tab:point_combined} are built from a per-object quantity
$g_i = g(\mathbf{x}_i, z_i, \hat{f})$.  Examples include the CDE-loss contribution,
the log density at the true redshift, the CRPS contribution, the $90\%$
coverage indicator, the signed residual for bias, the squared error for RMSE,
and the outlier indicators.  Because each method is fit on $R=5$ independent
training-set repetitions, we first compute the per-object component separately
for each repetition and then average it across repetitions,
\begin{equation}
\bar g_i \;=\; \frac{1}{R}\sum_{r=1}^{R} g_i^{(r)}.
\label{eq:rep_avg}
\end{equation}
The table entry is then the weighted test-set average of these repetition-averaged
components, followed by any transformation needed to put the result on the
reported scale:
\begin{equation}
\hat m \;=\; \phi\!\left(\sum_{i=1}^{n} w_i\,\bar g_i \right),
\qquad
w_i = \frac{\hat{\beta}(\mathbf{x}_i)^{\alpha}}{\sum_{j=1}^{n} \hat{\beta}(\mathbf{x}_j)^{\alpha}}.
\label{eq:metric_estimator}
\end{equation}
Here $\hat{\beta}(\mathbf{x})$ is the estimated density ratio from
Sect.~\ref{sec:alpha_weights}.  The three table panels correspond to
$\alpha=0$ (unweighted, so $w_i=1/n$), the variance-controlled correction
$\alpha^\star \simeq 0.36$, and the raw density-ratio weights $\alpha=1$.
For RMSE, $g_i$ is the squared error and $\phi(u)=\sqrt{u}$; for the other
metrics covered by Eq.~\eqref{eq:metric_estimator}, $\phi$ is the identity.
Thus the reported mean is an average over the five fitted repetitions, while
the standard error below measures uncertainty from the finite test set after
that repetition average has been formed.

The reported uncertainty is a \emph{test-set bootstrap} standard error.  We draw
$B=100$ bootstrap resamples of the test set with a fixed random seed, using the
same resampling indices for all methods and metrics.  On each bootstrap resample
we recompute Eq.~\eqref{eq:metric_estimator}, including both the normalisation of
the selected importance weights and the transformation $\phi$.  Equivalently, if
$I_b$ is the multiset of test-set indices in bootstrap replicate $b$, then
\begin{equation}
\hat m^{(b)}
\;=\;
\phi\!\left(
    \frac{\sum_{i\in I_b}\hat{\beta}(\mathbf{x}_i)^\alpha \bar g_i}
         {\sum_{i\in I_b}\hat{\beta}(\mathbf{x}_i)^\alpha}
\right),
\end{equation}
with $\alpha=0$ giving the ordinary unweighted bootstrap mean.  The reported SE is
\begin{equation}
\mathrm{SE}(\hat m) \;=\; \sqrt{\frac{1}{B-1}\sum_{b=1}^{B}\left(\hat m^{(b)} - \overline{\hat m}\right)^{2}}.
\label{eq:boot_se}
\end{equation}

PIT-KS and NMAD are handled separately because they do not commute with the
per-object averaging in Eq.~\eqref{eq:rep_avg}.  PIT-KS is a supremum distance
between the empirical PIT CDF and the Uniform$(0,1)$ CDF, and NMAD is a double
median of the normalised residuals.  For these two metrics we compute the
statistic separately in each repetition that has the required cached PIT or
point-prediction arrays, take the mean across repetitions as the point estimate,
and bootstrap the test objects within each repetition.  For the importance-weighted
versions, PIT-KS uses the weighted empirical PIT CDF and NMAD uses weighted
medians.  The per-repetition bootstrap values are pooled and their sample
standard deviation gives the SE.

For metrics reported with an SE, boldface is assigned by a one-SE rule based on
the \emph{combined} SE for comparing the best method $A$ with a candidate method
$B$,
\begin{equation}
\mathrm{SE}_{\mathrm{comb}} \;=\; \sqrt{\mathrm{SE}(\hat m_A)^2 + \mathrm{SE}(\hat m_B)^2},
\end{equation}
where ``best'' uses the natural direction of the metric: lower is better for
CDE loss, CRPS, RMSE, NMAD, PIT-KS, and the outlier fractions, while higher is
better for log-likelihood.  All methods satisfying
$|\hat m_A-\hat m_B|\leq\mathrm{SE}_{\mathrm{comb}}$ are included in the bold
set.  For bias, ``best'' means smallest absolute bias, and the same rule is
applied to the difference in absolute bias.  For $90\%$ coverage, we instead
bold all methods whose value $\pm$ SE contains the nominal level $0.90$; if none
do, we bold the method closest to $0.90$.  If a method-metric combination lacks
the per-repetition arrays needed for the bootstrap, the cell shows the point
estimate without an SE and only the best value, including exact ties, is bolded.

\begin{figure*}[t]
\centering
\includegraphics[width=0.8\textwidth]{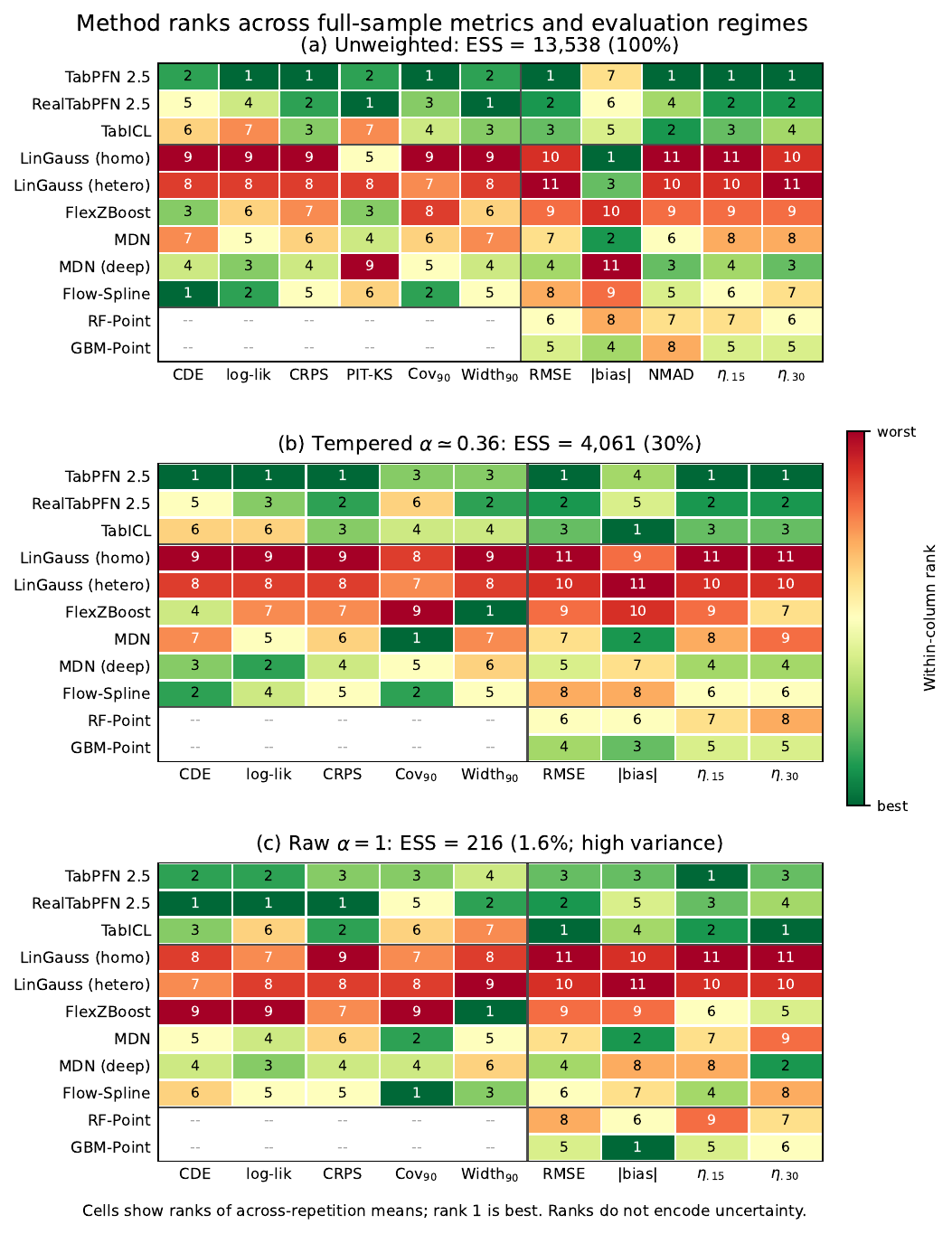}
\caption{\review{Ranks of the across-repetition mean performance at
$n_{\mathrm{train}}=121\,626$ across the reported metrics and evaluation
regimes. Rank 1 is best within each column; coverage is ranked by distance
from 0.90 and bias by absolute value. White cells denote metrics that do not
apply to point-only estimators. Ranks are descriptive and do not account for
uncertainty; inferential comparisons therefore use the standard errors and
bold sets in the main tables. The raw $\alpha=1$ regime has an effective
sample size of only 216 objects (1.6\%) and should be interpreted cautiously.}}
\label{fig:metric_rank_heatmap}
\end{figure*}

\review{Figure~\ref{fig:metric_rank_heatmap} supplements
Tables~\ref{tab:density_combined} and~\ref{tab:point_combined} by displaying
the within-metric ranks of all methods under the unweighted, tempered, and raw
importance-weighting regimes. The panel headings also show the corresponding
effective sample sizes. Because these ranks use only the across-repetition
means and do not represent uncertainty, statistical comparisons should still
be based on the reported SEs and bold sets. In particular, the raw
$\alpha=1$ results have an ESS of only 216 objects (1.6\%) and are treated as
a high-variance sensitivity analysis.}

\section{Sensitivity to the weight estimator and grid resolution}
\label{app:weight_estimator}

\review{The importance weights of Sect.~\ref{sec:alpha_weights} are obtained from a
probabilistic classifier that separates spectroscopic from photometric objects
(Eq.~\eqref{eq:weight_classifier}), and the main analysis uses TabICL for this
classifier.  Because the classification approach to density-ratio estimation is
agnostic to the choice of classifier \citep{Bickel2009}, we verify that the
weighted conclusions do not depend on this particular choice by re-estimating
the density ratio $\widehat{\beta}(\mathbf{x})$ with a structurally unrelated
classifier and repeating the full weighted evaluation.}

\review{We replace the TabICL discriminator with a gradient-boosted decision-tree
classifier (scikit-learn's \texttt{HistGradientBoostingClassifier}), which
handles the missing photometric measurements through its native missing-value
support.}

\review{At the level of the weights themselves, the two estimators agree closely.  The
exponent that brings the effective sample size to $30\%$ of the test set is
$\alpha^{\star} \simeq 0.36$ for the gradient-boosted weights, matching the value
obtained with TabICL, and the raw ($\alpha = 1$) weights collapse the ESS to
$1.8\%$ of the test set, close to the $1.6\%$ of the main analysis.  Across the
$n = 13\,538$ test objects the two weight vectors are strongly correlated, with a
Pearson correlation of $0.94$ between the log-weights; the Spearman rank
correlation is $0.77$, reduced by the many objects pinned at the lower clipping
bound.  The two classifiers therefore up- and down-weight the same objects.}

\review{Tables~\ref{tab:density_combined_gbm} and~\ref{tab:point_combined_gbm} reproduce
the density- and point-prediction results with these gradient-boosted weights, in
the same three-panel layout as Tables~\ref{tab:density_combined}
and~\ref{tab:point_combined}.  The importance-weighted rankings are essentially
unchanged.  On the variance-controlled panel ($\alpha^{\star} \simeq 0.36$), the
best method under every metric is the same as in the main analysis, and the
method ordering is preserved: it is identical for the weighted CDE loss and the
weighted CRPS, and identical among the top three for the weighted RMSE.  The
differences between the two weight estimators are at most $0.04$ in weighted CDE
loss, $0.003$ in weighted CRPS, and $0.01$ in weighted RMSE; for every method and
metric they are smaller than the combined test-set bootstrap standard error
[Eq.~\eqref{eq:boot_se}], and far smaller than the gaps between methods, so the
highlighted methods are unaffected.  We conclude that the importance-weighted
results reported in the main text are robust to the choice of weight estimator.}

\input{table_density_gbm.tex}
\input{table_point_gbm.tex}

\begin{figure*}[t]
\centering
\includegraphics[width=\textwidth]{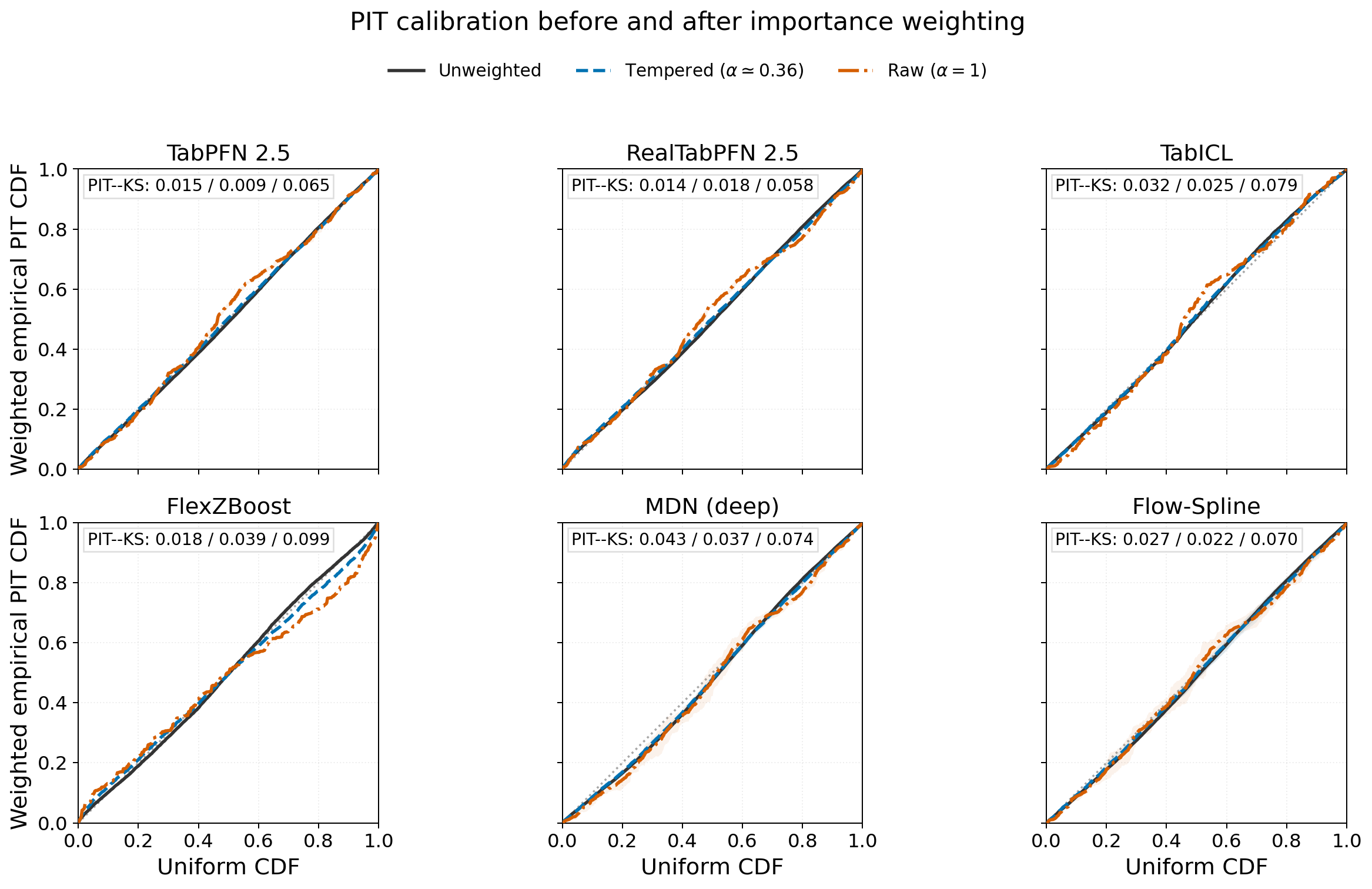}
\caption{\review{PIT P--P curves at $n_{\mathrm{train}}=121\,626$ before and after
importance weighting.  The black,
blue, and orange curves correspond to the unweighted, tempered
($\alpha\simeq0.36$), and raw ($\alpha=1$) evaluations, respectively.
The dotted diagonal represents exact PIT uniformity, and the PIT--KS values
in each panel follow the legend order. This comparison diagnoses sensitivity
to reweighting over the observed covariates.}}
\label{fig:pit_weight_comparison}
\end{figure*}

\paragraph{Grid-resolution sensitivity.}
\review{We tested numerical sensitivity by coarsening the retained 200-point
density curves to 100 equally spaced redshift values over the same support,
renormalising them, and recomputing the metrics. Across three complete
full-training repetitions, the mean CRPS changed by at most $1.1\times
10^{-3}$, the $90\%$ coverage by $0.010$, the PIT--KS statistic by $0.021$,
and the posterior-mean RMSE by $1.3\times10^{-4}$. The CRPS and RMSE
rankings were unchanged. CDE loss was more sensitive, changing by as much
as $0.51$, and the ordering of the two leading methods, TabPFN and
Flow-Spline, was reversed. We therefore find that the principal CDF-based
and point-prediction conclusions are robust to this coarsening, while
density-local metrics require greater caution.  }

\section{Difficult-case density examples}
\label{app:difficult_densities}

\begin{figure*}[t]
\centering
\includegraphics[width=\textwidth]{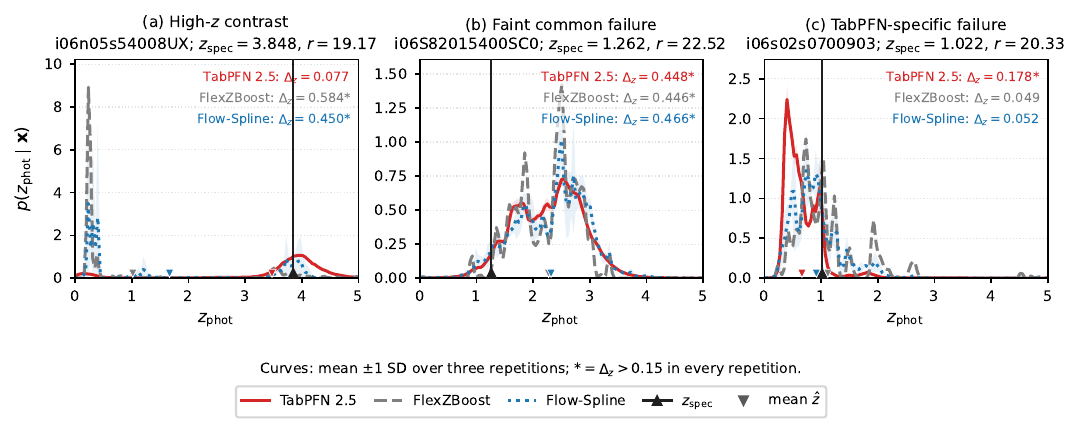}
\caption{\review{Targeted difficult photo-$z$ cases selected using the catastrophic
point-prediction criterion
$\Delta_z=|\hat z-z_{\mathrm{spec}}|/(1+z_{\mathrm{spec}})>0.15$.
Panel (a) shows a high-redshift object for which TabPFN~2.5 is
non-catastrophic while FlexZBoost and Flow-Spline are catastrophic; panel (b)
shows a faint object for which all three methods are catastrophic; and panel
(c) shows a case for which TabPFN~2.5 is catastrophic while the other two
methods are not. One object was selected uniformly with a fixed seed from each
predefined category. Curves and shaded regions
show the mean and $\pm1$ standard deviation across these repetitions. The
upward black marker denotes $z_{\mathrm{spec}}$, coloured downward markers
denote the mean point predictions, and an asterisk marks methods that are
catastrophic in every repetition.}}
\label{fig:difficult_densities}
\end{figure*}

\review{To complement the randomly selected magnitude--redshift examples in
Figure~\ref{fig:density_examples_spline}, Figure~\ref{fig:difficult_densities}
shows three cases selected using the catastrophic point-prediction criterion.
The examples were selected from predefined failure categories and are intended
to illustrate different behaviours, not their population frequencies.}

\section{Methods and tuning details}
\label{app:methods}

We give the implementation details for the eleven benchmark entries used in the main tables.   All tuning is repeated separately for each training-set size and each of the five training repetitions.  Unless noted otherwise, density methods are evaluated on a grid of 200 redshift values spanning the training-set redshift range expanded by $10\%$ on each side, and point predictions for density methods are posterior means on that grid.

\paragraph{Feature handling.}
\review{The preprocessed tables store missing measurements using the sentinel value
99. For the tree-based methods (FlexZBoost, RF-Point, and GBM-Point), this
value was retained before scaling. For the tabular foundation models
(TabPFN~2.5, RealTabPFN~2.5, and TabICL), as well as the other non-tree
methods, the sentinel was replaced by 0. All features were then standardized
using means and standard deviations fitted on the corresponding training
subset, and the same transformation was applied to the test set. No further
imputation, clipping, feature selection, dimensionality reduction, data
augmentation, or task-specific fine-tuning was performed.}

\paragraph{Random-search protocol.}  MDN, Flow-Spline, and GBM-Point are tuned by random search over the Cartesian grids listed below.  For each method we draw 200 unique configurations without replacement, using the repetition seed.  Each configuration is evaluated with shuffled 3-fold cross-validation on the current spectroscopic training subset.  The selection criterion is the unweighted CDE loss for density estimators and the unweighted mean squared error for GBM-Point.  Failed configurations are assigned infinite loss.  After selection, the winning configuration is refit on the current training subset and evaluated on the fixed test set.

\paragraph{LinGauss-Homo-Ridge and LinGauss-Hetero-Ridge.}  These are conditional Gaussian density estimators with a linear mean.  The homoscedastic model estimates a single residual standard deviation from 5-fold out-of-fold residuals.  The heteroscedastic model estimates both the mean and an input-dependent log-variance by penalized Gaussian maximum likelihood.  The ridge grid is
\[
    \alpha \in \{10^{-4}, \ldots, 10^{4}\},
\]
with 20 logarithmically spaced values and leave-one-out \texttt{RidgeCV}.  This is not part of the random-search protocol.

\paragraph{FlexZBoost.}  FlexZBoost is the FlexCode conditional density estimator \citep{IzbickiLee2017,dalmasso2020conditional} with a cosine basis and XGBoost regressors for the basis coefficients.  The XGBoost settings are fixed at 100 trees, maximum depth 4, and learning rate 0.1.  Three-fold CV selects the number of basis terms
\[
    J \in \{1,\ldots,120\}
\]
using the FlexCode loss.  A post-hoc sharpening exponent is then selected on the same out-of-fold predictions from
\[
    a \in \{0.5, 0.6, \ldots, 3.0\},
\]
that is, 26 equally spaced values.  The density is projected to be non-negative and renormalized after basis reconstruction and after sharpening.  FlexZBoost uses its internal redshift grid range, expanded by $5\%$ beyond the training redshift range.

\paragraph{MDN.}  The MDN is a one-hidden-layer mixture-density network \citep{Bishop1994}.  It standardizes the response internally, uses a ReLU hidden layer, and outputs mixture weights, means, and standard deviations for a Gaussian mixture.  Training uses Adam and early stopping patience 30.  The random-search grid is
\[
\begin{split}
K &\in \{1,2,3,5,8,12,16\},\\
h &\in \{8,16,32,64,128,256\},\\
\eta &\in \{5{\times}10^{-4},10^{-3},2{\times}10^{-3},5{\times}10^{-3},10^{-2},2{\times}10^{-2},5{\times}10^{-2}\},\\
T &\in \{200,300,500,800,1200,1600\},
\end{split}
\]
where $K$ is the number of mixture components, $h$ the hidden width, $\eta$ the learning rate, and $T$ the maximum number of epochs.

\paragraph{MDN-deep.}  MDN-deep uses the same Gaussian-mixture likelihood but a fixed deeper architecture, not random-search tuning.  The settings are $K=7$, hidden widths $(256,196,196)$, dropout 0.10, AdamW with learning rate $10^{-3}$ and weight decay $10^{-4}$, batch size 1024, maximum 120 epochs, gradient clipping at 1.0,  early-stopping patience 25, and a reduce-on-plateau learning-rate scheduler with patience 10 and factor 0.5.

\paragraph{Flow-Spline.}  Flow-Spline is a conditional rational-quadratic spline flow \citep{DurkanEtAl2019} with a standard Gaussian base density.  The response is standardized before fitting; the spline tail bound is chosen from the standardized training redshifts.  Training uses   early-stopping patience 12, and at most 120 epochs, with an epoch cap for large training subsets.  The random-search grid is
\[
\begin{split}
B &\in \{2,4,8,12,16,24,32\},\\
L &\in \{1,2,3,4,5,6\},\\
h &\in \{16,32,64,128,256,512\},\\
\eta &\in \{2{\times}10^{-4},5{\times}10^{-4},10^{-3},2{\times}10^{-3},5{\times}10^{-3},10^{-2},2{\times}10^{-2}\},\\
\lambda &\in \{10^{-8},10^{-7},10^{-6},10^{-5},10^{-4},10^{-3}\},
\end{split}
\]
where $B$ is the number of spline bins, $L$ the number of flow layers, $h$ the conditioner hidden width, $\eta$ the learning rate, and $\lambda$ the weight decay.

\paragraph{RF-Point.}
RF-Point is a scikit-learn random forest regressor used exclusively for point prediction. We do not tune its hyperparameters, since random forests are generally robust to hyperparameter choices. 

\paragraph{GBM-Point.}  GBM-Point is an XGBoost regressor when XGBoost is available, with a scikit-learn gradient-boosting fallback.  It is tuned with the random-search protocol above, using the grid
\[
\begin{split}
N_{\mathrm{tree}} &\in \{25,50,100,200,400,800,1200\},\\
d_{\max} &\in \{2,3,4,6,8,10\},\\
\eta &\in \{0.005,0.01,0.02,0.05,0.1,0.2,0.3\},\\
s &\in \{0.4,0.5,0.6,0.8,1.0\},
\end{split}
\]
where $s$ is the row subsampling fraction.

\paragraph{TabPFN~2.5 and RealTabPFN~2.5.}  TabPFN~2.5   is used with the public version-2.5 regressor checkpoint; RealTabPFN~2.5 uses the corresponding real-data checkpoint \texttt{tabpfn-v2.5-regressor-v2.5\_real.ckpt}.  Both are used without additional training or hyperparameter tuning.  We set \texttt{n\_estimators=8}, enable \texttt{ignore\_pretraining\_limits}, and use the same repetition seed as the model random state.  Densities are extracted from the model's full bar distribution when available: logits are converted to probabilities, divided by bar widths, interpolated to the 200-point redshift grid, and renormalized.  If the full bar distribution API is unavailable, the implementation falls back to 99 quantile levels in $[0.01,0.99]$ and numerically differentiates the implied CDF.

\paragraph{TabICL.}
TabICL is used as a tabular in-context regressor with
\texttt{n\_estimators=8}, automatic Flash Attention 3/offload settings,
and no benchmark-specific hyperparameter tuning. Densities are obtained
from quantile predictions at 199 levels in $[0.005,0.995]$: the quantiles
are sorted to enforce monotonicity, linearly interpolated as an empirical
CDF on the redshift grid, differentiated, clipped to be non-negative, and
renormalized. This reconstruction can introduce grid-scale artefacts,
particularly in local density height. We assessed the resulting densities
using held-out log-likelihood and the other density diagnostics, together
with visual inspection of representative TabICL posteriors. In the
grid-coarsening check, the changes for TabICL were 0.0047 in log-likelihood,
$6.2\times10^{-4}$ in CRPS, 0.016 in PIT--KS, and 0.007 in $90\%$ coverage.
CDE loss was more sensitive, changing by 0.305; consequently, metrics that
depend directly on local density height should be interpreted with greater
caution.

%% file: table_density_gbm.tex
\begin{table*}[t]
\centering
\caption{\review{Weight-estimator sensitivity analysis. Importance weights are here estimated with a gradient-boosted decision-tree classifier of spectroscopic versus photometric objects, in place of the TabICL classifier used in Table~\ref{tab:density_combined}; all other choices are identical. This isolates the effect of the weight-estimator choice on the importance-weighted metrics. Density-estimation metrics on the DR6 test set at the largest training-set size ($n_{\mathrm{train}} = 121\,626$), reported as mean $\pm$ standard error over five repetitions. The three panels correspond to three evaluation settings, and \emph{rankings should be compared only within each panel}: (a) unweighted metrics; (b) importance-weighted metrics with recalibrated weights $\hat{\beta}(\mathbf{x})^{\alpha}$ chosen to target an effective sample size of $30\%$ ($\alpha \simeq 0.36$); (c) importance-weighted metrics at $\alpha = 1$ (untransformed weights). For each metric the best method is shown in bold, together with all methods whose value lies within one combined standard error of the best.}}
\label{tab:density_combined_gbm}
\begin{tabular}{lcccccc}
\toprule
Method & CDE loss & log-lik & CRPS & PIT KS & Cov$_{90}$ & Width$_{90}$ \\
\midrule
\multicolumn{7}{l}{\textbf{(a)~No weights (unweighted metrics)}} \\
\cmidrule(lr){1-7}
\multicolumn{7}{l}{\quad\textit{\small Foundation models}} \\
\quad TabPFN~2.5 & \textbf{-2.561 $\pm$ 0.032} & \textbf{0.190 $\pm$ 0.013} & \textbf{0.188 $\pm$ 0.002} & 0.015 & \textbf{0.900 $\pm$ 0.002} & \textbf{1.045 $\pm$ 0.006} \\
\quad RealTabPFN~2.5 & -2.366 $\pm$ 0.030 & 0.131 $\pm$ 0.012 & \textbf{0.191 $\pm$ 0.002} & \textbf{0.015} & 0.895 $\pm$ 0.002 & \textbf{1.045 $\pm$ 0.006} \\
\quad TabICL & -2.271 $\pm$ 0.027 & -0.090 $\pm$ 0.024 & 0.192 $\pm$ 0.002 & 0.032 & 0.907 $\pm$ 0.002 & 1.091 $\pm$ 0.006 \\
\addlinespace[1pt]
\multicolumn{7}{l}{\quad\textit{\small Other methods}} \\
\quad LinGauss (homo) & -0.431 $\pm$ 0.003 & -1.029 $\pm$ 0.009 & 0.375 $\pm$ 0.003 & 0.026 & 0.921 $\pm$ 0.002 & 2.212 $\pm$ 0.000 \\
\quad LinGauss (hetero) & -0.461 $\pm$ 0.003 & -0.951 $\pm$ 0.008 & 0.368 $\pm$ 0.003 & 0.033 & 0.916 $\pm$ 0.002 & 2.100 $\pm$ 0.004 \\
\quad FlexZBoost & -2.539 $\pm$ 0.034 & -0.034 $\pm$ 0.022 & 0.217 $\pm$ 0.003 & 0.018 & 0.883 $\pm$ 0.003 & 1.200 $\pm$ 0.005 \\
\quad MDN & -2.188 $\pm$ 0.026 & 0.058 $\pm$ 0.012 & 0.210 $\pm$ 0.002 & 0.024 & 0.912 $\pm$ 0.002 & 1.274 $\pm$ 0.006 \\
\quad MDN (deep) & -2.454 $\pm$ 0.027 & 0.144 $\pm$ 0.012 & 0.194 $\pm$ 0.002 & 0.043 & 0.912 $\pm$ 0.002 & 1.131 $\pm$ 0.007 \\
\quad Flow-Spline & \textbf{-2.593 $\pm$ 0.034} & \textbf{0.175 $\pm$ 0.013} & 0.206 $\pm$ 0.003 & 0.027 & 0.905 $\pm$ 0.002 & 1.190 $\pm$ 0.006 \\
\addlinespace[6pt]
\multicolumn{7}{l}{\textbf{(b)~Importance-weighted, ESS target $30\%$ ($\alpha \simeq 0.36$)}} \\
\cmidrule(lr){1-7}
\multicolumn{7}{l}{\quad\textit{\small Foundation models}} \\
\quad TabPFN~2.5 & \textbf{-1.867 $\pm$ 0.030} & \textbf{-0.178 $\pm$ 0.023} & \textbf{0.239 $\pm$ 0.006} & -- & 0.894 $\pm$ 0.005 & 1.315 $\pm$ 0.010 \\
\quad RealTabPFN~2.5 & -1.740 $\pm$ 0.028 & -0.219 $\pm$ 0.024 & \textbf{0.241 $\pm$ 0.006} & -- & 0.883 $\pm$ 0.005 & 1.269 $\pm$ 0.009 \\
\quad TabICL & -1.676 $\pm$ 0.026 & -0.426 $\pm$ 0.040 & \textbf{0.241 $\pm$ 0.006} & -- & 0.907 $\pm$ 0.005 & 1.399 $\pm$ 0.010 \\
\addlinespace[1pt]
\multicolumn{7}{l}{\quad\textit{\small Other methods}} \\
\quad LinGauss (homo) & -0.438 $\pm$ 0.005 & -1.044 $\pm$ 0.022 & 0.374 $\pm$ 0.006 & -- & 0.923 $\pm$ 0.004 & 2.213 $\pm$ 0.000 \\
\quad LinGauss (hetero) & -0.458 $\pm$ 0.005 & -0.989 $\pm$ 0.021 & 0.369 $\pm$ 0.005 & -- & 0.922 $\pm$ 0.004 & 2.193 $\pm$ 0.008 \\
\quad FlexZBoost & -1.756 $\pm$ 0.036 & -0.452 $\pm$ 0.043 & 0.265 $\pm$ 0.007 & -- & 0.840 $\pm$ 0.007 & \textbf{1.251 $\pm$ 0.005} \\
\quad MDN & -1.586 $\pm$ 0.025 & -0.293 $\pm$ 0.022 & 0.259 $\pm$ 0.006 & -- & \textbf{0.904 $\pm$ 0.004} & 1.512 $\pm$ 0.009 \\
\quad MDN (deep) & -1.772 $\pm$ 0.026 & -0.216 $\pm$ 0.019 & \textbf{0.245 $\pm$ 0.006} & -- & 0.910 $\pm$ 0.004 & 1.426 $\pm$ 0.010 \\
\quad Flow-Spline & \textbf{-1.852 $\pm$ 0.031} & -0.225 $\pm$ 0.027 & 0.255 $\pm$ 0.006 & -- & \textbf{0.896 $\pm$ 0.004} & 1.413 $\pm$ 0.008 \\
\addlinespace[6pt]
\multicolumn{7}{l}{\textbf{(c)~Importance-weighted, $\alpha = 1$ (untransformed weights)}} \\
\cmidrule(lr){1-7}
\multicolumn{7}{l}{\quad\textit{\small Foundation models}} \\
\quad TabPFN~2.5 & \textbf{-0.593 $\pm$ 0.037} & \textbf{-0.855 $\pm$ 0.074} & \textbf{0.321 $\pm$ 0.020} & -- & \textbf{0.893 $\pm$ 0.017} & 1.773 $\pm$ 0.018 \\
\quad RealTabPFN~2.5 & \textbf{-0.586 $\pm$ 0.037} & \textbf{-0.864 $\pm$ 0.080} & \textbf{0.321 $\pm$ 0.021} & -- & 0.868 $\pm$ 0.021 & 1.636 $\pm$ 0.014 \\
\quad TabICL & \textbf{-0.589 $\pm$ 0.028} & \textbf{-0.973 $\pm$ 0.122} & \textbf{0.320 $\pm$ 0.020} & -- & \textbf{0.910 $\pm$ 0.018} & 1.959 $\pm$ 0.015 \\
\addlinespace[1pt]
\multicolumn{7}{l}{\quad\textit{\small Other methods}} \\
\quad LinGauss (homo) & -0.471 $\pm$ 0.020 & -1.000 $\pm$ 0.069 & 0.353 $\pm$ 0.020 & -- & 0.938 $\pm$ 0.014 & 2.216 $\pm$ 0.000 \\
\quad LinGauss (hetero) & -0.469 $\pm$ 0.020 & -1.031 $\pm$ 0.074 & 0.354 $\pm$ 0.020 & -- & 0.939 $\pm$ 0.014 & 2.362 $\pm$ 0.021 \\
\quad FlexZBoost & -0.381 $\pm$ 0.064 & -1.082 $\pm$ 0.115 & \textbf{0.337 $\pm$ 0.023} & -- & 0.757 $\pm$ 0.027 & \textbf{1.295 $\pm$ 0.009} \\
\quad MDN & \textbf{-0.545 $\pm$ 0.031} & \textbf{-0.904 $\pm$ 0.067} & \textbf{0.329 $\pm$ 0.020} & -- & \textbf{0.895 $\pm$ 0.017} & 1.923 $\pm$ 0.017 \\
\quad MDN (deep) & \textbf{-0.575 $\pm$ 0.027} & \textbf{-0.846 $\pm$ 0.057} & \textbf{0.323 $\pm$ 0.020} & -- & \textbf{0.913 $\pm$ 0.016} & 1.926 $\pm$ 0.013 \\
\quad Flow-Spline & -0.535 $\pm$ 0.038 & \textbf{-0.905 $\pm$ 0.080} & \textbf{0.325 $\pm$ 0.020} & -- & \textbf{0.887 $\pm$ 0.017} & 1.795 $\pm$ 0.012 \\
\bottomrule
\end{tabular}
\end{table*}

%% file: table_point_gbm.tex
\begin{table*}[t]
\centering
\caption{\review{Weight-estimator sensitivity analysis. Importance weights are here estimated with a gradient-boosted decision-tree classifier; all other choices are identical. This isolates the effect of the weight-estimator choice on the importance-weighted metrics. Point-prediction metrics on the DR6 test set at the largest training-set size, reported as mean $\pm$ standard error. The three panels correspond to three evaluation settings, and \emph{rankings should be compared only within each panel}. }}
\label{tab:point_combined_gbm}
\begin{tabular}{lccccc}
\toprule
Method & RMSE & Bias & NMAD & Out$_{0.15}$ & Out$_{0.30}$ \\
\midrule
\multicolumn{6}{l}{\textbf{(a)~No weights (unweighted metrics)}} \\
\cmidrule(lr){1-6}
\multicolumn{6}{l}{\quad\textit{\small Foundation models}} \\
\quad TabPFN~2.5 & \textbf{0.4435 $\pm$ 0.0078} & \textbf{+0.0059 $\pm$ 0.0040} & \textbf{0.0852} & \textbf{0.2222 $\pm$ 0.0032} & \textbf{0.0664 $\pm$ 0.0022} \\
\quad RealTabPFN~2.5 & \textbf{0.4475 $\pm$ 0.0078} & \textbf{+0.0054 $\pm$ 0.0040} & 0.0887 & 0.2269 $\pm$ 0.0032 & \textbf{0.0682 $\pm$ 0.0022} \\
\quad TabICL & \textbf{0.4497 $\pm$ 0.0078} & \textbf{-0.0040 $\pm$ 0.0040} & 0.0867 & 0.2296 $\pm$ 0.0033 & 0.0707 $\pm$ 0.0022 \\
\addlinespace[1pt]
\multicolumn{6}{l}{\quad\textit{\small Other density methods}} \\
\quad LinGauss (homo) & 0.6788 $\pm$ 0.0064 & \textbf{+0.0005 $\pm$ 0.0065} & 0.2537 & 0.5513 $\pm$ 0.0042 & 0.2294 $\pm$ 0.0035 \\
\quad LinGauss (hetero) & 0.6835 $\pm$ 0.0062 & \textbf{+0.0030 $\pm$ 0.0065} & 0.2458 & 0.5338 $\pm$ 0.0043 & 0.2317 $\pm$ 0.0035 \\
\quad FlexZBoost & 0.5047 $\pm$ 0.0085 & +0.0099 $\pm$ 0.0044 & 0.1177 & 0.2846 $\pm$ 0.0036 & 0.0858 $\pm$ 0.0024 \\
\quad MDN & 0.4814 $\pm$ 0.0074 & \textbf{-0.0012 $\pm$ 0.0043} & 0.1037 & 0.2657 $\pm$ 0.0031 & 0.0844 $\pm$ 0.0022 \\
\quad MDN (deep) & 0.4571 $\pm$ 0.0078 & +0.0118 $\pm$ 0.0041 & 0.0878 & 0.2334 $\pm$ 0.0031 & 0.0706 $\pm$ 0.0021 \\
\quad Flow-Spline & 0.4829 $\pm$ 0.0084 & +0.0096 $\pm$ 0.0041 & 0.0992 & 0.2564 $\pm$ 0.0031 & 0.0805 $\pm$ 0.0022 \\
\addlinespace[1pt]
\multicolumn{6}{l}{\quad\textit{\small Point estimators only}} \\
\quad RF-Point & 0.4700 $\pm$ 0.0075 & \textbf{-0.0071 $\pm$ 0.0044} & 0.1077 & 0.2568 $\pm$ 0.0031 & 0.0795 $\pm$ 0.0024 \\
\quad GBM-Point & 0.4621 $\pm$ 0.0075 & \textbf{-0.0037 $\pm$ 0.0043} & 0.1099 & 0.2539 $\pm$ 0.0033 & 0.0754 $\pm$ 0.0023 \\
\addlinespace[6pt]
\multicolumn{6}{l}{\textbf{(b)~Importance-weighted, ESS target $30\%$ ($\alpha \simeq 0.36$)}} \\
\cmidrule(lr){1-6}
\multicolumn{6}{l}{\quad\textit{\small Foundation models}} \\
\quad TabPFN~2.5 & \textbf{0.5212 $\pm$ 0.0168} & \textbf{+0.0055 $\pm$ 0.0091} & -- & \textbf{0.2765 $\pm$ 0.0073} & \textbf{0.0812 $\pm$ 0.0049} \\
\quad RealTabPFN~2.5 & \textbf{0.5245 $\pm$ 0.0170} & \textbf{+0.0079 $\pm$ 0.0091} & -- & \textbf{0.2822 $\pm$ 0.0073} & \textbf{0.0821 $\pm$ 0.0049} \\
\quad TabICL & \textbf{0.5257 $\pm$ 0.0172} & \textbf{+0.0023 $\pm$ 0.0092} & -- & \textbf{0.2829 $\pm$ 0.0072} & \textbf{0.0825 $\pm$ 0.0048} \\
\addlinespace[1pt]
\multicolumn{6}{l}{\quad\textit{\small Other density methods}} \\
\quad LinGauss (homo) & 0.6883 $\pm$ 0.0142 & +0.0337 $\pm$ 0.0110 & -- & 0.5149 $\pm$ 0.0076 & 0.2001 $\pm$ 0.0055 \\
\quad LinGauss (hetero) & 0.6890 $\pm$ 0.0139 & +0.0456 $\pm$ 0.0109 & -- & 0.4995 $\pm$ 0.0075 & 0.1963 $\pm$ 0.0053 \\
\quad FlexZBoost & 0.5687 $\pm$ 0.0172 & +0.0322 $\pm$ 0.0095 & -- & 0.3241 $\pm$ 0.0073 & 0.0935 $\pm$ 0.0049 \\
\quad MDN & 0.5525 $\pm$ 0.0163 & \textbf{-0.0010 $\pm$ 0.0092} & -- & 0.3128 $\pm$ 0.0066 & 0.0979 $\pm$ 0.0048 \\
\quad MDN (deep) & \textbf{0.5337 $\pm$ 0.0172} & +0.0205 $\pm$ 0.0089 & -- & 0.2893 $\pm$ 0.0069 & \textbf{0.0831 $\pm$ 0.0046} \\
\quad Flow-Spline & 0.5566 $\pm$ 0.0173 & +0.0194 $\pm$ 0.0091 & -- & 0.3035 $\pm$ 0.0068 & 0.0920 $\pm$ 0.0047 \\
\addlinespace[1pt]
\multicolumn{6}{l}{\quad\textit{\small Point estimators only}} \\
\quad RF-Point & \textbf{0.5415 $\pm$ 0.0159} & \textbf{-0.0086 $\pm$ 0.0097} & -- & 0.3073 $\pm$ 0.0071 & 0.0948 $\pm$ 0.0050 \\
\quad GBM-Point & \textbf{0.5333 $\pm$ 0.0163} & \textbf{-0.0032 $\pm$ 0.0094} & -- & 0.3008 $\pm$ 0.0068 & \textbf{0.0879 $\pm$ 0.0050} \\
\addlinespace[6pt]
\multicolumn{6}{l}{\textbf{(c)~Importance-weighted, $\alpha = 1$ (untransformed weights)}} \\
\cmidrule(lr){1-6}
\multicolumn{6}{l}{\quad\textit{\small Foundation models}} \\
\quad TabPFN~2.5 & \textbf{0.6114 $\pm$ 0.0465} & \textbf{+0.0133 $\pm$ 0.0397} & -- & \textbf{0.3454 $\pm$ 0.0293} & \textbf{0.0916 $\pm$ 0.0183} \\
\quad RealTabPFN~2.5 & \textbf{0.6126 $\pm$ 0.0468} & \textbf{+0.0221 $\pm$ 0.0400} & -- & \textbf{0.3526 $\pm$ 0.0294} & \textbf{0.0913 $\pm$ 0.0183} \\
\quad TabICL & \textbf{0.6125 $\pm$ 0.0480} & \textbf{+0.0254 $\pm$ 0.0398} & -- & \textbf{0.3420 $\pm$ 0.0281} & \textbf{0.0888 $\pm$ 0.0182} \\
\addlinespace[1pt]
\multicolumn{6}{l}{\quad\textit{\small Other density methods}} \\
\quad LinGauss (homo) & \textbf{0.6578 $\pm$ 0.0467} & +0.1005 $\pm$ 0.0443 & -- & 0.4233 $\pm$ 0.0312 & 0.1205 $\pm$ 0.0208 \\
\quad LinGauss (hetero) & \textbf{0.6539 $\pm$ 0.0470} & +0.1289 $\pm$ 0.0426 & -- & 0.3950 $\pm$ 0.0298 & \textbf{0.0989 $\pm$ 0.0173} \\
\quad FlexZBoost & \textbf{0.6284 $\pm$ 0.0467} & +0.0816 $\pm$ 0.0403 & -- & \textbf{0.3750 $\pm$ 0.0304} & \textbf{0.0907 $\pm$ 0.0184} \\
\quad MDN & \textbf{0.6206 $\pm$ 0.0453} & \textbf{+0.0158 $\pm$ 0.0394} & -- & \textbf{0.3501 $\pm$ 0.0277} & \textbf{0.1017 $\pm$ 0.0177} \\
\quad MDN (deep) & \textbf{0.6159 $\pm$ 0.0465} & \textbf{+0.0515 $\pm$ 0.0396} & -- & \textbf{0.3523 $\pm$ 0.0273} & \textbf{0.0875 $\pm$ 0.0167} \\
\quad Flow-Spline & \textbf{0.6201 $\pm$ 0.0460} & \textbf{+0.0477 $\pm$ 0.0394} & -- & \textbf{0.3490 $\pm$ 0.0275} & \textbf{0.0946 $\pm$ 0.0173} \\
\addlinespace[1pt]
\multicolumn{6}{l}{\quad\textit{\small Point estimators only}} \\
\quad RF-Point & \textbf{0.6241 $\pm$ 0.0455} & \textbf{-0.0009 $\pm$ 0.0408} & -- & \textbf{0.3574 $\pm$ 0.0290} & \textbf{0.0994 $\pm$ 0.0185} \\
\quad GBM-Point & \textbf{0.6161 $\pm$ 0.0456} & \textbf{+0.0140 $\pm$ 0.0405} & -- & \textbf{0.3463 $\pm$ 0.0274} & \textbf{0.0954 $\pm$ 0.0190} \\
\bottomrule
\end{tabular}
\end{table*}

%% file: ApJ.bib
@article{2024A&A...689A.249H,
  author = {Herpich, F. and Almeida-Fernandes, F. and Schwarz, G. B. O. and Lima, E. and Nakazono, L. and Alonso-Garc'ia, J. and Fonseca-Faria, M. and Sartori, M. J. and Bolutavicius, G. F. and Souza, G. F. D. and others},
  title = {{The Fourth S-PLUS Data Release: 12-filter photometry covering  3000 square degrees in the southern hemisphere}},
  journal = {Astronomy \& Astrophysics},
  year = {2024},
  volume = {689},
  pages = {A249},
  doi = {10.1051/0004-6361/202449725},
  publisher = {EDP Sciences},
}

@inproceedings{BenDavid2010,
  author    = {Ben-David, Shai and Lu, Tyler and Luu, Teresa and P{\'a}l, D{\'a}vid},
  title     = {{Impossibility Theorems for Domain Adaptation}},
  booktitle = {Proceedings of the Thirteenth International Conference on
               Artificial Intelligence and Statistics},
  series    = {Proceedings of Machine Learning Research},
  volume    = {9},
  pages     = {129--136},
  year      = {2010},
  url       = {https://proceedings.mlr.press/v9/david10a.html}
}

@inproceedings{GulrajaniHashimoto2022,
  author    = {Gulrajani, Ishaan and Hashimoto, Tatsunori},
  title     = {{Identifiability Conditions for Domain Adaptation}},
  booktitle = {Proceedings of the 39th International Conference on Machine Learning},
  series    = {Proceedings of Machine Learning Research},
  volume    = {162},
  pages     = {7982--7997},
  year      = {2022},
  url       = {https://proceedings.mlr.press/v162/gulrajani22a.html}
}

@article{2016arXiv161100036D,
  author = {{DESI Collaboration} and Aghamousa, Amir and Aguilar, J. and Ahlen, S. and Alam, S. and Allen, L. and Prieto, C. and Annis, J. and Bailey, S. and Balland, C. and Ballester, O. and others},
  title = {{The DESI Experiment Part I: Science, Targeting, and Survey Design}},
  journal = {arXiv e-prints},
  year = {2016},
}

@article{2026AJ....171..285D,
  author = {Abdul Karim,  M. and Adame, A. G. and Aguado, D. and Aguilar, J. and Ahlen, S. and Alam, S. and Aldering, G. and Alexander, D. and Alfarsy, R. and Allen, L. and others},
  title = {{Data Release 1 of the Dark Energy Spectroscopic Instrument}},
  volume={171},
  number={5},
  journal = {Astronomical Journal},
  year = {2026},
  doi = {10.3847/1538-3881/ae4c43},
}

@article{2000AJ....120.1579Y,
  author = {York, D. and Adelman, Jennifer K. and Anderson, J. and Anderson, S. and Annis, J. and Bahcall, N. and Bakken, J. and Barkhouser, R. and Bastian, S. and Berman, E. and others},
  title = {{The Sloan Digital Sky Survey: Technical Summary}},
  journal = {The Astronomical Journal},
  year = {2000},
  volume = {120},
  number = {3},
  pages = {1579-1587},
  doi = {10.1086/301513},
  publisher = {American Astronomical Society},
}

@article{cenarro+19,
  author = {Cenarro, A. J. and Molés, M. and Crist'obal-Hornillos, D. and Marín-Franch, A. and Ederoclite, A. and Varela, J. and L'opez-Sanjuan, C. and Hern'andez-Monteagudo, C. and Angulo, R. and Rami'o, H. V. and others},
  title = {{J-PLUS: The Javalambre Photometric Local Universe Survey}},
  volume={622},
  pages={A176},
  journal = {Astronomy \& Astrophysics},
  year = {2019},
  doi = {10.1051/0004-6361/201833036},
}

@article{2023MNRAS.524.5109R,
  author = {Rau, M. and Dalal, R. and Zhang, Tianqing and Li, Xiangchong and Nishizawa, A. and More, S. and Mandelbaum, R. and Miyatake, H. and Strauss, M. and Takada, M.},
  title = {{Weak lensing tomographic redshift distribution inference for the Hyper Suprime-Cam Subaru Strategic Program three-year shape catalogue}},
  journal = {Monthly notices of the Royal Astronomical Society},
  year = {2023},
  volume={524},
  number={4},
  doi = {10.1093/mnras/stad1962},
}

@article{2009MNRAS.399.2279M,
  author = {Myers, A. and White, M. and Ball, N.},
  title = {{Incorporating photometric redshift probability density information into real-space clustering measurements}},
  journal = {Monthly Notices of the Royal Astronomical Society},
  year = {2009},
  volume = {399},
  number = {4},
  pages = {2279-2287},
  doi = {10.1111/j.1365-2966.2009.15432.x},
  publisher = {Oxford University Press (OUP)},
}

@article{2016MNRAS.459.1293A,
  title={Galaxy clustering with photometric surveys using PDF redshift information},
  author={Asorey, J and Carrasco Kind, M and Sevilla-Noarbe, I and Brunner, RJ and Thaler, J},
  journal={Monthly Notices of the Royal Astronomical Society},
  volume={459},
  number={2},
  pages={1293--1309},
  year={2016},
  publisher={Oxford University Press}
}

@article{2021MNRAS.507.5847N,
  author = {Nakazono, L. and Oliveira, C. Mendes de and Hirata, N. and Jeram, S. and Queiroz, C. and Eikenberry, S. and Gonzalez, A. and Abramo, R. and Overzier, R. and Espadoto, M. and others},
  title = {{On the discovery of stars, quasars, and galaxies in the Southern Hemisphere with S-PLUS DR2}},
  journal = {Monthly notices of the Royal Astronomical Society},
  year = {2021},
  volume = {507},
  number = {4},
  pages = {5847-5868},
  doi = {10.1093/mnras/stab1835},
  publisher = {Oxford University Press (OUP)},
}

@article{2017ApJ...849...53H,
  author = {Hickox, R. and Myers, A. and Greene, J. and Hainline, K. and Zakamska, N. and Dipompeo, M.},
  title = {{Composite Spectral Energy Distributions and Infrared-Optical Colors of Type 1 and Type 2 Quasars}},
  journal = {The Astrophysical Journal},
  year = {2017},
  volume = {849},
  number = {1},
  pages = {53},
  doi = {10.3847/1538-4357/aa8c77},
  publisher = {American Astronomical Society},
}

@article{2001AJ....121.2308R,
  title={Colors of 2625 quasars at 0< z< 5 measured in the sloan digital sky survey photometric system},
  author={Richards, Gordon T and Fan, Xiaohui and Schneider, Donald P and Vanden Berk, Daniel E and Strauss, Michael A and York, Donald G and Anderson, Jr, John E and Anderson, Scott F and Annis, James and Bahcall, Neta A and others},
  journal={The Astronomical Journal},
  volume={121},
  number={5},
  pages={2308--2330},
  year={2001}
}

@article{2016arXiv160808016P,
  author = {Polsterer, K. and D'Isanto, Antonio and Gieseke, Fabian},
  title = {{Uncertain Photometric Redshifts}},
  journal = {arXiv e-prints},
  year = {2016},
}

@article{2001ARA&A..39...19L,
  author = {Loeb, A. and Barkana, R.},
  title = {{The Reionization of the Universe by the First Stars and Quasars}},
  journal = {\araa},
  volume={39},
  year = {2001},
  doi = {10.1146/annurev.astro.39.1.19},
}

@article{2014ARA&A..52..589H,
  author = {Heckman, T. and Best, P.},
  title = {{The Coevolution of Galaxies and Supermassive Black Holes: Insights from Surveys of the Contemporary Universe}},
  journal = {Annual Review of Astronomy and Astrophysics},
  year = {2014},
  volume = {52},
  number = {1},
  pages = {589-660},
  doi = {10.1146/annurev-astro-081913-035722},
  publisher = {Annual Reviews},
}

@inproceedings{2016ASSL..423..187M,
  author = {Mortlock, D.},
  title = {{Quasars as Probes of Cosmological Reionization}},
  booktitle = {Understanding the Epoch of Cosmic Reionization: Challenges and Progress},
  year = {2016},
  editor = {{Mesinger}, Andrei},
  series = {Astrophysics and Space Science Library},
  volume = {423},
  pages = {187--226},
  doi = {10.1007/978-3-319-21957-8_7},
  archiveprefix = {arXiv},
  eprint = {1511.01107},
  primaryclass = {astro-ph.CO},
  adsurl = {https://ui.adsabs.harvard.edu/abs/2016ASSL..423..187M},
}

@article{2023ARA&A..61..373F,
  author = {Fan, Xiaohui and Bañados, E. and Simcoe, R.},
  title = {{Quasars and the Intergalactic Medium at Cosmic Dawn}},
  journal = {Annual Review of Astronomy and Astrophysics},
  year = {2023},
  volume={61},
  pages={373--426},
  doi = {10.1146/annurev-astro-052920-102455},
}

@book{Izbicki2025,
  author    = {Rafael Izbicki},
  title     = {Machine Learning Beyond Point Predictions: Uncertainty Quantification},
  edition   = {1st},
  year      = {2025},
  pages     = {260},
  isbn      = {978-65-01-20272-3}
}

@article{2019NatAs...3..212S,
  author = {Salvato, M. and Ilbert, O. and Hoyle, B.},
  title = {{The many flavours of photometric redshifts}},
  journal = {Nature Astronomy},
  year = {2019},
  volume = {3},
  number = {3},
  pages = {212-222},
  doi = {10.1038/s41550-018-0478-0},
  publisher = {Springer Science and Business Media LLC},
}

@article{dalmasso2020conditional,
  title={Conditional density estimation tools in python and R with applications to photometric redshifts and likelihood-free cosmological inference},
  author={Dalmasso, Niccol{\`o} and Pospisil, Taylor and Lee, Ann B and Izbicki, Rafael and Freeman, Peter E and Malz, Alex I},
  journal={Astronomy and Computing},
  volume={30},
  pages={100362},
  year={2020},
  publisher={Elsevier}
}

@article{izbicki2016nonparametric,
  title={Nonparametric conditional density estimation in a high-dimensional regression setting},
  author={Izbicki, Rafael and Lee, Ann B},
  journal={Journal of Computational and Graphical Statistics},
  volume={25},
  number={4},
  pages={1297--1316},
  year={2016},
  publisher={Taylor \& Francis}
}

@inproceedings{zhao2021diagnostics,
  title={Diagnostics for conditional density models and Bayesian inference algorithms},
  author={Zhao, David and Dalmasso, Niccol{\`o} and Izbicki, Rafael and Lee, Ann B},
  booktitle={Uncertainty in Artificial Intelligence},
  pages={1830--1840},
  year={2021},
  organization={PMLR}
}

@article{dey2025towards,
  title={Towards instance-wise calibration: local amortized diagnostics and reshaping of conditional densities (LADaR)},
  author={Dey, Biprateep and Zhao, David and Andrews, Brett H and Newman, Jeffrey A and Izbicki, Rafael and Lee, Ann B},
  journal={Machine Learning: Science and Technology},
  volume={6},
  number={4},
  pages={045058},
  year={2025},
  publisher={IOP Publishing}
}

@article{bordoloi2010photo,
  title={Photo-z performance for precision cosmology},
  author={Bordoloi, Rongmon and Lilly, Simon J and Amara, Adam},
  journal={Monthly Notices of the Royal Astronomical Society},
  volume={406},
  number={2},
  pages={881--895},
  year={2010},
  publisher={The Royal Astronomical Society}
}

@software{erik_v_2025_15127060,
  author       = {Erik Lima},
  title        = {ErikVini/specz\_compilation: Southern Hemisphere
                   Spectrocopic Redshift Compilation
                  },
  month        = apr,
  year         = 2025,
  publisher    = {Zenodo},
  version      = {2025.03.27},
  doi          = {10.5281/zenodo.15127060},
  url          = {https://doi.org/10.5281/zenodo.15127060},
  swhid        = {swh:1:dir:07c93903017d7302d2813ad26bbc965db289c19e
                   ;origin=https://doi.org/10.5281/zenodo.11641314;vi
                   sit=swh:1:snp:a702ab03a31e1d97dfbeac21afde27345a20
                   67f9;anchor=swh:1:rel:8fc976db163f9cee382b2bc5fb27
                   50aa50892ba6;path=ErikVini-
                   specz\_compilation-210abc5
                  },
}

@article{Weinstein2004,
  author = {{Weinstein}, Michael A. and {Richards}, Gordon T. and
            {Schneider}, Donald P. and {Younger}, Joshua D. and
            {Strauss}, Michael A. and {Hall}, Patrick B. and
            {Budav{\'a}ri}, Tam{\'a}s and {Gunn}, James E. and
            {York}, Donald G. and {Brinkmann}, J.},
  title = {{An Empirical Algorithm for Broad-band Photometric Redshifts
            of Quasars from the Sloan Digital Sky Survey}},
  journal = {\apjs},
  year = {2004},
  volume = {155},
  pages = {243--256},
  doi = {10.1086/425355},
}

@article{2024MNRAS.531..327N,
  author = {Nakazono, L. and Valencca, R. R. and Soares, G. and Izbicki, R. and Ivezi'c, vZeljko and Lima, E. and Hirata, N. S. T. and Sodré, L. and Overzier, R. and Almeida-Fernandes, F. and others},
  title = {{The Quasar Catalogue for S-PLUS DR4 (QuCatS) and the estimation of photometric redshifts}},
  journal = {Monthly notices of the Royal Astronomical Society},
  year = {2024},
  volume = {531},
  number = {1},
  pages = {327-339},
  doi = {10.1093/mnras/stae971},
  publisher = {Oxford University Press (OUP)},
}

@article{splus,
	title = {The {Southern} {Photometric} {Local} {Universe} {Survey} ({S}-{PLUS}): improved {SEDs}, morphologies, and redshifts with 12 optical filters},
	volume = {489},
	copyright = {https://academic.oup.com/journals/pages/open\_access/funder\_policies/chorus/standard\_publication\_model},
	issn = {0035-8711, 1365-2966},
	shorttitle = {The {Southern} {Photometric} {Local} {Universe} {Survey} ({S}-{PLUS})},
	url = {https://academic.oup.com/mnras/article/489/1/241/5543943},
	doi = {10.1093/mnras/stz1985},
	language = {en},
	number = {1},
	urldate = {2025-01-18},
	journal = {Monthly Notices of the Royal Astronomical Society},
	author = {Mendes de Oliveira, C and Ribeiro, T and Schoenell, W and Kanaan, A and Overzier, R A and Molino, A and Sampedro, L and Coelho, P and Barbosa, C E and Cortesi, A and Costa-Duarte, M V and Herpich, F R and Hernandez-Jimenez, J A and Placco, V M and Xavier, H S and Abramo, L R and Saito, R K and Chies-Santos, A L and Ederoclite, A and Lopes de Oliveira, R and Gonçalves, D R and Akras, S and Almeida, L A and Almeida-Fernandes, F and Beers, T C and Bonatto, C and Bonoli, S and Cypriano, E S and Vinicius-Lima, E and de Souza, R S and Fabiano de Souza, G and Ferrari, F and Gonçalves, T S and Gonzalez, A H and Gutiérrez-Soto, L A and Hartmann, E A and Jaffe, Y and Kerber, L O and Lima-Dias, C and Lopes, P A A and Menendez-Delmestre, K and Nakazono, L M I and Novais, P M and Ortega-Minakata, R A and Pereira, E S and Perottoni, H D and Queiroz, C and Reis, R R R and Santos, W A and Santos-Silva, T and Santucci, R M and Barbosa, C L and Siffert, Beatriz B and Sodré, L and Torres-Flores, S and Westera, P and Whitten, D D and Alcaniz, J S and Alonso-García, Javier and Alencar, S and Alvarez-Candal, A and Amram, P and Azanha, L and Barbá, R H and Bernardinelli, P H and Borges Fernandes, M and Branco, V and Brito-Silva, D and Buzzo, M L and Caffer, J and Campillay, A and Cano, Z and Carvano, J M and Castejon, M and Cid Fernandes, R and Dantas, M L L and Daflon, S and Damke, G and de la Reza, R and de Melo de Azevedo, L J and De Paula, D F and Diem, K G and Donnerstein, R and Dors, O L and Dupke, R and Eikenberry, S and Escudero, Carlos G and Faifer, Favio R and Farías, H and Fernandes, B and Fernandes, C and Fontes, S and Galarza, A and Hirata, N S T and Katena, L and Gregorio-Hetem, J and Hernández-Fernández, J D and Izzo, L and Jaque Arancibia, M and Jatenco-Pereira, V and Jiménez-Teja, Y and Kann, D A and Krabbe, A C and Labayru, C and Lazzaro, D and Lima Neto, G B and Lopes, Amanda R and Magalhães, R and Makler, M and de Menezes, R and Miralda-Escudé, J and Monteiro-Oliveira, R and Montero-Dorta, A D and Muñoz-Elgueta, N and Nemmen, R S and Nilo Castellón, J L and Oliveira, A S and Ortíz, D and Pattaro, E and Pereira, C B and Quint, B and Riguccini, L and Rocha Pinto, H J and Rodrigues, I and Roig, F and Rossi, S and Saha, Kanak and Santos, R and Schnorr Müller, A and Sesto, Leandro A and Silva, R and Smith Castelli, Analia V and Teixeira, R and Telles, E and Thom de Souza, R C and Thöne, C and Trevisan, M and de Ugarte Postigo, A and Urrutia-Viscarra, F and Veiga, C H and Vika, M and Vitorelli, A Z and Werle, A and Werner, S V and Zaritsky, D},
	month = oct,
	year = {2019},
	pages = {241--267}
}

@article{landsgesell2026distributional,
  author = {Landsgesell, Jonas and Knoll, Pascal},
  title = {{Distributional Regression with Tabular Foundation Models: Evaluating Probabilistic Predictions via Proper Scoring Rules}},
  journal = {arXiv e-prints},
  year = {2026},
  eid = {arXiv:2603.08206},
  pages = {arXiv:2603.08206},
  doi = {10.48550/arXiv.2603.08206},
  archiveprefix = {arXiv},
  eprint = {2603.08206},
  primaryclass = {cs.LG},
}

@article{izbicki2026benchmarking,
  author = {Izbicki, R. and Rodrigues, Pedro L. C.},
  title = {{Benchmarking Tabular Foundation Models for Conditional Density Estimation in Regression}},
  journal = {arXiv e-prints},
  year = {2026},
  eid = {arXiv:2603.26611},
  pages = {arXiv:2603.26611},
  doi = {10.48550/arXiv.2603.26611},
  archiveprefix = {arXiv},
  eprint = {2603.26611},
  primaryclass = {stat.ML},
}

@article{schmidt2020evaluation,
  author = {Schmidt, S. and Malz, A. and Soo, J. and Almosallam, I. and Brescia, M. and Cavuoti, S. and Cohen-Tanugi, J. and Connolly, A. and DeRose, J. and others},
  title = {{Evaluation of probabilistic photometric redshift estimation approaches for The Rubin Observatory Legacy Survey of Space and Time (LSST)}},
  journal = {Monthly notices of the Royal Astronomical Society},
  year = {2020},
  volume = {499},
  number = {2},
  pages = {1587--1606},
  doi = {10.1093/mnras/staa2799},
  archiveprefix = {arXiv},
  eprint = {2001.03621},
  primaryclass = {astro-ph.IM},
  publisher = {Oxford University Press (OUP)},
}

@article{mandelbaum2008precision,
  author = {Mandelbaum, R. and Seljak, U.   and Hirata, C. and Bardelli, S. and Bolzonella, M. and Bongiorno, A. and Carollo, M. and Contini, T. and Cunha, C. E. and others},
  title = {{Precision photometric redshift calibration for galaxy--galaxy weak lensing}},
  journal = {\mnras},
  year = {2008},
  volume = {386},
  number = {2},
  pages = {781--806},
  doi = {10.1111/j.1365-2966.2008.12947.x},
  archiveprefix = {arXiv},
  eprint = {0709.1692},
  primaryclass = {astro-ph},
}

@article{izbicki2017photo,
  author = {Izbicki, Rafael and Lee, Ann B. and Freeman, P.},
  title = {{Photo-z estimation: An example of nonparametric conditional density estimation under selection bias}},
  journal = {Annals of Applied Statistics},
  year = {2017},
  volume = {11},
  number = {2},
  pages = {698--724},
  doi = {10.1214/16-AOAS1013},
}

@article{cunha2009estimating,
  title={Estimating the redshift distribution of photometric galaxy samples--II. Applications and tests of a new method},
  author={Cunha, Carlos E and Lima, Marcos and Oyaizu, Hiroaki and Frieman, Joshua and Lin, Huan},
  journal={Monthly Notices of the Royal Astronomical Society},
  volume={396},
  number={4},
  pages={2379--2398},
  year={2009},
  publisher={The Royal Astronomical Society}
}

@article{freeman2017unified,
  author = {Freeman, P. and Izbicki, Rafael and Lee, A. B.},
  title = {{A unified framework for constructing, tuning and assessing photometric redshift density estimates in a selection bias setting}},
  journal = {Monthly notices of the Royal Astronomical Society},
  year = {2017},
  volume = {468},
  number = {4},
  pages = {4556--4565},
  doi = {10.1093/mnras/stx764},
  archiveprefix = {arXiv},
  eprint = {1703.09242},
  primaryclass = {astro-ph.IM},
  publisher = {Oxford University Press (OUP)},
}

@article{Shimodaira2000,
  author = {Shimodaira, Hidetoshi},
  title = {{Improving predictive inference under covariate shift by weighting the log-likelihood function}},
  journal = {Journal of Statistical Planning and Inference},
  year = {2000},
  volume = {90},
  number = {2},
  pages = {227--244},
  doi = {10.1016/S0378-3758(00)00115-4},
  publisher = {Elsevier BV},
}

@article{IzbickiLee2017,
  author = {Izbicki, Rafael and Lee, Ann B.},
  title = {{Converting High-Dimensional Regression to High-Dimensional Conditional Density Estimation}},
  journal = {Electronic Journal of Statistics},
  year = {2017},
  volume = {11},
  number = {2},
  pages = {2800--2831},
  doi = {10.1214/17-EJS1302},
  publisher = {Institute of Mathematical Statistics},
}

@article{Bishop1994,
  title={Mixture density networks},
  author={Bishop, Christopher M},
  year={1994},
  publisher={Aston University}
}

@article{Hollmann2025,
  author = {Hollmann, Noah and Müller, Samuel G. and Purucker, Lennart and Krishnakumar, Arjun and Körfer, Max and Hoo, Shi Bin and Schirrmeister, R. T. and Hutter, Frank},
  title = {{Accurate Predictions on Small Data with a Tabular Foundation Model}},
  journal = {Nature},
  year = {2025},
  volume = {637},
  pages = {319--326},
  doi = {10.1038/s41586-024-08328-6},
  number = {8045},
  publisher = {Springer Science and Business Media LLC},
}

@article{Qu2025,
  author = {Qu, Jingang and Holzmüller, David and Varoquaux, G. and Morvan, Marine Le},
  title = {{TabICL: A Tabular Foundation Model for In-Context Learning on Large Data}},
  journal = {International Conference on Machine Learning},
  year = {2025},
  eid = {arXiv:2502.05564},
  pages = {arXiv:2502.05564},
  doi = {10.48550/arXiv.2502.05564},
  archiveprefix = {arXiv},
  eprint = {2502.05564},
  primaryclass = {cs.LG},
}

@inproceedings{DurkanEtAl2019,
  author = {Durkan, Conor and Bekasov, Artur and Murray, Iain and Papamakarios, G.},
  title = {{Neural Spline Flows}},
  booktitle = {Neural Information Processing Systems},
  year = {2019},
  archiveprefix = {arXiv},
  eprint = {1906.04032},
  journal = {Neural Information Processing Systems},
}

@article{Bickel2009,
  author = {Bickel, S. and Brückner, Michael and Scheffer, T.},
  title = {{Discriminative Learning Under Covariate Shift}},
  journal = {Journal of Machine Learning Research},
  year = {2009},
  volume = {10},
  pages = {2137--2155},
  doi = {10.5555/1577069.1755858},
}

@article{GneitingRaftery2007,
  author = {Gneiting, T. and Raftery, A.},
  title = {{Strictly Proper Scoring Rules, Prediction, and Estimation}},
  journal = {Journal of the American Statistical Association},
  year = {2007},
  volume = {102},
  number = {477},
  pages = {359--378},
  doi = {10.1198/016214506000001437},
}

@inproceedings{lundberg2017unified,
  title = {A Unified Approach to Interpreting Model Predictions},
  author = {Lundberg, Scott M. and Lee, Su-In},
  booktitle = {Neural Information Processing Systems},
  volume = {30},
  year = {2017},
  journal = {Neural Information Processing Systems},
}

@inproceedings{muschalik2024shapiq,
  author = {Muschalik, Maximilian and Baniecki, Hubert and Fumagalli, Fabian and Kolpaczki, Patrick and Hammer, Barbara and Hüllermeier, Eyke},
  title = {{shapiq: Shapley Interactions for Machine Learning}},
  booktitle = {Neural Information Processing Systems},
  volume = {37},
  year = {2024},
  url = {https://proceedings.neurips.cc/paper_files/paper/2024/hash/eb3a9313405e2d4175a5a3cfcd49999b-Abstract-Datasets_and_Benchmarks_Track.html},
  journal = {Neural Information Processing Systems},
  pages = {130324-130357},
  doi = {10.48550/arXiv.2410.01649},
  publisher = {Neural Information Processing Systems Foundation, Inc. (NeurIPS)},
}

@inproceedings{priorlabs_tabpfn_interpretability,
  title={Interpretable machine learning for TabPFN},
  author={Rundel, David and Kobialka, Julius and von Crailsheim, Constantin and Feurer, Matthias and Nagler, Thomas and R{\"u}gamer, David},
  booktitle={World Conference on Explainable Artificial Intelligence},
  pages={465--476},
  year={2024},
  organization={Springer}
}

@article{2008ApJ...683...12B,
  author = {Ball, N. and Brunner, Robert J. and Myers, A. and Strand, N. E. and Alberts, Stacey L. and Tcheng, D.},
  title = {{Robust Machine Learning Applied to Astronomical Data Sets. III. Probabilistic Photometric Redshifts for Galaxies and Quasars in the SDSS and GALEX}},
  journal = {The Astrophysical Journal},
  year = {2008},
  volume = {683},
  number = {1},
  pages = {12-21},
  doi = {10.1086/589646},
  publisher = {American Astronomical Society},
}

@article{Grinsztajn2025TabPFN25,
  author = {Grinsztajn, L'eo and Floge, Klemens and Key, Oscar and Birkel, Felix and Jund, Philipp and Roof, Brendan and Jager, Benjamin and Safaric, Dominik and Alessi, Simone and Hayler, A. and others},
  title = {{TabPFN-2.5: Advancing the State of the Art in Tabular Foundation Models}},
  journal = {arXiv.org},
  year = {2025},
  eid = {arXiv:2511.08667},
  pages = {arXiv:2511.08667},
  doi = {10.48550/arXiv.2511.08667},
  archiveprefix = {arXiv},
  eprint = {2511.08667},
  primaryclass = {cs.LG},
}

@article{Garg2025RealTabPFN,
  author = {Garg, Anurag and Ali, Muhammad and Hollmann, Noah and Purucker, Lennart and Müller, Samuel G. and Hutter, Frank},
  title = {{Real-TabPFN: Improving Tabular Foundation Models via Continued Pre-training With Real-World Data}},
  journal = {arXiv.org},
  year = {2025},
  eid = {arXiv:2507.03971},
  pages = {arXiv:2507.03971},
  doi = {10.48550/arXiv.2507.03971},
  archiveprefix = {arXiv},
  eprint = {2507.03971},
  primaryclass = {cs.LG},
}

@inproceedings{izbicki2014high,
  title={High-dimensional density ratio estimation with extensions to approximate likelihood computation},
  author={Izbicki, Rafael and Lee, Ann and Schafer, Chad},
  booktitle={Artificial intelligence and statistics},
  pages={420--429},
  year={2014},
  organization={PMLR}
}

@article{Breiman2001RandomForests,
  author = {Breiman, L.},
  title = {{Random Forests}},
  journal = {Machine Learning},
  year = {2001},
  volume = {45},
  number = {1},
  pages = {5--32},
  doi = {10.1023/A:1010933404324},
}

@inproceedings{ChenGuestrin2016XGBoost,
  author = {Chen, Tianqi and Guestrin, Carlos},
  title = {{XGBoost: A Scalable Tree Boosting System}},
  booktitle = {Knowledge Discovery and Data Mining},
  year = {2016},
  pages = {785--794},
  doi = {10.1145/2939672.2939785},
  archiveprefix = {arXiv},
  eprint = {1603.02754},
  primaryclass = {cs.LG},
  journal = {Knowledge Discovery and Data Mining},
}

@article{mar,
	title = {{MAR}: {A} {Multiband} {Astronomical} {Reduction} package},
	volume = {51},
	issn = {22131337},
	shorttitle = {{MAR}},
	url = {https://linkinghub.elsevier.com/retrieve/pii/S2213133724001148},
	doi = {10.1016/j.ascom.2024.100899},
	language = {en},
	urldate = {2025-01-18},
	journal = {Astronomy and Computing},
	author = {Schwarz, G.B. Oliveira and Herpich, F. and Almeida-Fernandes, F. and Nakazono, L. and Cardoso, N.M. and Machado-Pereira, E. and Schoenell, W. and Perottoni, H.D. and Menéndez-Delmestre, K. and Sodré, L. and Kanaan, A. and Ribeiro, T.},
	month = apr,
	year = {2025},
	pages = {100899},
}

@article{Martin2005,
  author  = {Martin, D.~C. and others},
  title   = {{The Galaxy Evolution Explorer: A Space Ultraviolet Survey Mission}},
  journal = {\apjl},
  year    = {2005},
  volume  = {619},
  pages   = {L1--L6},
  doi     = {10.1086/426387}
}

@article{Morrissey2007,
  author  = {Morrissey, P. and others},
  title   = {{The Calibration and Data Products of GALEX}},
  journal = {\apjs},
  year    = {2007},
  volume  = {173},
  pages   = {682--697},
  doi     = {10.1086/520512}
}

@article{Bianchi2017,
  author  = {Bianchi, L. and others},
  title   = {{Revised Catalog of GALEX Ultraviolet Sources. I. The All-Sky Survey: GUVcat\_AIS}},
  journal = {\apjs},
  year    = {2017},
  volume  = {230},
  pages   = {24},
  doi     = {10.3847/1538-4365/aa7053}
}

@article{Wright2010,
  author  = {Wright, E.~L. and others},
  title   = {{The Wide-field Infrared Survey Explorer (WISE): Mission Description and Initial On-orbit Performance}},
  journal = {\aj},
  year    = {2010},
  volume  = {140},
  pages   = {1868--1881},
  doi     = {10.1088/0004-6256/140/6/1868}
}

@article{Schlafly2019,
  author  = {Schlafly, E.~F. and others},
  title   = {{The unWISE Catalog: Two Billion Infrared Sources from Five Years of WISE Imaging}},
  journal = {\apjs},
  year    = {2019},
  volume  = {240},
  pages   = {30},
  doi     = {10.3847/1538-4365/aafbea}
}

@ARTICLE{stern2012,
       author = {{Stern}, Daniel and {Assef}, Roberto J. and {Benford}, Dominic J. and {Blain}, Andrew and {Cutri}, Roc and {Dey}, Arjun and {Eisenhardt}, Peter and {Griffith}, Roger L. and {Jarrett}, T.~H. and {Lake}, Sean and {Masci}, Frank and {Petty}, Sara and {Stanford}, S.~A. and {Tsai}, Chao-Wei and {Wright}, E.~L. and {Yan}, Lin and {Harrison}, Fiona and {Madsen}, Kristin},
        title = "{Mid-infrared Selection of Active Galactic Nuclei with the Wide-Field Infrared Survey Explorer. I. Characterizing WISE-selected Active Galactic Nuclei in COSMOS}",
      journal = {\apj},
         year = 2012,
        month = jul,
       volume = {753},
       number = {1},
          eid = {30},
        pages = {30},
          doi = {10.1088/0004-637X/753/1/30},
archivePrefix = {arXiv},
       eprint = {1205.0811},
 primaryClass = {astro-ph.CO},
       adsurl = {https://ui.adsabs.harvard.edu/abs/2012ApJ...753...30S}
}

@ARTICLE{assef2013,
       author = {{Assef}, R.~J. and {Stern}, D. and {Kochanek}, C.~S. and {Blain}, A.~W. and {Brodwin}, M. and {Brown}, M.~J.~I. and {Donoso}, E. and {Eisenhardt}, P.~R.~M. and {Jannuzi}, B.~T. and {Jarrett}, T.~H. and {Stanford}, S.~A. and {Tsai}, C.-W. and {Wu}, J. and {Yan}, L.},
        title = "{Mid-infrared Selection of Active Galactic Nuclei with the Wide-field Infrared Survey Explorer. II. Properties of WISE-selected Active Galactic Nuclei in the NDWFS Bo{\"o}tes Field}",
      journal = {\apj},
         year = 2013,
        month = jul,
       volume = {772},
       number = {1},
          eid = {26},
        pages = {26},
          doi = {10.1088/0004-637X/772/1/26},
archivePrefix = {arXiv},
       eprint = {1209.6055},
 primaryClass = {astro-ph.CO},
       adsurl = {https://ui.adsabs.harvard.edu/abs/2013ApJ...772...26A}
}

@article{Brammer+2008,
doi = {10.1086/591786},
url = {https://doi.org/10.1086/591786},
year = {2008},
month = {oct},
publisher = {},
volume = {686},
number = {2},
pages = {1503},
author = {Brammer, Gabriel B. and van Dokkum, Pieter G. and Coppi, Paolo},
title = {EAZY: A Fast, Public Photometric Redshift Code},
journal = {The Astrophysical Journal},
}

@article{Hildebrandt+2010,
  title = {PHAT: PHoto-z Accuracy Testing},
  volume = {523},
  ISSN = {1432-0746},
  url = {http://dx.doi.org/10.1051/0004-6361/201014885},
  DOI = {10.1051/0004-6361/201014885},
  journal = {Astronomy \& Astrophysics},
  publisher = {EDP Sciences},
  author = {Hildebrandt,  H. and Arnouts,  S. and Capak,  P. and Moustakas,  L. A. and Wolf,  C. and Abdalla,  F. B. and Assef,  R. J. and Banerji,  M. and Benítez,  N. and Brammer,  G. B. and Budavári,  T. and Carliles,  S. and Coe,  D. and Dahlen,  T. and Feldmann,  R. and Gerdes,  D. and Gillis,  B. and Ilbert,  O. and Kotulla,  R. and Lahav,  O. and Li,  I. H. and Miralles,  J.-M. and Purger,  N. and Schmidt,  S. and Singal,  J.},
  year = {2010},
  month = Nov,
  pages = {A31}
}

@article{Brescia2013,
  author  = {{Brescia}, M. and {Cavuoti}, S. and {D'Abrusco}, R. and
             {Longo}, G. and {Mercurio}, A.},
  title   = {{Photometric Redshifts for Quasars in Multi-band Surveys}},
  journal = {The Astrophysical Journal},
  year    = {2013},
  volume  = {772},
  number  = {2},
  pages   = {140},
  doi     = {10.1088/0004-637X/772/2/140}
}

@article{DiPompeo_2015,
   title={Quasar probabilities and redshifts from<i>WISE</i>mid-IR through<i>GALEX</i>UV photometry},
   volume={452},
   ISSN={1365-2966},
   url={http://dx.doi.org/10.1093/mnras/stv1562},
   DOI={10.1093/mnras/stv1562},
   number={3},
   journal={Monthly Notices of the Royal Astronomical Society},
   publisher={Oxford University Press (OUP)},
   author={DiPompeo, M. A. and Bovy, J. and Myers, A. D. and Lang, D.},
   year={2015},
   month=July, pages={3124–3138} }

@ARTICLE{Benitez2014,
author = {{Ben{'\i}tez}, N. and {Dupke}, R. and {Moles}, M. and et al.},
title = "{J-PAS: The Javalambre-Physics of the Accelerating Universe Astrophysical Survey}",
journal = {arXiv e-prints},
year = 2014,
eprint = {1403.5237},
}
